\documentclass[a4paper,12pt]{article}
\pdfoutput=1
\usepackage{amssymb}
\usepackage{amsmath}
\usepackage{epsfig}
\usepackage{graphicx}
\usepackage{xcolor,ulem}
\usepackage{verbatim}

\usepackage[numbers,sort&compress]{natbib}  

\makeatletter
\renewcommand{\theequation}{\thesection.\arabic{equation}}
\@addtoreset{equation}{section}
\makeatother
\def\be{\begin{equation}}
\def\ee{\end{equation}}
\def\bea{\begin{eqnarray}}
\def\eea{\end{eqnarray}}
\def\eq{\begin{eqnarray}}
\def\eqx{\end{eqnarray}}

\def\({\left(}
\def\){\right)}
\def\<{\left<}
\def\>{\right>}

\def\tr{{\mbox{tr}}}
\def\be{\begin{equation}}
\def\ee{\end{equation}}
\def\ben{\begin{eqnarray}}
\def\een{\end{eqnarray}}
\def\({\left(}
\def\){\right)}
\def\<{\left<}
\def\>{\right>}
\def\!{\right|}
\def\|{\left|}

\def\[{\left[}
\def\]{\right]}

\def\+{\bar}

\def\tr{{\mbox{tr}}}

\def\A{{\cal{A}}}
\def\B{{\cal{B}}}

\def\rmd{{\rm d}}

\def\E{{\cal{E}}}

\def\tth{{\negthinspace}}

\usepackage{hyperref}
\hypersetup{colorlinks,%
  citecolor=blue,%
  linkcolor=blue,%
  urlcolor=black,%
  filecolor=red,%
 pdftex}

\begin{document}

\begin{titlepage}
\vskip1cm
\begin{flushright}
\end{flushright}
\vskip0.25cm
\centerline{
\bf \large 
Unitarity of Entanglement and Islands in Two-Sided Janus Black Holes
} 
\vskip1cm \centerline{ \textsc{
 Dongsu Bak,$^{ \negthinspace  a,d}$  Chanju Kim,$^{ \negthinspace b}$ Sang-Heon Yi,$^{\negthinspace a}$ Junggi Yoon$^{\,  c}$} }
\vspace{1cm} 
\centerline{\sl  a) Physics Department,
University of Seoul, Seoul 02504 \rm KOREA}
 \vskip0.3cm
 \centerline{\sl b) Department of Physics, Ewha Womans University,
  Seoul 03760 \rm KOREA}
   \vskip0.3cm
 \centerline{\sl c) School of Physics, Korea Institute for Advanced Study}
 \centerline{\sl 85 Hoegiro, Dongdaemun-ku, Seoul 02455 \rm KOREA}
   \vskip0.3cm
 \centerline{\sl d) Natural Science Research Institute,
University of Seoul, 
Seoul 02504 \rm KOREA}
 \vskip0.4cm
 \centerline{
\tt{(\small dsbak@uos.ac.kr,\,cjkim@ewha.ac.kr,\,shyi704@uos.ac.kr,\,junggiyoon@kias.re.kr})} 
  \vspace{2cm}
\centerline{ABSTRACT} \vspace{0.75cm} 
{
\noindent 
	We explore the entanglement evolution of  boundary intervals in 
eternal Janus black holes  that can be embedded consistently into string theory in the 
low-energy limit. By studying the geodesics we show that there is a transition  in the entanglement characteristic around the Page time, which manifests the unitarity of the evolution. We reproduce and reinterpret these bulk results from 
two different lower-dimensional perspectives: first as 
an interface CFT  in the usual AdS/CFT correspondence and second as an effective gravity theory in one lower dimension coupled to a radiation background. In the limit where the number of interface degrees of freedom becomes large, we obtain an effective theory 
 on appropriate branes that replace the deep interior region in the bulk, coined 
the shadow region. In this effective  theory, we also identify the island of the radiation entanglement wedge and verify the newly proposed quantum extremization  method. 
Our model clarifies that  double holography with gravity in two higher dimensions can be realized in a concrete and consistent way and that  the occurrence of islands is natural in  one higher dimension. Furthermore, our model reveals that there can be a transitional behavior of  the Page curve before the Page time, which is related  to the emergence of new matter degrees of freedom on the branes. 

}

\end{titlepage}


\section{Introduction}
There has, for a  long time, been a riddle in black hole physics named the {\it black hole information paradox}, which was embarked on by    Hawking's semi-classical computation on the particle creation in the black hole background~\cite{Hawking:1974sw,Hawking:1974rv}. This issue was initially incurred by the  apparent result that black holes behave as thermal objects with the erased information of in-falling matter forming the black holes.  Though there have been numerous attempts to resolve this issue, the complete settlement of the paradox has not yet been achieved and  the consensus of the status of the problem has not even been reached.  In fact, opposite opinions have been made on  whether the information can be destroyed or not. See~\cite{tHooft:1996rdg,Mathur:2005zp,Polchinski:2016hrw,Unruh:2017uaw,Ashtekar:2020ifw,Maldacena:2020ady,Almheiri:2020cfm} for a review. 

Even though some physicists~\cite{Hawking:1982dj,Jacobson:2003vx,Unruh:2017uaw} argue that gravity or the curved spacetime allows the evolution of the pure state to the mixed one, most string theorists and  AdS/CFT practitioners prefer the preservation of unitarity of  quantum mechanics. Based on the validity of the AdS/CFT correspondence, the bulk physics is argued to be unitary since it is equivalent to a unitary boundary theory.  However, this statement does not provide a clear picture  of what happens with the locality assumption in field theory, which is taken as a valid approximation near the horizon far from the black hole singularity. In other words, the present understanding of the AdS/CFT correspondence does not provide an answer to what is going wrong in  Hawking's results nor  how the bulk locality can be realized in the AdS/CFT context.

From time to time,  this information problem is reincarnated  in disguise.  One of the recent reformulation is based on the entanglement characteristic of the Hawking radiation. Roughly speaking, the clash between  unitarity  and the semi-classical approximation  in the near horizon region may be phrased as the seeming bigamy of the late Hawking radiation with the behind the horizon degrees of freedom and with the early  Hawking radiation, which violates the monogamy of the entanglement in quantum mechanics. Proposed  resolutions of this problem are to either abandon the semi-classical features by introducing a  high energy curtain (firewall)~\cite{Almheiri:2012rt,Almheiri:2013hfa}, or to preserve the semi-classical picture but to make a bit bold and clever  identification between the behind horizon degrees of freedom and  the   emitted Hawking radiation (ER=EPR)~\cite{Maldacena:2013xja}.  Though these proposals evade the apparent contradiction, the information loss problem  is still far from understood. For instance, the Page curve~\cite{Page:1993wv} for the entanglement evolution of the Hawking radiation needs to be explained in an appropriate way. 

Very recently,  an explicit computation for the entanglement of the Hawking radiation in the Jackiw-Teitelboim (JT) model~\cite{Jackiw:1984je,Teitelboim:1983ux}   was performed and the entanglement evolution  was shown to follow the Page curve by using the  quantum extremal surface~(QES) prescription~\cite{Lewkowycz:2013nqa,Faulkner:2013ana,Engelhardt:2014gca} for the entanglement wedge~\cite{Almheiri:2019psf} (See also~\cite{Penington:2019npb}). In a more recent work~\cite{Almheiri:2019hni}, the missing ingredient in  gravity theory to the entanglement computation was clearly identified  and called as the island of the entanglement wedge.  A new prescription was proposed  to extremize  the generalized entropy including the contribution of islands.
Furthermore,  the QES was argued to become the ordinary Ryu-Takayanagi surface~\cite{Ryu:2006bv,Hubeny:2007xt} in one-higher-dimensional holographic setup. 

In this paper, we take a consistent top-down approach to a one-higher-dimensional holographic model, known as the holographic dual to interface CFT (ICFT)~\cite{Bak:2003jk}. This model contains solutions known as the  Janus black holes~\cite{Bak:2011ga}, which is our main concern in the following. Since this model may be embedded consistently into  string/M-theory as a low energy limit, our analysis might be extended to the full string theory level. We describe the low dimensional gravity coupled to the thermal radiation from the viewpoint of an effective reduction of a higher dimensional gravity in the limit of a large number of interface degrees of freedom.   Our effective 2d gravity description is realized by introducing a  boundary action of a surface in the deep interior  and by replacing its behind-the-interior region, coined   the shadow region, by 
a
brane-like boundary surface.
    Our model provides a different perspective  compared to the bottom-up brane models~\cite{Rozali:2019day,Balasubramanian:2020hfs,Sully:2020pza,Geng:2020qvw,Chen:2020uac} in the sense that the usual holography method is applied  with some additional machinery.    

Our paper is organized as follows. In Section~\ref{sec2}, we provide some details of our model and of Janus black hole solutions. Especially,  various coordinates in our setup are introduced for later convenience. In Section~\ref{sec3}, the entanglement entropy is briefly reviewed and the holographic entanglement entropy~(HEE)  is computed for RR/LL   geodesics. In Section~\ref{sec4}, we compute the HEE for RL geodesics. We present the Page curve of our black hole model and the unitarity  of the entanglement  in Section~\ref{sec5}. In this section, we also comment on the evolution of the mutual information of the Hawking radiation in our setup. In Section~\ref{sec6}, we provide the two-dimensional gravity interpretation  and reproduce our results from the generalized entropy extremization procedure.  In Section~\ref{sec7}, we provide another perspective on our results by using ICFT, which is the boundary viewpoint of our model.  In Section~\ref{sec8}, we provide  interesting new aspects to the entanglement island picture by using our model. In this section, we provide clues to a slope change behavior of the Page curve before the Page time and interpret our computation as the emergence of new matter degrees of freedom on the branes or the appearance of its corresponding boundary operators in ICFT. We also provide another computation supporting this interpretation. 
In the conclusion, we summarize our results and present some future directions. In Appendices~\ref{AppA} and~\ref{AppB},  we provide some detailed formulae and explain the effective CFT$_{2}$ viewpoint, respectively. 
 Some details of our 2d effective gravity description of the shadow region are relegated to Appendix~\ref{AppC}.

\section{Janus black holes in three dimensions}\label{sec2}
In this section, we shall investigate  holography of thermo-field double~(TFD)  of  Janus ICFT$_2$, whose gravity dual 
is the two-sided version of a 3d Janus black hole.  
It is known that the Janus  geometry arises as a classical solution to the system of Einstein gravity with a negative cosmological constant coupled to a minimal massless scalar field
whose action is given by 
\be
I_{\rm gravity} 
= \frac{1}{16\pi G}\int_{M_{d+1}} d^{d+1} x \left[ R -g^{ab} \partial_a \phi \partial_b \phi + \frac{d (d-1)}{\ell^2} \right] \,.
\label{einstein}
\ee
The Janus geometry can be found for arbitrary dimensions. For $(d+1) = 3$ and $5$, this system can be consistently embedded into  Type IIB supergravity and hence, via the AdS/CFT correspondence, the microscopic understanding of the dual ICFT$_d$ system can be obtained \cite{Bak:2003jk, Bak:2007jm}. The scalar field here originates from the dilaton field of the underlying Type IIB supergavity and hence it is holographically dual to the CFT$_d$ Lagrangian density.  The equations of motion read
\begin{align}    \label{}
& g^{ab} \nabla_a \nabla_b \phi = 0 \,,\nonumber \\
& R_{ab} = -\frac{d}{\ell^2} g_{ab} +\partial_a \phi \partial_b \phi\,. \label{ein}
\end{align}
The vacuum solution is  AdS$_{d+1}$ space with curvature radius $\ell$ and an everywhere constant scalar field. The Janus geometry is a nontrivial domain-wall solution in which the scalar field and metric approach those of the vacuum solutions. Below, 
we specialize to  three dimensions for simplicity.

The three-dimensional Janus solution is given by \cite{Bak:2007jm}
%
%
\begin{align}   
 \rmd s^2 = &~ \ell^2 \left[ \rmd y^2 + f(y) \, \rmd s^2_{\rm AdS_2}\right]\,, \nonumber \\
 \phi(y)  =  &~ \frac{1}{\sqrt{2}} \ln\left( 
\frac{1+\sqrt{1-2\gamma^2} +\sqrt{2} \gamma \tanh y}
{1+\sqrt{1-2\gamma^2} -\sqrt{2} \gamma \tanh y }
\right)\,,
\label{3djanus}
\end{align}
where 
\bea
f(y)= \frac{1}{2}\(1+\sqrt{1-2\gamma^2} \cosh 2y \) \qquad \mbox{with} \qquad
\gamma < \frac{1}{\sqrt{2}}. \label{eq: def f}
\eea
As $y \rightarrow \pm \infty$, the value of the scalar field approaches $\pm \phi_{\rm as}$ where $\phi_{\rm as} = \frac{1}{\sqrt{2}} {\rm arctanh} \sqrt{2}\gamma$.
For our two-sided Janus black hole,  we choose the AdS$_2$ part as the global AdS$_2$
\bea 
\rmd s^2_{\rm AdS_2}= \frac{d\lambda^2-d\tau^2}{\cos^2 \lambda} =
d q^2 -\cosh^2 q \, d\tau^2\,,   \qquad \frac{1}{\cos\lambda} = \cosh q\,, \label{qads2}
\eea 
where 
$\lambda \in (-\lambda_\infty, \lambda_\infty)$ with $\lambda_\infty$ ranged over $[0,\frac{\pi}{2})$ and $q \in (-q_\infty, q_\infty)$ with  $q_{\infty}$ over $[0,\infty)$.
The  R and L   boundaries are at $|y|=\infty$ with  R/L boundary coordinates $(\tau,  +/- \lambda_\infty)$   (See below for the details). 
One may introduce the coordinate $\mu$ defined by 
\bea
\rmd \mu = \frac{\rmd y}{\sqrt{f(y)}}\,, 
\label{muy}
\eea
which is ranged over $[-\mu_0, \mu_0]$ with the boundary value 
$\mu_0 = \int^\infty_0  \frac{dy}{\sqrt{f(y)}}$.
One can evaluate the integral exactly to
\bea
\mu_0 =
\frac{1}{\kappa_+}
{\bf K}\( 
\frac{\kappa_-^2}{\kappa_+^2}
\)=\frac{\pi}{2} \(1+ \frac{3}{8}\gamma^2 +{\cal O}(\gamma^4) 
\)\,,
\eea
where ${\bf K}(x)$ is the first kind of complete elliptic integral 
and
$\kappa^2_\pm =\frac{1}{2}\(1\pm\sqrt{1-2\gamma^2}\)$.
%
In the last equality, we presented its Taylor expansion with respect to the deformation parameter $\gamma$. From this, one finds $\mu_0 \ge \frac{\pi}{2}$, which is a consequence of the deformation. In this coordinate system, the metric becomes
\bea
 \rmd s^2=\ell^2 f (\mu)\left[ \rmd \mu^2 +  \, \rmd s^2_{\rm AdS_2}\right]  \,.                  
\label{3djmetric}
\eea  

For the  Rindler-type solution for the right/left  wedge, we perform a coordinate transformation
\bea
w= \frac{\cos \tau}{\cos \lambda},\ \ \ \ \tanh \frac{L t}{\ell^2}= \frac{\sin \tau}{\sin \lambda}\,,
\label{twcoor}
\eea
and then the AdS$_2$ metric is replaced by the Rindler metric
\bea 
\rmd s^2_{\rm AdS_2}= -\frac{(w^2-1) L^2}{\ell^4} dt^2 +\frac{dw^2}{w^2-1}\,.
\label{wads2}
\eea 
This leads to the Rindler-type Janus black hole solution \cite{Bak:2011ga}
where the horizon is located at $w=1$ with the horizon size $L$ and $w \in [1, \infty)$ describes the region outside horizon.

\begin{figure}[htb!]
\centering  
\includegraphics[height=6.5cm]{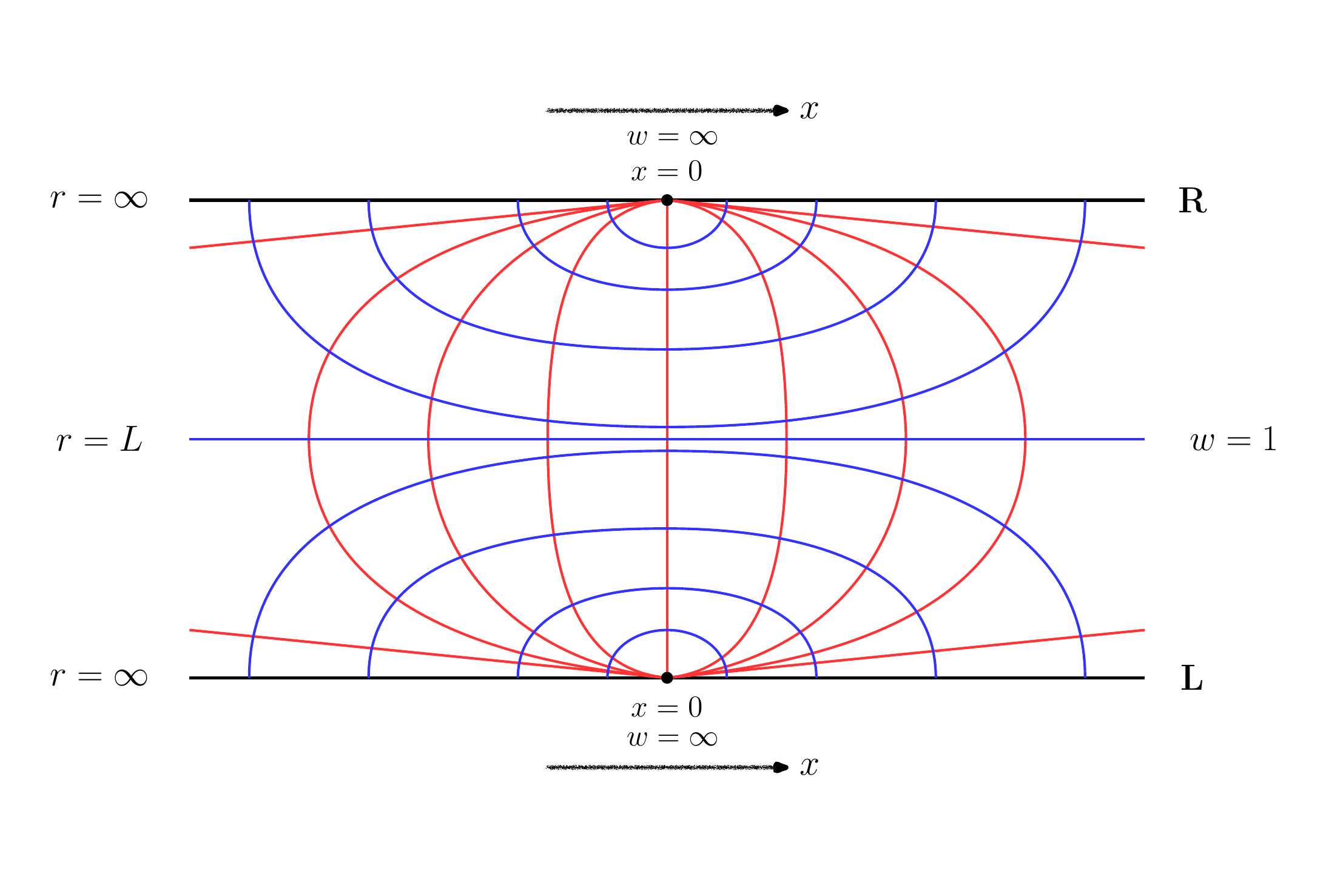} 
\caption{\small 
We draw the constant $t$ section of the BTZ spacetime where we show $(\mu, w)$ together with $(r,x)$ coordinates. The middle line with $w=1$ and $r=L$ represents the horizon. The red lines are representing constant $\mu$ surfaces whereas the blue lines are  constant $w$ surfaces. The top/ bottom line represents the spatial direction of  the R/L boundary respectively.
}
\label{fig1}
\end{figure}

Without deformation, $\gamma =0$, one has the standard planar BTZ black hole~\cite{Banados:1992wn},  
given by the geometry
\bea
\rmd s^2= \ell^2 \left[ \rmd y^2 + \cosh^2 y \, \rmd s^2_{AdS_2}\right] \,,
\label{btz1}
\eea
and, by integrating (\ref{muy}), one has
\be 
\cos \mu = \frac{1}{\cosh y}\,.
\ee
Indeed, through the coordinate transformation
\bea
 \frac{L}{r} = \frac{\cos \mu}{\sqrt{w^2- \sin^2\mu}}\,, \qquad 
\sinh \frac{Lx}{\ell^2}&=& \frac{\sin \mu}{\sqrt{w^2- \sin^2\mu}}\,,
\eea
the $\gamma = 0$ geometry is reduced to the conventional form of the planar BTZ metric 
\bea
\rmd s^2= -\frac{(r^2-L^2)}{\ell^2} \rmd t^2 + \frac{ \ell^2}{r^2-L^2}\rmd r^2 +\frac{ r^2 }{\ell^2} \rmd x^2  .
\label{btz2}
\eea
One finds that both 
coordinates $\tau$ and $\mu$ are ranged over 
$[-\frac{\pi}{2}, \frac{\pi}{2}]$
and 
$x$ can be compactified as $x \sim x + L_s$ since the system possesses a translational symmetry in the $x$  direction.  With the Janus deformation breaking the translational symmetry, we shall be concerned with the planar Janus black holes 
where the spatial extent is noncompact as the size $L_s$ goes to infinity. 

In  Figure \ref{fig1}, we depict the shape of a constant $t$ slice of the undeformed BTZ geometry where we show $(\mu, w)$ together with $(r,x)$ coordinates. The top/bottom line represents the spatial direction of  the R/L boundary spacetime. 
The R/L boundaries are parametrized either by $(t,x)$ or by $(\tau, \pm\lambda_{\infty})$ where the two coordinate systems are related by
\bea
&& \tanh  \frac{Lx}{\ell^2} = \frac{{\epsilon}(x)}{w}= {\epsilon}(x) 
\frac{\cos\lambda_\infty}{\cos \tau}\,,  \cr
&& \tanh  \frac{Lt}{\ell^2}\, = \pm 
\frac{\sin \tau \,\,}{\sin\lambda_\infty} \,,
\label{txbcoor}
\eea
where ${\epsilon}(x)$ denotes the sign function of $x$ and the two signs  $+/-$ refer to the R/L boundary spacetime, respectively. These relations
may be inverted as
\bea
\tan \lambda_\infty = \frac{\cosh  \frac{Lt}{\ell^2}\,\,}{\sinh  \frac{L|x|}{\ell^2}}, \qquad  \tan \tau = \frac{\sinh  \frac{Lt}{\ell^2}\,\,}{\cosh  \frac{L x}{\ell^2}} \,,
\label{invb}
\eea
which will be useful in the following.
The middle 
line corresponds to the horizon location with  $w=1$ (or $r=L$). The red lines are denoting constant $\mu$
trajectory where the $\mu$ coordinate runs over $[-\pi/2, \pi/2]$ for the  BTZ geometry. The blue curves represent constant $w$ surfaces. A few comments are in order.  The first is the well-known time translation isometry of the BTZ geometry. 
This leads to the boundary time translational symmetry $t \rightarrow t + dt$ with $t_{R/L}= \pm t$ where $t_{R/L}$ denotes the time coordinate of R/L boundary respectively, and both go in the positive direction in our choice.
Secondly, the region of $x\rightarrow \infty$ merges to a point at the spatial section of the boundary of global AdS$_3$ geometry and the same is true for the region of $x\rightarrow -\infty$. Hence the R and L boundaries  form a single boundary that is the boundary of the global 
AdS$_3$ spacetime \cite{Hartman:2013qma}. Nonetheless, 
the R and L boundary theories are causally disconnected  from each other completely.

\begin{figure}[htb!]
\centering  
\includegraphics[height=6.5cm]{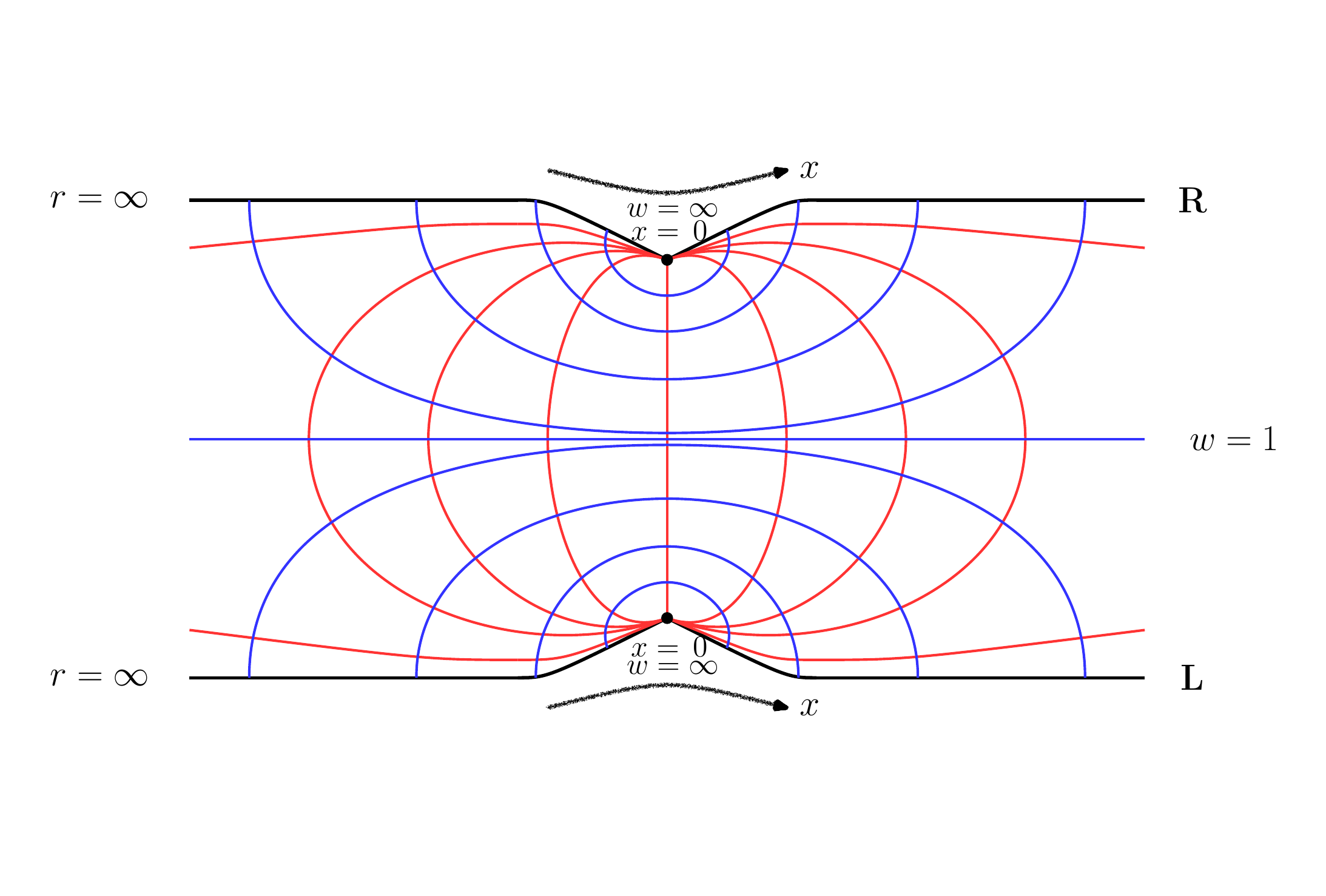} 
\caption{\small 
We draw the constant $t$ section of the two-sided  Janus black hole where we show $(\mu, w)$ together with $(r,x)$ coordinates. The middle line with $w=1$ represents the horizon. The red lines are representing constant $\mu$ surfaces whereas the blue lines are  constant $w$ surfaces. The $\mu$ coordinate is ranged over $[-\mu_0,\mu_0]$ with $\mu_0 > \pi/2$. This leads to the $x=0$ angled-joints 
of the R-L boundaries. 
}
\label{fig2}
\end{figure}

In Figure \ref{fig2}, we draw also  a constant $t$ spatial section of the Janus 
black hole spacetime. This geometry is asymptotically AdS;  one may map  its asymptotic region 
to that of 
the BTZ spacetime. The coordinates in this region can be identified as
\bea 
 \frac{r}{L}\ \ &\simeq& \sqrt{(w^2-1)f+ 1}\,,\cr
\sinh \frac{Lx}{\ell^2}&\simeq& {\epsilon}(x)\frac{\sqrt{f-1}}{\sqrt{(w^2- 1)f+1}}\,, \label{cutoffRel}
\eea
with $t$ and $w$ coordinates defined by (\ref{twcoor}). The boundary coordinates
$(t, x)$ are defined by (\ref{txbcoor}) from which the inverse  in (\ref{invb})  follows. In Figure \ref{fig2}, the red lines are for the constant $\mu$ surface where $\mu$ is ranged over $[-\mu_0,\mu_0]$ with 
$\mu_0 >\pi/2$ as a result of deformation. This leads to the angled-joints
at $x=0$ of Fig.~\ref{fig2}, where  each interface of ICFT$_{R/L}$ is located respectively \cite{Bak:2007jm, Bak:2011ga}. 
 
 The Gibbons-Hawking temperature of the Janus black hole can be identified from
the Euclidean version of the solution  obtained by Wick rotation $t= -i t_E$. By requiring the regularity of this Euclidean geometry at $w=1$, 
 one finds 
\be  \label{beta}
T= \frac{L}{2\pi \ell^2}\,,
\ee
which agrees with that of the undeformed BTZ black hole. The mass of the system can be obtained by studying the holographic stress-energy tensor leading to \cite{Bak:2011ga}
\be 
E= \frac{c}{6} \pi T^2 L_s\,,
\ee
where $c=\frac{3\ell}{2G}$ is the central charge of the boundary ICFT and we take 
the system size $L_s$ large enough.  For simplicity, we take an interval $x \in [-L_s/2, L_s/2]$
with $x=0$ the place where the interface is located.
Similarly, the Bekenstein-Hawking entropy 
of the system can be obtained as \cite{Bak:2011ga}
\be 
S=  \frac{c}{3} \pi T L_s + S_I\,,
\label{3entropy}
\ee 
where the interface contribution $S_I$ is
\be
S_I = \frac{c}{6} \ln A \,. 
\ee
Here, $A$ denotes a bulk parameter defined by
\begin{equation} \label{ConA} 
A \equiv \frac{1}{\sqrt{1- 2\gamma^2}}=\cosh \sqrt{2}\phi_{\rm as} \,.
\end{equation}
 This interface entropy is a measure of  the interface QM degrees of freedom and  
the corresponding  number of ground states  is given by $e^{S_I}$ \cite{Bak:2011ga}.

\begin{figure}[htb!]
\centering  
\includegraphics[height=6cm]{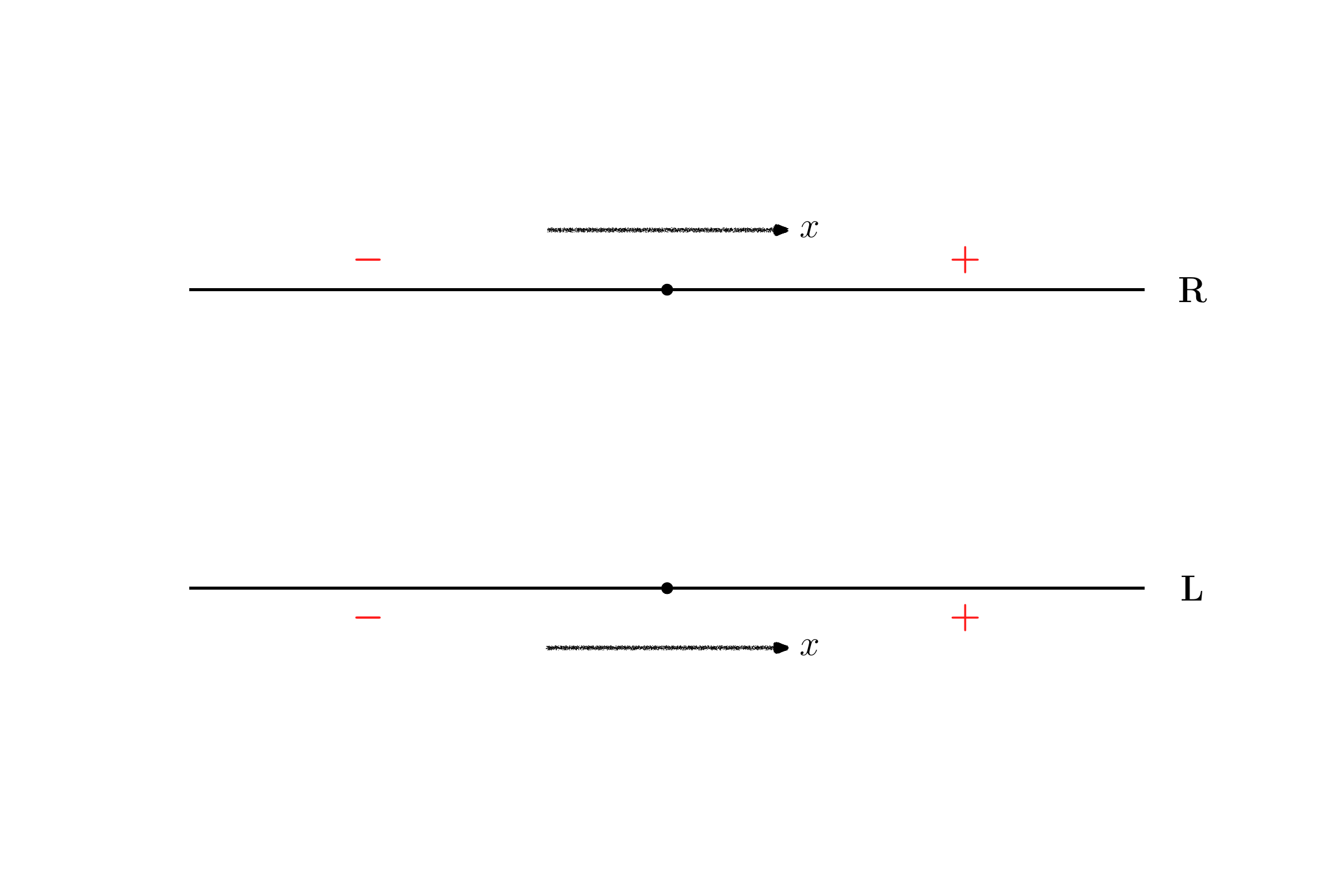} 
\caption{\small 
We draw ICFT$\, \times \,$ICFT living on the L and R boundaries of our two-sided Janus black hole. Each ICFT consists of  three components of CFT$_- \times {\rm QM}_0 \times {\rm CFT}_+$, which preserves 1d conformal symmetries of $SO(1,2)$.
}
\label{fig3}
\end{figure}

Let us now describe the dual field-theory side. The bulk scalar field is dual to an exactly marginal 
scalar operator $O(x,t)$. The boundary value of the scalar field implies turning on the 
operator $O$ with a source term: The CFT$_2$ is deformed by the perturbation  $\int d^2x \,  g({\epsilon}(x)\phi_{\rm as})  O(x,t)$ which breaks $x$ translation invariance explicitly.   One has in general 
$g(z) = z + O(z^2)$, which can be identified to all orders in our 
AdS/CFT correspondence. This basically leads to an ICFT  
\be \label{ICFTdecom}
{\rm ICFT} = {\rm CFT}_- 
\times {\rm QM}_0 
\times {\rm CFT}_+\,,
\ee
where QM$_0$ denotes the quantum mechanical system of the interface degrees of freedom. This system preserves 
the one dimensional  conformal symmetries of $SO(1,2)$.  See Figure \ref{fig3}.

For our two-sided Janus black hole, one has R and L ICFT theories at the same time
ICFT$ \, \times \, $ICFT, which are initially entangled in a particular manner.
Following the Hartle-Hawking construction of the wave function \cite{Maldacena:2001kr}, one gets a  TFD initial state
\be
|\psi (0,0) \rangle  = \frac{1}{\sqrt{Z}}\sum_{n} e^{-\frac{\beta}{2} E_n} | n\rangle \otimes | n\rangle\,,
\label{initial}
\ee
where 
$|n\rangle $ is the energy eigenstate of the ICFT Hamiltonian $H$ with the energy eigenvalue $E_n$.
The subsequent Lorentzian time evolution is then given by
\bea
|\psi (t_{\rm L}, t_{\rm R})\rangle= e^{-i (t_{L} H\otimes I  +t_{R} I \otimes H )} |\psi (0,0)\rangle \,.
\label{tfda}
\eea
This gives a desired TFD  of our ICFT, which will serve as our main framework in the field theory side. It is clear that the state with $t_R=-t_L =t$ is $t$-independent which is consistent with the time-like Killing symmetry  of our black hole geometry. In this work we shall be interested in the time evolution  with $t_L=t_R= t$ which indeed becomes nontrivial.

\begin{figure}[htb!]
\centering  
\includegraphics[height=5cm]{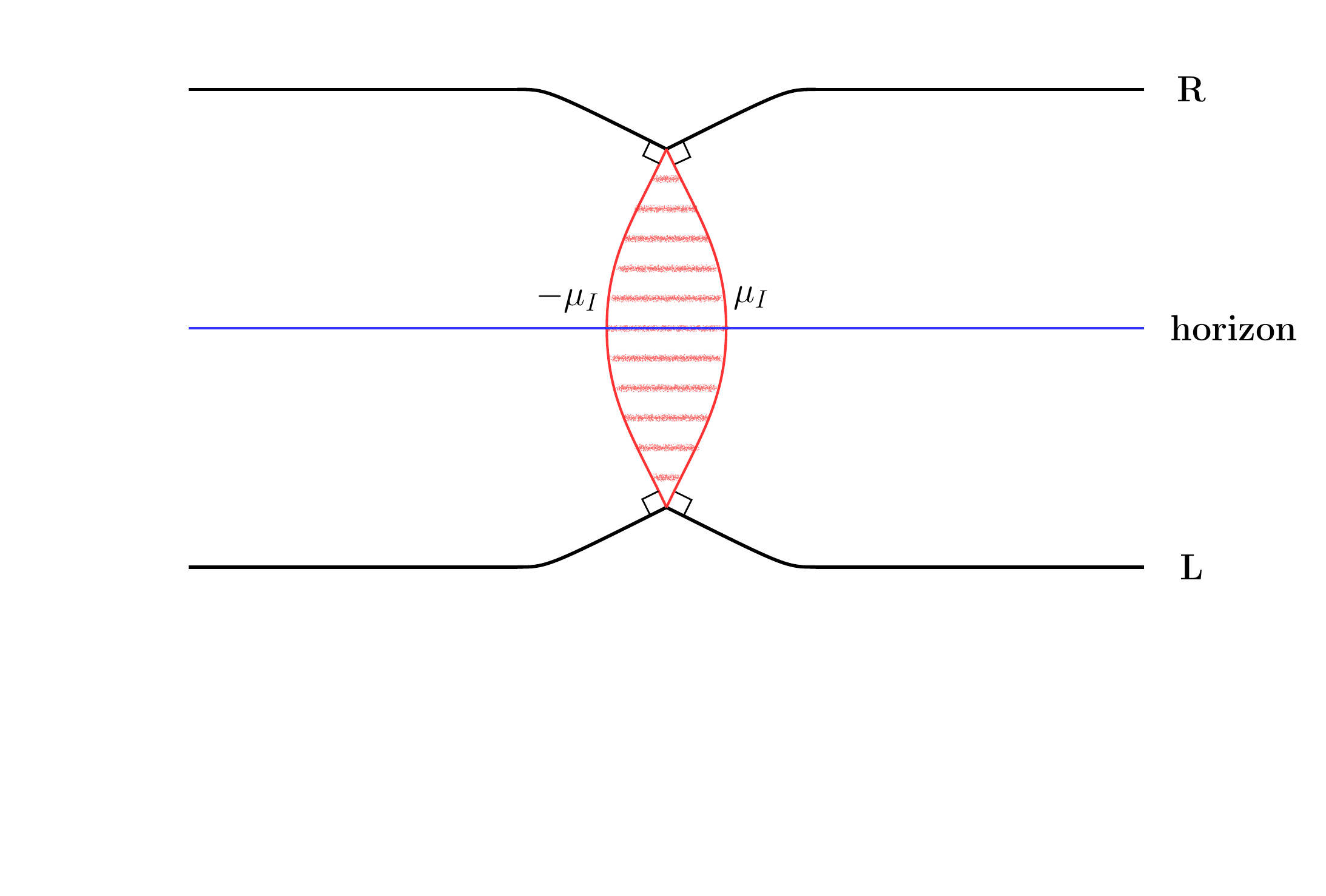} 
\vskip-1.2cm
\caption{\small 
We draw the shadow region specified by $\mu = constant$ slices ranged over
$\mu \in [-\mu_I, \mu_I]$. One may integrate out the bulk degrees of freedom in this shadow region and view the resulting  2d gravity theories as living on  the two slices  at $\mu=\pm \mu_I$, respectively.
}
\label{fig4}
\end{figure}

Finally let us  briefly comment upon a shadow region of interfaces in the bulk. Note that, in our Janus black hole 
system, the $x$-translational symmetry is 
broken by the interfaces 
and the ``entropy density'' becomes
$x$-dependent. In our geometrical setup, the entropy is defined on the horizon side and, hence, 
one needs a map which relates the boundary coordinate $x$ to the horizon coordinate. We use here the boundary horizon map based on null geodesics emanating from the boundary in a hypersurface-orthogonal manner, whose details are
described in \cite{Bak:2011ga}. For a given boundary point $x$, the horizon coordinate can be identified as \cite{Bak:2011ga}
\be
\mu_H = \left(\mu_0 -\frac{\pi}{2}\right) \epsilon(x) + {\rm arctan} \sinh \frac{2\pi x}{\beta}\,.
\ee
With this boundary horizon map, one finds there is an excluded region specified by
$-\mu_I\le \mu \le \mu_I$ where $\mu_I$ equals to $\mu_0 -\frac{\pi}{2} = \frac{3\pi}{16}\gamma^2 + O(\gamma^4)$. This excluded region may be regarded as
an extra bulk space created and affected by the interfaces, which shall be dubbed as the  shadow
of the interfaces. However, this shadow region is not sharply defined as we shall discuss further below. For our later purpose, we shall choose $\mu_I$ (and the corresponding shadow) as\footnote{This choice will be used in later sections where we are mainly interested in the regime $A \gg 1$. Especially in Section \ref{sec6}, we shall compare the 3d off-shell description of extremal curves with  the 2d one  and have a good agreement of two results in the large $A$ limit with this 
particular choice.}
\be 
\mu_I = \mu_0 -\left[\frac{\pi}{2} +
\left(  \sqrt{2}\ln(e+\sqrt{e^2-1})- \frac{\pi}{2} \right) \tanh^2 (A-1)
\right]\,,
\label{mui}
\ee
instead of $\mu_0-\frac\pi2$.
We depict this shadow 
in Figure \ref{fig4}.  

One may integrate out the bulk degrees of freedom in this shadow to get 2d gravity theories Grav$_\pm$ defined on $\mu=\pm \mu_I$ slices whose dual quantum mechanical systems 
may be denoted by QM$_{\pm}$ respectively. Therefore, one may alternatively view our ICFT as
\be 
{\rm ICFT} = 
{\rm CFT}_- \times {\rm QM}_-\times {\rm QM}_+ \times {\rm CFT}_+\,.
\ee  
In this manner, one may get a picture of 2d  gravities coupled to 2d CFT's. As will be clarified later on, the separation of Grav$_+$ and Grav$_-$ becomes apparent only in the limit where $A$ becomes large.

\section{Entanglement of an interval}\label{sec3}

\subsection{Review: Entanglement Entropy}

In this section, we shortly review the entanglement entropy. Let us consider a bi-partite system $\mathcal{H}=\mathcal{H}_A\otimes \mathcal{H}_B$. From the reduced density matrix $\rho_A=\tr_{\tiny B} \rho$ of the subsystem A, the entanglement entropy of the subsystem A is given by
\begin{equation}
	S_{EE}=-\ln \bigg( \rho_A\ln \rho_A \bigg) \,.
\end{equation}
In general, it is difficult to evaluate this entanglement entropy because of the logarithm of the density matrix. Instead, we evaluate $\tr \rho_A^n$ and take a limit to get the entanglement entropy
\begin{equation}
	S_{EE}=- \lim_{n\rightarrow 1} {\partial\over \partial n} \tr (\rho_A^n)\,.
\end{equation}
The trace of the $n$th power of the density matrix can be evaluated by $n$ replicas of the original system~\cite{Calabrese:2004eu}. The boundary condition of the replica trick for the subsystem A can be incorporated by twist operators $\Phi_{n}^\pm(z)$, and $\tr \rho^n_A$ can be computed by inserting twist operators at the end of the interval ${\cal I}_A$ with length $L_0$
\begin{equation}
	\tr \rho_A^n =\langle \Phi^+_n(z)\Phi^-_n(w) \rangle={1\over \bigg[ {\beta\over 2\pi \varepsilon} \sinh \bigg({\pi L_0 \over \beta}\bigg) \bigg]^{2\Delta_n}}\,,
\end{equation}
where $\Delta_n$ is the conformal dimension of the twist operator $\Phi^n_\pm$ 
\begin{equation}
	\Delta_n={c\over 12} \bigg(n-{1\over n}\bigg)\,.
\end{equation}
Thus
 the entanglement entropy of the interval ${\cal I}_A$  is found to be
\begin{equation}
	S_{EE}= S_{\varepsilon}+ {c\over 3} \ln \bigg[ 2 \sinh \bigg({\pi L_{0} \over \beta}\bigg) \bigg]\,,
\end{equation}
where $S_{\varepsilon}$ corresponds to the contribution of the short distance degrees of freedom with the cutoff scale $\varepsilon$ as 
\begin{equation}
	S_{\varepsilon}\equiv {c\over 3} \ln \bigg[{\beta\over 4\pi \varepsilon}\bigg]\,.
\end{equation}
In a boundary CFT~(BCFT) or ICFT, the degrees of freedom living on the boundary or the interface give a contribution to the entanglement entropy as
\begin{equation}
	S_{EE}=S_{\varepsilon}+{c\over 3} \ln \bigg[ 2 \sinh \bigg({\pi L_0 \over \beta}\bigg) \bigg]+  \ln g\,,
\end{equation}
where $\ln g$ is the boundary entropy~\cite{Affleck:1991tk}. In the Janus ICFT, the boundary entropy can be evaluated from the two point function of the twist operators by the conformal perturbation
\begin{equation}
	\langle \Phi_+(z)\Phi_-(w)  \rangle_{\gamma}= \langle \Phi_+(z)\Phi_-(w)  \rangle +\gamma \int d^2x\;  \epsilon(x^1) \langle \Phi_+(z)\Phi_-(w)  \mathcal{O}(x) \rangle + \mathcal{O}(\gamma^2)\,.
\end{equation}
The leading contribution of order $\mathcal{O}(\gamma)$ is universal up to OPE coefficient because of the universal form of three point function
\begin{align}
	&\gamma \int d^2x\;  \epsilon(x^1) \langle \Phi_+(z)\Phi_-(w)  \mathcal{O}(x) \rangle = \gamma \int {d^2x\;  \epsilon(x^1) \mathcal{C}_{\Phi_+\Phi_-\mathcal{O} }\over (z-w)^{2\Delta_n-2}(z-x)^2(w-x)^2}\,.
\end{align}
However, this gives a correction of order $\mathcal{O}(c^0)$ to the conformal dimension and the normalization of the two point function of the twist operators. Hence, in large $c$ limit, the non-zero correction to the entanglement entropy is of order $\mathcal{O}(\gamma^2)$.
\begin{equation}
	S_{EE} = S_{\varepsilon}+{c\over 3} \ln \bigg[ 2 \sinh \bigg({\pi L_0 \over \beta}\bigg) \bigg] +\mathcal{O}(\gamma^2c) \,.
\end{equation}
Though the $\mathcal{O}(\gamma^2)$ correction is not universal, one can deduce it from the conformal perturbation of the free energy of the Janus ICFT
\begin{align}
	\beta F=-\ln Z= \beta \mathcal{F} -\ln g=\beta F_0 + \gamma^2 \beta F_2+\cdots\,.
\end{align}
The $\mathcal{O}(\gamma^2)$ correction is found to be~\cite{Bak:2011ga}
\begin{align}
	\gamma^2\beta F_2 = {1\over 2}\gamma^2\int d^2x d^2y \; \epsilon(x^1)\epsilon(x^1) \langle \mathcal{O}(x)\mathcal{O}(y) \rangle = - {\ell  \over 4 G}\gamma^2= - {c\over 6}\gamma^2\,,
\end{align}
and, this leads to
%
\begin{equation}
	\ln g= {c\over 6}\gamma^2 +\mathcal{O}(c\gamma^4)\,.
\end{equation}
Therefore, we have
\begin{equation}
	S_{EE}=S_{\varepsilon}+{c\over 3} \ln \bigg[ 2 \sinh \bigg({\pi L_0 \over \beta}\bigg) \bigg]+ {c\over 6}\gamma^2 +\mathcal{O}(\gamma^4)\label{eq: ee bcft}\,.
\end{equation}

\subsection{Holographic Entanglement Entropy}
\label{secaa}

Now, we will study the entanglement entropy of a single interval $\mathcal{I}$ from the bulk geometry by using the AdS/CFT correspondence. Holographically, the entanglement entropy can be evaluated from the area of the Ryu-Takayanagi surface~\cite{Ryu:2006bv} whose boundary is the interval $\mathcal{I}$ 
\begin{equation}
    S_{EE}={\mbox{Area}\over 4G}\,.
\end{equation}
In AdS$_3$/CFT$_2$, the area of the Ryu-Takayanagi surface corresponds to the  distance of the geodesic connecting to both ends of the interval $\mathcal{I}$. For this, we consider the metric in \eqref{3djmetric} of the three-dimensional Janus black hole solution~\cite{Bak:2011ga}
with the coordinate transformation $w= \cosh \rho$ in (\ref{3djanus})  
\begin{equation}
	ds^{2} =  \ell^2\Bigg[ \rmd y^2 +   f (y) \Big(  -\frac{L^2}{\ell^4}\sinh^2 \rho\; dt^2 +d\rho^2 \bigg)\Bigg]\,.\label{eq:metric}
\end{equation}
For simplicity, let us consider a geodesic on the constant  time slice
\begin{equation}
	t=\mbox{constant}\,.
\end{equation}
The rest of the geodesic equations are given by
\begin{equation}
	f(y) {d \rho\over ds}={\E\over \ell }\,, \qquad \qquad \bigg({dy\over ds}\bigg)^2+ {\E^2\over \ell^2  f(y)}={1\over \ell^2}\,,\label{eq: geodesic eq}
\end{equation}
where $\E$ is a constant. Let us consider the simplest case
\begin{equation}
	\E=0 \,.
\end{equation}
This corresponds to a geodesic with constant $t$ and $\rho$
\begin{equation}
	t=\mbox{constant}\quad,\quad \rho=\mbox{constant}\,.
\end{equation}
Note that such a geodesic is presented as a blue line\footnote{Recall that Figure~\ref{fig2} is a constant $t$ slice and the blue line denotes the constant $\mu$ curve which is identical with the constant $\rho$ curve.} in Figure~\ref{fig2}. Now, one can easily integrate \eqref{eq: geodesic eq} to obtain the geodesic distance between two points corresponding to $y_{\infty}$ and $-y_{\infty}$ on the boundary 
\begin{equation}
	s=2ly_{\infty}\,.
\end{equation}
Note that because $y_\infty$ goes to infinity as we approach  the boundary, the geodesic distance between these two points on the boundary diverges.  
 To obtain the HEE, we need  appropriate variables in the bulk to match the boundary values, which would be the so-called Fefferman-Graham coordinates or simply the Poincar\`{e} ones in our case. Hence, we introduce cut-off ${1\over \varepsilon}$ along the radial direction in the bulk in terms of the coordinates $r$ as
\begin{equation} \label{rcutoff}
    {r\over L}={1\over \varepsilon}\,.
\end{equation}
From \eqref{cutoffRel}, one can also obtain the asymptotic behavior of $y_\infty$
\begin{equation} \label{qRel}
    \sinh\frac{Lx}{\ell^2} \simeq  \varepsilon \sqrt{f(y_{\infty})-1} \simeq  \varepsilon {(1-2\gamma^2)^{1\over 4}\over 2}  e^{y_{\infty}} \,.
\end{equation}
This gives
\begin{equation}
	y_{\infty}=\ln {1\over \varepsilon} +\ln \bigg[2 \sinh{2\pi x\over \beta}\bigg]  + {1\over 2 }\ln{1\over \sqrt{1-2\gamma^2}}   + {\cal O}(\varepsilon)\,, \label{eq: asymptotic y}
\end{equation}
where we used \eqref{beta}.
%
%
Note that the geodesic distance has $2\ell \ln {1\over \varepsilon} $ divergence as $\varepsilon\rightarrow 0$. Hence, we subtract this divergence to define the renormalized geodesic distance $s_R$. Then, the entanglement entropy of the interval $[-x,x]$ on the boundary is found to be
\begin{equation} \label{LLEE}
    S_{HEE}={s_R\over 4G}={c\over 3}\ln \bigg[2 \sinh{2\pi x \over \beta}\bigg]+{c\over 6} \ln A\,,
\end{equation}
where we used $c={3\ell \over 2G}$ together with $A$ given in (\ref{ConA}). Note that the last term corresponds to the contribution of the boundary entropy, and its small $\gamma$ expansion reads
\begin{equation}
	 S_{HEE}\simeq {c\over 3}\ln \bigg[2 \sinh{2 \pi x \over \beta}\bigg]+{c\gamma^2\over 6}+\mathcal{O}(c\gamma^4)\,,
\end{equation}
which agrees with \eqref{eq: ee bcft}.

\section{Entanglement of double RL intervals}\label{sec4}
\begin{figure}[t]   
\begin{center}
\includegraphics[width=0.5\textwidth]{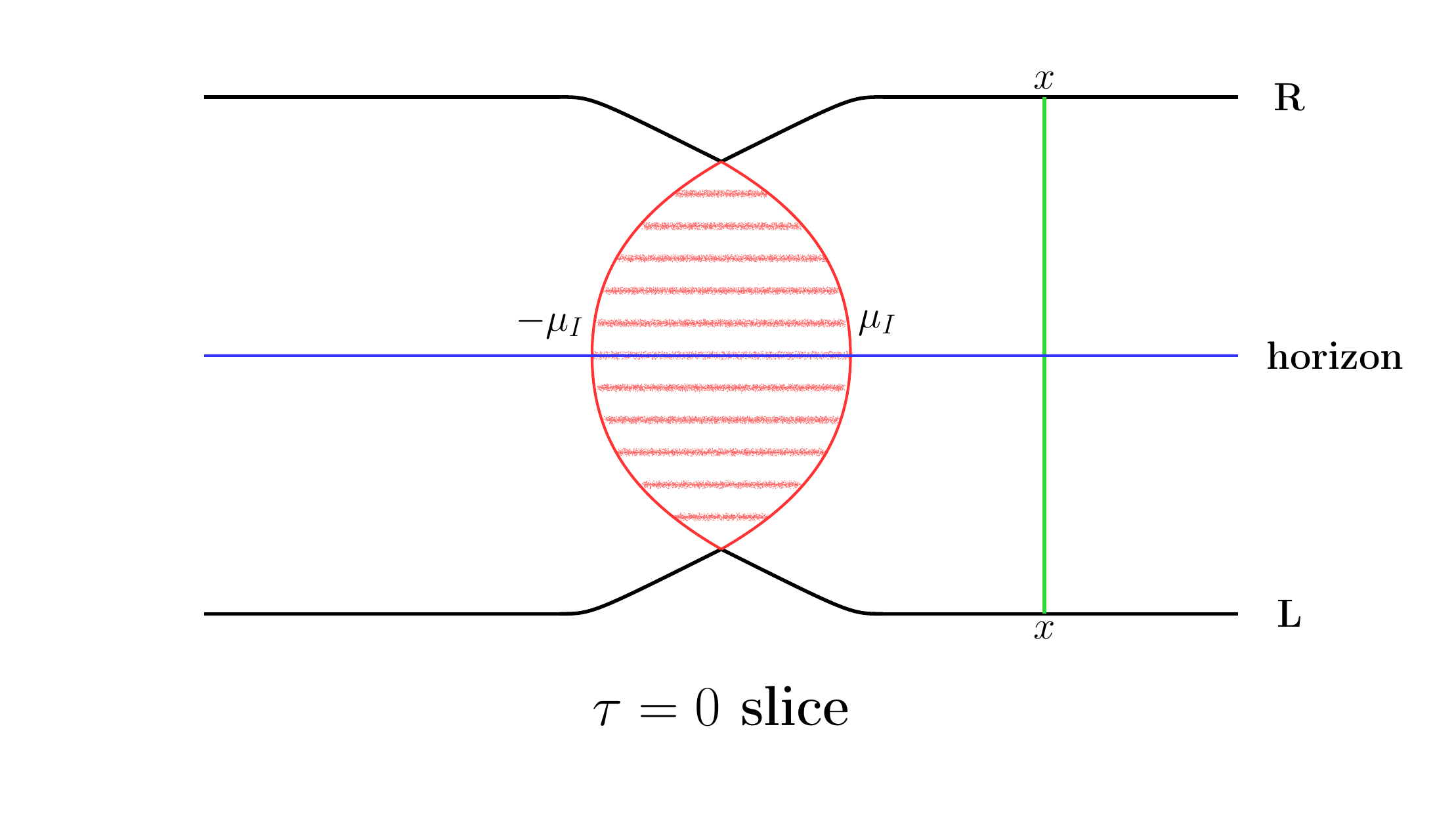}\includegraphics[width=0.4\textwidth]{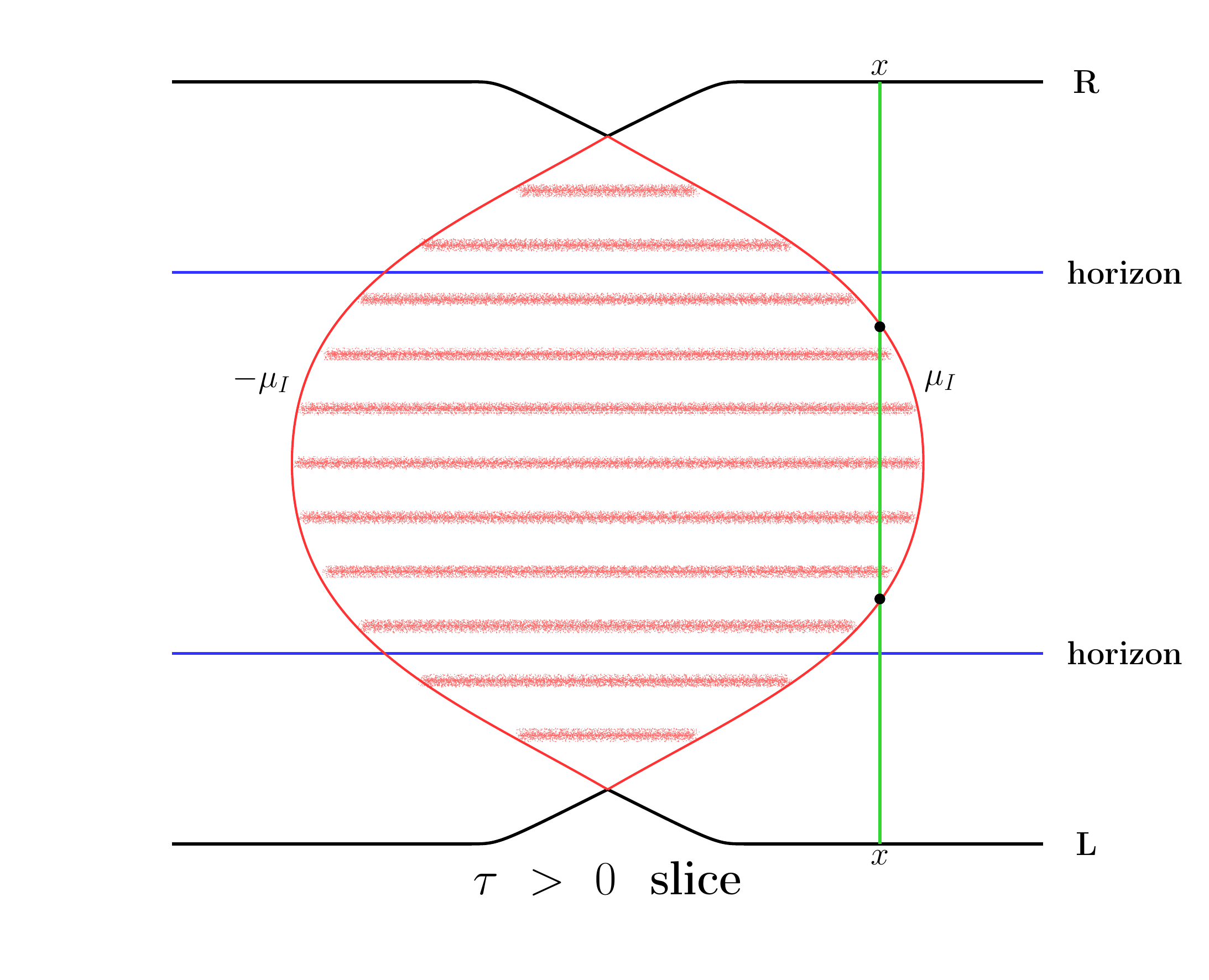}
	\caption{\small We have depicted the RL geodesic at $\tau=0$ and at  a later time $\tau >0$ with a shadow region. Though  we have depicted a constant $\tau$ slice, the curve $\mu=\pm\mu_{I}$ takes nearly the same form as in Figure~\ref{fig2}. }
\label{LRgeo}
\end{center}
\end{figure}

In this section, we provide some details about the HEE of (double) RL intervals on the two-sided Janus black holes  reviewed in Section~\ref{sec2}.  As is done in the previous section, the HEE can be obtained by  the geodesic distance in this case, too.  For the  geodesics connecting the R and L sides of Janus black holes, we will focus on the constant time slice $\tau=constant$ using the following form of the metric
\begin{equation} \label{}
ds^{2} =  \ell^2\left[ \rmd y^2 +   f (y) \Big(  \rmd q^{2} -\cosh^{2}q \rmd \tau^{2} \Big)\right]\,,
\end{equation}
where $f(y)$ was introduced in~\eqref{eq: def f} and the AdS$_{2}$ part is taken by the metric form given in \eqref{qads2}. 
In the following, we consider a single geodesic whose boundary position is taken by the same coordinate values as $(x,t)$ with $x>0$ on the R and L sides, first (See Figure~\ref{LRgeo}). And then the double geodesics will be taken into consideration to obtain the relevant HEE. For simplicity, the boundary locations of these double geodesics are taken symmetrically as $(x,t)$ and $(-x,t)$ and they will be called the doubled geodesic.

The geodesic equation in the above $(y, q, \tau)$ coordinate system\footnote{One easy way to deduce this expression may utilize the Hamiltonian conservation of the Lagrangian $L=\sqrt{\dot{y}^{2} +f(y)}$.} may be integrated  as
\begin{equation} \label{JGeo1}
\dot{y}^{2} + f = \frac{f^{2}}{E^{2}}\,, \qquad  \dot{}~ \equiv \frac{\rmd}{\rmd q}\,,
\end{equation}
where $E$ is an integration constant that turns out to be related to the boundary position of the geodesic.   This form of the  geodesic equation can be integrated in terms of the incomplete elliptic integral of the first kind, as 
\begin{equation} \label{qexp}
q-q_{0}= {\textstyle \frac{A+B}{E} }\int^{y_{max}}_{y_{min}}\frac{dy}{ \sqrt{\cosh2y +A}{\sqrt{\cosh 2y -B}} }= \sqrt{A} \sqrt{m}F(\varphi\,|\, m)\,,
\end{equation}
where  $q_{0}$ is another integration constant\footnote{This constant $q_{0}$ will be dropped in the following, since it may be set to zero by shifting the origin of the coordinates.} and $y_{min} = \frac{1}{2} \textrm{arccosh}\, B$. Here, the constant $A$ has been introduced before in (\ref{ConA}) and  the constants $B$ and $m$ are defined, respectively, by  
\begin{equation} \label{Constm}
   B \equiv \frac{2E^{2}-1}{\sqrt{1-2\gamma^{2}}}\,, \qquad m\equiv  {\textstyle \frac{2(A+B)}{(A+1)(B+1)} }\,,
\end{equation}
while  the so-called amplitude $\varphi$  denotes
\begin{equation} \label{}
\sin\varphi \equiv  \textstyle{\sqrt{\frac{A+1}{2(A+B)}} \frac{\sqrt{\cosh 2y_{max} -B}}{\sinh y_{max}} } \,.
\end{equation}
Eventually, we will take $y_{max}$ to infinity  which corresponds to the position of the AdS boundary in these coordinates (See Figure~\ref{fig2} for the R/L boundaries, which may also be interpreted as denoting constant $\tau$ surface with $(y, q)$ coordinates). As usual in the holographic computation, this infinity may be controlled by an appropriate cutoff  in the AdS space as in the previous section.

In terms of the geodesic distance $s$, the geodesic equation may also be written as
\begin{equation} \label{}
\dot{s} = \sqrt{\dot{y}^{2} +f} = \frac{f}{E}\,.
\end{equation}
Using \eqref{JGeo1}  in the above geodesic distance expression and integrating  with respect to the $y$-coordinate,  one can deduce that the geodesic distance may be written in terms of the $y$-coordinate as
\begin{equation} \label{GeoInt}
s-s_{0} = \int^{y_{max}}_{y_{min}} dy \frac{\sqrt{\cosh 2y +A}}{\sqrt{\cosh 2y -B}}\,,\end{equation}
where $s_{0}$ is an integration constant\footnote{By taking the origin of the proper distance in such a way that $s=0$ when $y_{max}=y_{min}$, we  set $s_{0}=0$   in the following.}.  We would like to emphasize that the geodesic distance between the R and L boundaries should be twice of the above geodesic distance $s$ with $y_{max}=y_{\infty}$, since $y_{min}$ may be understood as located in the middle of the  R and L boundaries. To see this,  recall that the coordinate $y$ is related directly to $\mu$ by \eqref{muy}.  

It is straightforward to integrate  the above equation  to the form of
\begin{equation} \label{ProDisEllip}
s  =  \frac{1}{\sqrt{A+1}\sqrt{B+1}} \bigg[ (A+1) F(\varphi\,|\, m) + (B-1)\Pi(\nu\,;\, \varphi\, |\, m) \bigg]\,,  \qquad \nu \equiv \textstyle{\frac{A+B}{A+1}}\,,
\end{equation}
where  $\Pi$ is the incomplete elliptic integral of the third kind, whose properties are summarized in  Appendix~\ref{AppA}. 
To proceed  to the HEE computation, one needs to introduce the cutoff as in the previous sections. By introducing the cutoff  as in~\eqref{rcutoff} and using the relation in~\eqref{qRel} and \eqref{eq: asymptotic y},  
%
the renormalized geodesic distance can be obtained by removing the cutoff part. To this purpose,  consider the behavior of the large $y_{max}(=y_{\infty})$ limit as in the previous section. It is straightforward to check, from the integral expression in \eqref{GeoInt}, that $s_{\infty} = y_{\infty} + finite$ as $\varepsilon\rightarrow 0$. Hence, it is useful to  introduce $Q(A,B)$ as follows:
\begin{equation} \label{}
Q(A,B) = s_{\infty} - y_{\infty} + {\cal O}\Big(\frac{1}{y_{\infty}}\Big)\,,
\end{equation}
which should be a finite quantity by construction and independent of the cutoff in the limit of $\epsilon\rightarrow 0$ (or $y_{\infty}\rightarrow\infty$). 
Then the renormalized geodesic distance, $s_{R}$ is taken in this case to be\footnote{Here, $s_{\infty}$ corresponds to the half of the (unrenormalized) proper distance of the RL geodesic, since we are taking the integration range of $y$  from $y_{min}=\frac{1}{2} \textrm{arccosh}\, B$ to $y_{max}=y_{\infty}$.} 
\begin{align}    
s_{R} \equiv  \Big[s_{\infty} - \ell \ln \textstyle{\frac{1}{\varepsilon}}\Big]_{\varepsilon\rightarrow 0} &=  \ln 2 \sinh\frac{2\pi x}{\beta}  + \frac{1}{2}\ln A +    Q(A,B)  \label{PagetimeQ}\,, \\
&= \ln 2 \cosh \frac{2\pi t}{\beta}  + \frac{1}{2}\ln A +    Q(A,B) -P(A,B)   \label{sQP} \,,
\end{align}
where $P$ is defined by 
\begin{equation} \label{}
P(A,B) \equiv  \ln  {\textstyle \frac{ \cosh \frac{2\pi t}{\beta} }{\sinh\frac{2\pi x}{\beta} } }\,.
\end{equation}
At this stage, one may be perplexed by our notation where  $P$ depends on the constants $A$ and $B$. This notation is related to   our choice of  the $q$-coordinate in \eqref{qads2} and its value $q_{\infty}$ for the emanating position of the geodesic at the boundary. To see this, 
recall that the coordinate $y_{max}=y_{\infty}\rightarrow \infty$ is  related to the $q_{\infty}$ coordinate as in  \eqref{qexp} for the RL   geodesics and  that the $q_{\infty}$ (or $\lambda_{\infty}$) coordinate is one of the boundary coordinates (See \eqref{invb}). By using  the relation  in the asymptotic region given by \eqref{cutoffRel}, one may set
\begin{equation} \label{btq}
P(A,B) \equiv \ln \sinh q_{\infty} =  \ln {\textstyle \frac{\cosh  \frac{2\pi t}{\beta}}{ \sinh  \frac{2\pi x}{\beta}} }\,, \qquad q_{\infty} = q_{\infty}(A,B)\,,
\end{equation}
where $q_{\infty}(A,B)$ denotes the boundary end point position of the RL geodesic. Then one can see that the $q_{\infty}$  value    itself depends on the constant $A$ and $B$ and so does $P(A,B)$.  As will be clear in the following, $P(A,B)$ characterizes the approximation for the matching of the bulk expression  to the boundary results.

Now, we present some steps leading to the elliptic integral representation of $Q(A,B)$. 
First, note that    the argument, $\sin\varphi$ of  the incomplete elliptic integrals in \eqref{ProDisEllip},  becomes  in the large $y_{\infty}$ limit
\begin{equation} \label{}
\sin\varphi \stackrel{y_{\infty}\rightarrow\infty }{\longrightarrow} \textstyle{\sqrt{\frac{A+1}{A+B}}} \equiv \sin\varphi_{\infty} = \frac{1}{\sqrt{\nu}}\,.
\end{equation}
In this large $y_{\infty}$ limit, by using the asymptotic expansion in \eqref{ASEJ}, one can also see that  %
\begin{align}    \label{}
& R_{J}(\cos^{2}\varphi,\, 1-m\sin^{2}\varphi,\,1,\, 1-\nu\sin^{2}\varphi) \nonumber \\  
& = \textstyle{\frac{3\sqrt{(A+B)(B+1)}}{B-1} }y_{\varepsilon} 
+ \bigg[\frac{3}{2}\frac{1}{\sqrt{xyz}} \ln \frac{2xyz}{(B-1)\sigma^{2}} + 2R_{J}(x+\sigma,y+\sigma,z+\sigma, \sigma)\bigg] + {\cal O}(y_{\infty}e^{-y_{\infty}})\,,
\end{align}
where $x,y,z$ and $\sigma$ are defined as
\begin{equation} \label{}
x\equiv \textstyle{\frac{B-1}{A+B}}\,, \qquad y \equiv \textstyle{\frac{B-1}{B+1}} \,, \qquad z\equiv 1\,, \qquad \sigma \equiv \sqrt{xy} + \sqrt{yz} + \sqrt{zx}\,.
\end{equation}
Finally, using the symmetric elliptic integral\footnote{See Appendix \ref{AppA} for some details of symmetric elliptic integrals.}, one can see that
\begin{align}    \label{Qexp}
Q(A,B) &= \textstyle{\sqrt{\frac{A+B}{2}}}\sqrt{m}~ F(\varphi_{\infty}\,|\, m)  -\ln \Big[\textstyle{\sqrt{\frac{B-1}{2}} + \sqrt{\frac{A+B}{2}} + \sqrt{\frac{B+1}{2}} }\Big] \nonumber \\
& \qquad \qquad \qquad \qquad  + \frac{2}{3}\sqrt{xyz}R_{J}(x+\sigma,y+\sigma,z+\sigma,\sigma)\,,
\end{align}
which is indeed a finite expression. 
Note also that 
\begin{equation} \label{qinfty}
q_{\infty} (A,B)=  \sqrt{A} \sqrt{m}F(\varphi_{\infty}\,|\, m)=  \sqrt{A} \sqrt{m} R_{F} \Big({\textstyle  \frac{B-1}{A+1},\frac{A+B}{A+1}\frac{B-1}{B+1},\frac{A+B}{A+1} } \Big)\,,
\end{equation}
which justifies our notation $P(A,B)$ in the above since this reveals the dependence on $A$ and $B$, explicitly.

Though we have obtained the closed form of the relevant quantities in terms of the bulk constants\footnote{Recall that $A$ is the parameter for the Janus background geometry and $B$ is the one for the geodesic.} $A$ and  $B$ (or equivalently constants $\gamma$ and $E$), it is quite involved to compute the HEE in this form. Rather than the bulk constants,  the HEE needs to be described by the renormalized geodesic distance related to the appropriate  boundary position  or the end points of geodesics.  In our case, the relevant boundary position needs to be written in terms of boundary coordinates  $(t,x)$  in \eqref{txbcoor}, not in terms of $A$ and $B$. In order to represent  $q_{\infty}$ and $Q(A,B)$ in terms of these boundary quantities, it is quite useful to consider some limiting regimes. To this end,  let us consider  two regimes  $q_{\infty} \ll 1$ and $q_{\infty} \gg 1$ respectively, depending on the influence of the interface. By using (\ref{btq}), these regimes can be represented in terms of boundary variables $x$ and $t$.
In these regimes, one can rewrite the expressions,  for instance $Q(A,B)$,  in terms of $q_{\infty}$ instead of $B$.  In later sections, one will encounter the same regimes from the boundary ICFT consideration. On the other hand, from the asymptotic expansion of the symmetric elliptic integrals, the useful limiting regimes correspond to  the cases of $B\gg A-1$ and $A\gg B-1$. In the following, we show that the appropriate regimes may be obtained from the limiting cases in the symmetric elliptic integral expressions.

\vskip0.3cm
\noindent $\bullet$ {\bf Regime 1}:  \underline{$q_{\infty} \ll 1$}\\
This regime will turn out to be related to the 
case of bulk constants $B\gg A-1$. 
This limit corresponds to the case where any effect of the interface degrees of freedom becomes negligible. For example, the RT surface lies far away from the the shadow region ({\it i.e.} $E\gg 1$), or the number of interface degrees of freedom is small enough ({\it i.e.} $\gamma\ll 1$).
First, note that $F(\varphi_{\infty}|m)$ reduces, in this bulk limit,  to
\begin{equation} \label{}
F(\varphi_{\infty}|m) = {\textstyle \sqrt {\frac{A+1}{B+1} }  } R_{F}( {\textstyle\frac{B-1}{B+1}}, {\textstyle \frac{B-1}{B+1} },1)  + {\cal O}({\textstyle \frac{A-1}{B+1} }) = {\textstyle \sqrt {\frac{A+1}{2} }  } \ln \Big[{\textstyle \sqrt{\frac{B+1}{B-1}} + \sqrt{\frac{2}{B-1}}}\Big]  + {\cal O}({\textstyle \frac{A-1}{B+1} }) \,,
\end{equation}
where we used \eqref{Rred} and \eqref{Rcexp}.  Secondly,  by using \eqref{RJtoRC} and \eqref{Rcexp},  the $R_{J}$ expression reduces to
\begin{equation} \label{}
 \frac{2}{3}\sqrt{xyz}R_{J}(x+\sigma,y+\sigma,z+\sigma,\sigma) =  \ln \Big[{\textstyle 1+ 2\sqrt{\frac{B+1}{B-1}} }\Big] - {\textstyle \sqrt{\frac{B+1}{2}} }\ln \Big[  {\textstyle \frac{\sqrt{B+1}+ \sqrt{2} } {B-1} } \Big]  + {\cal O}({\textstyle \frac{A-1}{B+1} })\,.
\end{equation}
%
%
As a result, $Q(A,B)$ becomes
\begin{equation} \label{}
Q(A,B) = \ln {\textstyle \sqrt{\frac{2}{B-1}}  }+ {\cal O}({\textstyle \frac{A-1}{B+1} })\,.
\end{equation}
And, the expression of $q_{\infty}$, given by $y_{\infty}\rightarrow\infty$ in \eqref{qexp}, reduces to 
\begin{equation} \label{}
q_{\infty}  = \sqrt{A} \ln \Big[{\textstyle \sqrt{\frac{B+1}{B-1}} + \sqrt{\frac{2}{B-1}}}\, \Big] +  {\cal O}({\textstyle \frac{A-1}{B+1} })\,,
\end{equation}
which leads to 
\begin{equation} \label{}
\sinh \frac{q_{\infty}}{\sqrt{A}} = {\textstyle \sqrt{\frac{2}{B-1}} } + \cdots\,,
\end{equation}
where $\cdots$ denotes 
small correction terms. 
It is clear that the $q_{\infty}\ll 1$ regime corresponds to the  $B\gg1$ case.  
 In the 
case of $B\gg A-1$, together with $P(A,B) = \ln \sinh q_{\infty}$, one obtains
\begin{equation} \label{QmP1}
Q(A,B) - P(A,B)  =\ln \frac{\sinh \frac{q_{\infty}}{\sqrt{A}}}{\sinh q_{\infty}} +  \cdots\,. 
\end{equation}
As a result, one can see, through \eqref{sQP},  that the renormalized geodesic distance becomes 
\begin{equation}   
s_{R} = 
\ln 2 \cosh \frac{2\pi t}{\beta}     \qquad    \textrm{for} \quad q_{\infty} \ll 1 \label{LREEa}\,.
\end{equation}
\\

\noindent $\bullet$ {\bf Regime 2}: \underline{$q_{\infty} \gg 1$}\\
 This regime turns out to be correspondent to the case of  $A \gg B-1$. In this bulk limit,  one may notice that $m\rightarrow \frac{2}{B+1} + {\cal O}(\frac{B-1}{A+1})$ and  $\sin\varphi_{\infty}= 1+{\cal O}(\frac{B-1}{A+1})$ and so  the expression of $q_{\infty}$ becomes 
\begin{equation} \label{}
q_{\infty} = {\scriptstyle  \sqrt{\frac{2A}{B+1}}}~{\bf K}( {\scriptstyle   \sqrt{\frac{2}{B+1}} } )  + {\cal O} ( {\textstyle \frac{B-1}{A+1} } )\,.
\end{equation}
Since  the $R_{J}$ expression reduces to
\begin{equation} \label{}
 \frac{2}{3}\sqrt{xyz}R_{J}(x+\sigma,y+\sigma,z+\sigma,\sigma) =  {\cal O}\Big( {\textstyle \frac{B-1}{\sqrt{A+1}} } \Big)   \,,
\end{equation}
one can see that
\begin{align}    \label{}
Q(A,B)  =  {\scriptstyle \sqrt{\frac{A+1}{B+1} } }~ {\bf K}( {\scriptstyle   \sqrt{\frac{2}{B+1}} } )  - \ln  \Big[ {\scriptstyle \sqrt{\frac{B+1}{2}} + \sqrt{\frac{B-1}{2}} + \sqrt{\frac{A+1}{2}}  }\Big]  + {\cal O}\Big( {\textstyle \frac{B-1}{\sqrt{A+1}} }  \Big)\,. 
\end{align}
More useful information may be obtained by taking a more specific case as  $A\gg 1$ or $B\rightarrow 1$. In these cases, one can see that $q_{\infty} \gg 1$ and so $P(A,B)=\ln\sinh q_{\infty} \simeq q_{\infty} -\ln 2$. In each case of $A\gg 1$ and $B\rightarrow 1$,
the renormalized proper distance is given by
\begin{equation} \label{LREEc}
s_{R} = %
\left\{ \begin{array}{ll}     
\frac{1}{\sqrt{2}}  \ln {\textstyle 2 \cosh \frac{2\pi t}{\beta} } +(1-\frac{1}{\sqrt{2}})  \ln {\textstyle  \sinh \frac{2\pi x}{\beta} }  + \frac{3}{2}\ln 2\,, & \textrm{for} \quad A\gg 1  \\
{\scriptstyle \sqrt{\frac{A+1}{2A} } } \ln {\textstyle 2 \cosh \frac{2\pi t}{\beta} }  + \frac{1}{2}\ln A + ( 1-{\scriptstyle \sqrt{\frac{A+1}{2A} } } ) \ln   {\textstyle  \sinh  \frac{2\pi x}{\beta}}  - \ln \frac{\scriptstyle  1+ \sqrt{\frac{ A+1}{2}} }{2} \,,  &     \textrm{for}\quad B\rightarrow 1
\end{array}  \right.   \,.
\end{equation}

It is interesting to observe that the above two regimes might be approached in a simple way by taking $\underline{B=A\alpha ~\&~ A \gg 1} $.
In this special case,  one may note that the parameter $m$ in \eqref{Constm} reduces as $m\rightarrow \frac{1+\alpha}{\alpha}\frac{2}{A}$. 
Using \eqref{qexp}, \eqref{Rcexp} and \eqref{EllipticF},    one can  see that  
\begin{equation} \label{q1}
q_{\infty} =  {\textstyle  \sqrt{2}\sqrt{\frac{1+\alpha}{\alpha}}\,  \textrm{arcsin} \frac{1}{\sqrt{1+\alpha}}  }\,.
\end{equation}
Note also that \eqref{Qexp}, \eqref{RJtoRC} and \eqref{Rcexp} lead to  
\begin{equation} \label{q2}
Q(A,B) = \frac{1}{\sqrt{\alpha}} \textrm{arcsin} {\textstyle \frac{1}{\sqrt{1+\alpha}}}  -\ln {\textstyle \sqrt{\frac{1+\alpha}{2}A} }\,. 
\end{equation}
In each case of $\alpha\gg 1$ ({\bf regime 1}) and  $\frac{1}{A}\ll\alpha\ll 1$ ({\bf regime 2}), one can obtain the $s_{R}$ expression in terms of the boundary variables by using the above expressions, which reproduce  the same forms of the expression in \eqref{LREEa} and the upper line expression in \eqref{LREEc}, respectively.  It is amusing to observe that the final results remain the same, although apparently different-looking functions appear through the different limiting procedures.

Before going ahead, one may consider the case of $A\rightarrow 1$, which may also be analyzed  in a definite analytic form by using\eqref{sQP} and \eqref{QmP1}.  In this case, the renormalized proper distance is given by
\begin{equation}     \label{LREEb}
s_{R} = 
\ln 2 \cosh \frac{2\pi t}{\beta}  + \frac{1}{2}\ln A + \cdots\,,   \qquad  \textrm{for} \quad     A\rightarrow 1 \,.
\end{equation}
In fact, one may obtain some analytic form beyond this $A\rightarrow 1$ limit.  From \eqref{GeoInt}, one may notice  that
\begin{align}    \label{}
{\textstyle \frac{\partial}{\partial A} } Q(A,B) 
= \int^{y_{\infty}}_{y_{min}}    \frac{dy}{\sqrt{\cosh 2y +A}\sqrt{\cosh 2y -B}} \bigg|_{     y_{\infty} \rightarrow \infty}   =  \frac{1}{2} \textstyle{\sqrt{\frac{m}{2(A+B)}}}~ F(\varphi_{\infty}\,|\, m) \,.
\end{align}
%
Then, the next order of $A = 1 + \gamma^{2} + {\cal O}(\gamma^{4}) $ in the expansion of $Q(A,B)$  may be obtained explicitly as
\begin{equation} \label{eq434}
Q(A,B) = Q(1,B) + \textstyle{ \frac{\gamma^{2}}{2} \sqrt{\frac{m}{2(A+B)}}}~ F(\varphi_{\infty}\,|\, m)    + {\cal O}( {\scriptstyle \gamma^{4} })\,, \qquad  Q(1,B) =\ln \textstyle{\sqrt{\frac{2}{B-1}}}\,,
\end{equation}
and the expression of $q_{\infty}$ in \eqref{qexp}  becomes 
\begin{equation} \label{}
q_{\infty} = \textrm{arctanh} {\scriptstyle{\sqrt{\frac{2}{B+1}}}  } + {\textstyle \frac{\gamma^{2}}{8}} \Big[ {\scriptstyle{\sqrt{\frac{2}{B+1}}}  } +
(4- {\scriptstyle{\frac{B-1}{B+1}}} ) \textrm{arctanh} {\scriptstyle{\sqrt{\frac{2}{B+1}}}  } \Big]  + {\cal O}( {\scriptstyle \gamma^{4} })\,.
\end{equation}
As before, one may see that  $ P(A,B) = \ln \sinh q_{\infty} = \ln {\scriptstyle \sqrt{\frac{2}{B-1}} } +{\cal O}(\gamma^{2})$ from the above expression and then one  obtains
\begin{equation} \label{QPsg}
Q(A,B) - P(A,B) =   {\textstyle \frac{\gamma^{2}}{8}} \Big[ -1+ ( {\scriptstyle{\sqrt{\frac{2}{B+1}}}  } 
-  3 {\scriptstyle \sqrt{\frac{B+1}{2}}} ) \textrm{arctanh} {\scriptstyle{\sqrt{\frac{2}{B+1}}}  } \Big]  + {\cal O}( {\scriptstyle \gamma^{4} })\,.
\end{equation}
Hence, the renormalized geodesic distance is given by
\begin{align}   
s_{R} &= \ln 2\cosh  \textstyle{\frac{2\pi t}{\beta}} + \frac{\gamma^{2}}{8} \Big[3 + ( {\scriptstyle{\sqrt{\frac{2}{B+1}}}  } 
-  3 {\scriptstyle \sqrt{\frac{B+1}{2}}} ) \textrm{arctanh} {\scriptstyle{\sqrt{\frac{2}{B+1}}}  } \Big]  + {\cal O}( {\scriptstyle \gamma^{4} })\,,  \nonumber \\
&=  \ln 2\cosh  \textstyle{\frac{2\pi t}{\beta}} + \frac{\gamma^{2}}{8} \Big[3 + \frac{1}{2}\Big(  
{\scriptstyle \frac{\cosh  \frac{2\pi t}{\beta}}{\sqrt{\sinh^{2}  \frac{2\pi x}{\beta} + \cosh^{2}  \frac{2\pi t}{\beta}}} } \nonumber \\
& \qquad \qquad  \qquad  \qquad \qquad  \quad 
-  3  {\scriptstyle \frac {\sqrt{\sinh^{2}  \frac{2\pi x}{\beta} + \cosh^{2}  \frac{2\pi t}{\beta}}} {\cosh  \frac{2\pi t}{\beta}} } \Big) \ln {\scriptstyle \frac{\sqrt{\sinh^{2}  \frac{2\pi x}{\beta} + \cosh^{2}  \frac{2\pi t}{\beta}} + \cosh  \frac{2\pi t}{\beta} }  {\sqrt{\sinh^{2}  \frac{2\pi x}{\beta} + \cosh^{2}  \frac{2\pi t}{\beta}} - \cosh  \frac{2\pi t}{\beta}}  }\Big]    + {\cal O}( {\scriptstyle \gamma^{4} })\,,   
\label{eq437}
\end{align}
where we used $P(A,B) = \ln {\scriptstyle \sqrt{\frac{2}{B-1}} } +{\cal O}(\gamma^{2})$.

In the limit of $ x \gg \frac{\beta}{2\pi}$, the RL geodesic may be drawn as a straight line,  since the interface  does not deform the shape of the geodesic significantly away from the BTZ limit.  It is instructive  to observe that the coordinate $(y, q)$ grid  or equivalently $(\mu, w)$ grid expands as time $\tau$ goes on.   The growth  of the shadow region, as time goes on, may be understood by the behavior of the coordinates in conjunction with the  shadow region  determination  formula in \eqref{mui}. Of course,  this growth of the shadow region is  reminiscent of  the growth of the spatial region inside the horizon along the time evolution.  See Figure~\ref{LRgeo}. The physical position in the boundary is denoted by $x$ in this figure.  
At the initial time $\tau=0$ with $ x \gg \frac{\beta}{2\pi}$, the straight line geodesic resides outside  the shadow region. However, 
as the time goes on,  the shadow region becomes larger and so the geodesic crosses  eventually the boundary  of the shadow region, which is  specified by $\mu= \mu_{I}$.

As was explained in the previous section,  
HEE may be read simply from the renormalized geodesic distance and the expression of HEE in the RL geodesic case becomes 
\begin{equation} \label{LREE}
S_{HEE} = \frac{c}{6} \times 2 s_{R} = \frac{c}{3}s_{R}\,,
\end{equation}
where the factor of 2 comes from the fact that the geodesic distance between the R and L boundaries is twice of our expression of $s$ in \eqref{GeoInt}. Furthermore, the final expression of the HEE for the region ${\cal B}_{L}\cup {\cal B}_{R}$ (See Figure~\ref{InfTrans}) should be multiplied by another factor of 2, since we have considered the symmetric  doubled geodesic of the same length in the $+$ and $-$ sides.  See the next section for a further interpretation of this doubled geodesic for the HEE.

\section{Unitarity, Page curve and Mutual Information}\label{sec5}
As is well-known, unitarity is one of the fundamental ingredients in quantum mechanics and the famous information loss problem of black hole physics is the clash between  the unitarity requirement and a semi-classical computation in the black hole geometry. Some time ago, Page has sharpened the clash by showing that the entanglement entropy of Hawking radiation (or that of black hole) should follow the so-called Page curve. On the other hand,  Hawking's semi-classical computation tells us that the  radiation is thermal so that it cannot follow the Page curve. An interesting picture on the behavior of the entanglement entropy for eternal black holes was given in~\cite{Hartman:2013qma}. Furthermore, very recent developments in this story~\cite{Penington:2019npb,Almheiri:2019psf}  are to explain the Page curve  by unveiling missing parts in the previous semi-classical reasoning and computation. In particular,  the island picture has been constructed~\cite{Almheiri:2019hni} and explicitly checked in the case of eternal black holes~\cite{Almheiri:2019yqk}. 

In this section, we provide  an interpretation of the results in the previous sections on the entanglement entropy for the three-dimensional Janus black holes. Basically, our interpretation is similar to that of~\cite{Hartman:2013qma}, but there are complications and new aspects, because of the  Janus deformation or the ICFT. The RR or LL geodesics correspond to the entanglement entropy viewed from one side when we trace out the other side, which is  time-independent  as  given in \eqref{LLEE}.  The additional term depending on $\gamma$ in these expressions, which  turns out to be temperature-independent, corresponds  to the additional entanglement  entropy from the interface QM degrees of freedom. 

On the other hand, the RL geodesic corresponds to time-dependent entanglement entropy of the radiation. The late time behavior of  this entropy  in \eqref{LREEa}, 
\eqref{LREEc} and~\eqref{LREE} becomes linear  and  corresponds to the usual deviation from the Page curve.   As was explored and explained in~\cite{Hartman:2013qma}, the prescription in HEE tells us that the actual entanglement entropy should be chosen to be the minimum among the extremal ones in the bulk. Therefore, the time-dependent part (or the RL geodesic) dominates at the initial stage of the black hole evaporation 
 while the time-independent one (or the RR/LL geodesics) becomes dominant after the Page time. This transition of the HEE configuration in the bulk is interpreted as the consequence of the existence of  entanglement islands  in eternal black holes~\cite{Almheiri:2019hni,Almheiri:2019yqk,Almheiri:2019qdq,Penington:2019kki}. In the case of our Janus  black holes, there are some additional features which are related to the Janus deformation given by the parameter~$\gamma$.  As shown in  \eqref{LREEc},    there is a term (third term in  the second line), which vanishes in the BTZ limit of $A\rightarrow 1$. This term  depends on  the end point $x$ of the geodesic on the R/L boundary,  while it is independent of time $t$.   We would like to interpret this $\gamma$-dependent contribution as the entanglement between the interface degrees of freedom and the radiation degrees of freedom living outside  the location $x$ in our ICFT\footnote{There are two outsides in CFT$_{-}$ and CFT$_{+}$.}. This entanglement shows us  the interplay of the interface degrees of freedom with the radiation ones.  At the end of this section, we provide some details on this interpretation, which uses a mutual information represented by the RL geodesic  expression.   

\begin{figure}[htbp]  
\begin{center}
\includegraphics[width=0.7\textwidth]{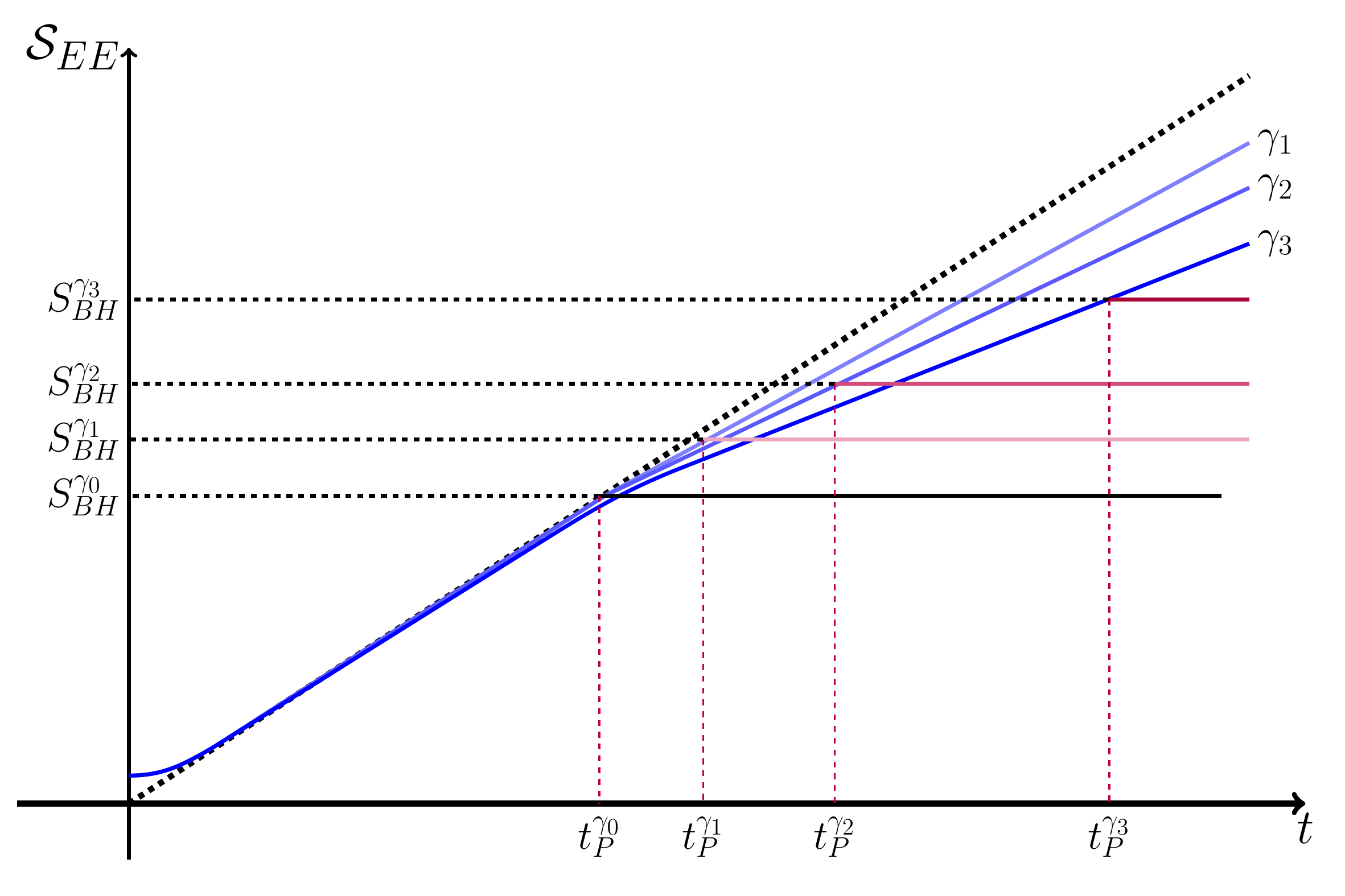}
\caption{\small We have depicted the Page curves for the parameters $0<\gamma_{1} < \gamma_{2} < \gamma_{3}< 1/\sqrt{2}$. $\gamma_0$ here refers to the $\gamma=0$ Page curve in the BTZ limit.}
\label{PageC}
\end{center}
\end{figure}

It is also interesting to observe that the behavior of the entanglement entropy before the Page time depends on the parameter $\gamma$.  In Figure~\ref{PageC}, we have depicted schematically the Page curves for various values of $\gamma$. 

According to the minimum choice prescription in the HEE, 
the Page time may be  determined to be the time when the RR/LL HEE in \eqref{LLEE} and the RL HEE in \eqref{PagetimeQ}  with (\ref{LREE}) become equal.  Note also that the integration constant $B$ is related to the boundary time $t$ (and the position $x=L_{0}/2$ in our setup) through \eqref{btq}. This tells us that one may write $B=B(t)$. This consideration leads to the following expression for the Page time:
\begin{equation} \label{p51}
Q\Big(A,B(t_{P}) \Big) =0\,.
\end{equation}
To obtain  an explicit expression of the Page time, let us first consider the case
 of $\gamma \ll 1$. In this regime we use the results in (\ref{eq434}), (\ref{QPsg}), and (\ref{p51}), 
which leads to
\begin{equation} \label{}
P(A,B(t_{P}) ) = \ln \frac{\cosh \frac{2\pi t_P}{\beta}}{\sinh \frac{\pi L_0}{\beta}}  =\frac{\gamma^2}{8}\left[1+\frac{5}{\sqrt{2}}\ln (1+\sqrt{2}) \right]+{\cal O}(\gamma^4).
\end{equation}
For $L_0 \gg \frac{\beta}{2\pi}$, 
this becomes
\begin{equation} \label{}
 t_P =\frac{L_0}{2}+\frac{\beta\gamma^2}{16\pi}\left[1+\frac{5}{\sqrt{2}}\ln (1+\sqrt{2}) \right]+\cdots.
\end{equation}
When $A$ becomes large, 
we use the result in (\ref{LREEc}) with (\ref{PagetimeQ}) to obtain
\begin{equation} \label{}
 \ln \frac{\cosh \frac{2\pi t_P}{\beta}}{\sinh \frac{\pi L_0}{\beta}}   = \frac{1}{\sqrt{2}} \ln \frac{A}{2}  -\ln 2 +{\cal O}(A^{-1}) \,.
\end{equation}
Again taking 
$L_0 \gg \frac{\beta}{2\pi}$, we are led to 
\begin{equation} \label{}
t_{P} = \frac{L_{0} }{2}+ \frac{\beta}{2\pi}\left[ \frac{1}{\sqrt{2}} \ln \frac{A}{2}  -\ln 2\right]+\cdots \,,
\end{equation}
which tells us that  the Page time becomes larger as $\gamma$ (or $A$) gets bigger.  This aspect is also depicted in Figure~\ref{PageC}.

\begin{figure}[htbp]   
\begin{center}
\includegraphics[width=0.7\textwidth]{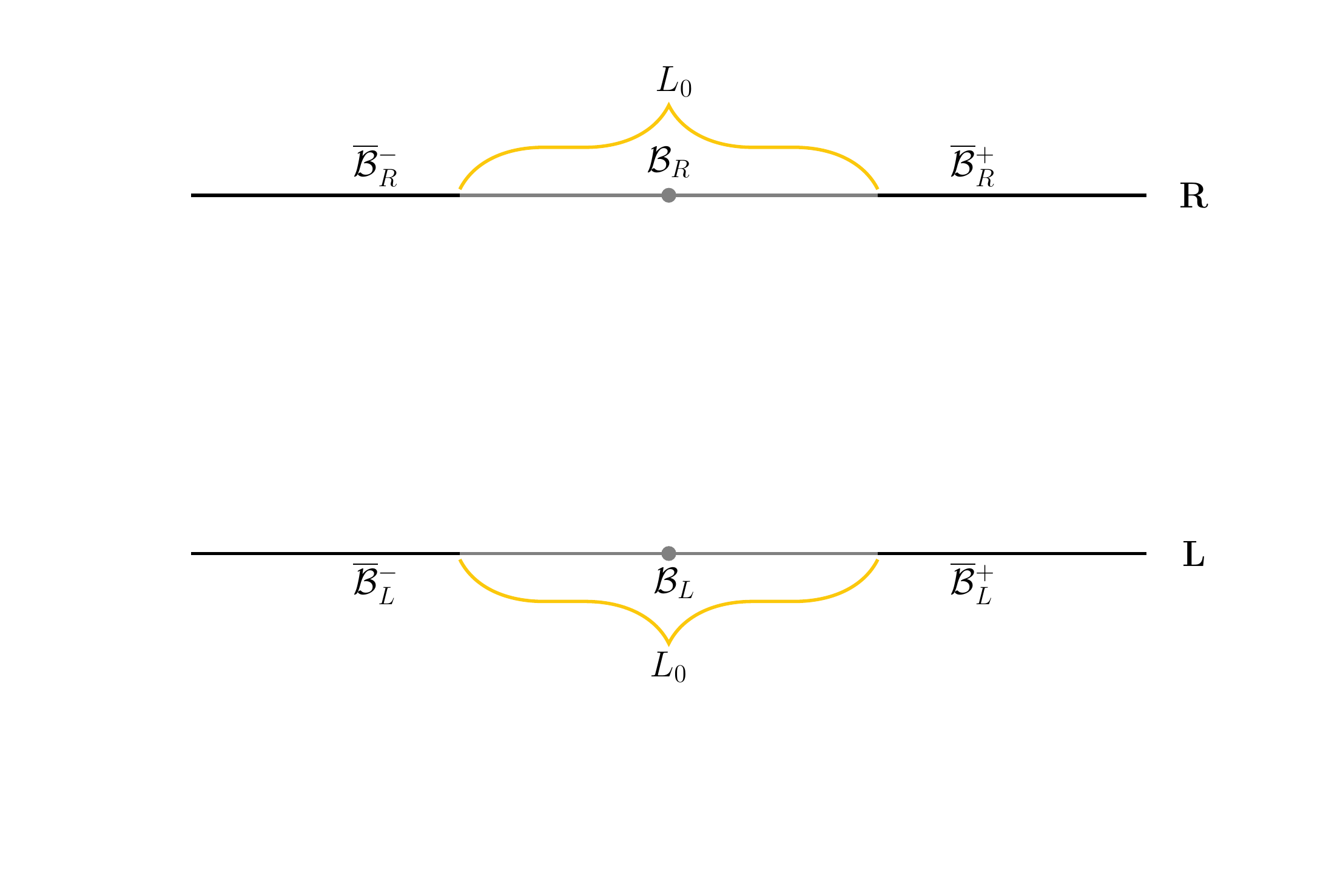}
\vskip-1.5cm
\caption{\small In this figure, our entanglement configuration is illustrated.}
\label{InfTrans}
\end{center}
\end{figure}

Before going ahead, let us consider the information transfer  from black holes to  radiations in our setup. Basically, this discussion is similar to the information transfer in  eternal BTZ black holes~\cite{Hartman:2013qma, Almheiri:2019yqk} but there are additional features because of the interface degrees of freedom. For a concrete discussion, let us denote the interval of our interest as ${\cal B}$ and its complement $\bar{\cal B}$, which  correspond to the black holes and radiations, respectively  from the two-dimensional gravity viewpoint. Our setup corresponds to the  two-sided black holes, and therefore it becomes a quadripartite system, ${\cal B}_{R}\cup {\cal B}_{L}\cup \bar{\cal B}_{R}\cup  \bar{\cal B}_{L}$.  The radiation parts may be further decomposed into the $\pm$ part as $\bar{\cal B}_{R/L} = \bar{\cal B}^{+}_{R/L} \cup   \bar{\cal B}^{-}_{R/L}$ in each R/L side, respectively, in this two dimensional case (See Figure~\ref{InfTrans}).  
For simplicity, we consider  the $+/-$ symmetric case with a R/L symmetric evolution.

 The initial TFD state in \eqref{initial}, which exhibits the maximal entanglement between the $\rm R$ and $\rm L$ sides, is pure. In our setup, we  begin with large entanglement between ${\cal B}_{R}$ and ${\cal B}_{L}$ and   also between $\bar{\cal B}_{R}$ and $\bar{\cal B}_{L}$\footnote{For  another initial entanglement case, see Ref.~\cite{Liu:2020gnp}.}. This may be achieved by taking a large length limit of the interval ${\cal B}$ as $\frac{2\pi}{\beta}L_{0} \gg 1$. In this limit, the initial entanglement between ${\cal B}_{R}$ and $\bar{\cal B}_{R}$  (${\cal B}_{L}$ and $\bar{\cal B}_{L}$) becomes very small, so  we may ignore it.  This initial setup may be phrased in terms of the mutual information as 
\begin{equation} \label{eq: mutual info br brb}
I({\cal B}_{R}, \bar{\cal B}_{R} ) (t=0) =I({\cal B}_{L}, \bar{\cal B}_{L} )  (t=0) \simeq 0\,. 
\end{equation}

Recall that the entanglement entropy $S({\cal B}_{R})=S({\cal B}_{L}) $ of the interval of  length $L_{0}$  can be obtained holographically by the RR or LL geodesics as given  in \eqref{LLEE}, which are time-independent.  In order to determine the entropy $2S(\bar{\cal B}^{+}_{R}\cup \bar{\cal B}^{+}_{L}) = 2S(\bar{\cal B}^{-}_{R}\cup \bar{\cal B}^{-}_{L})= S(\bar{\cal B}_{R}\cup \bar{\cal B}_{L})  =S({\cal B}_{R}\cup {\cal B}_{L})$ holographically, 
some care is needed since the  correct HEE should be taken as the minimum among the geodesics.
 In this regard, one may  rephrase    one version of the  information paradox~\cite{Hartman:2013qma,Caputa:2015waa} for eternal black holes in terms of  the mutual information. The mutual information of ${\cal B}_{R}$ and ${\cal B}_{L}$ is given by
\begin{equation} \label{}
I({\cal B}_{R},{\cal B}_{L}) = S({\cal B}_{R}) + S({\cal B}_{L}) - S({\cal B}_{R}\cup {\cal B}_{L})\,.
\end{equation}
If $S({\cal B}_{R}\cup {\cal B}_{L})$ is blindly taken  as the  doubled RL geodesic, the mutual information would become negative after the Page time, which  is a contradiction to the subadditivity. In other words, the non-negativity  of the mutual information  implies that  it  should be zero after the Page time and so the initial large entanglement between  ${\cal B}_{R}$ and ${\cal B}_{L}$ disappears after the Page time. In fact, we know that the correct  $S({\cal B}_{R}\cup {\cal B}_{L})$ needs to be taken by the combination of the  RR and LL  geodesics, as was done  above.  Concretely, one can obtain the explicit expression of $I({\cal B}_{R},{\cal B}_{L})$ from our bulk results as
\begin{equation} \label{}
I({\cal B}_{R},{\cal B}_{L})(t) = -\frac{2c}{3}Q\Big(A,B(t) \Big)\,.
\end{equation}
Indeed, since we have taken $L_{0}\gg \beta$, one can see that the large initial mutual information is given   by
\begin{equation} \label{}
I({\cal B}_{R},{\cal B}_{L})(t=0) = -\frac{2c}{3}Q\Big(A,B(t=0) \Big) = \frac{2c}{3}  \ln  \sinh{\pi L_0 \over \beta}+ \frac{c}{3}  \ln A \,.
\end{equation}

Now, one may wonder where the large initial entanglement goes after the Page time.  To see this,  
note that $S(\bar{\cal B}_{R/L})$ may also be obtained by the RR or LL geodesic and that it would be time-independent. Then, in conjunction with the {\it RL symmetry}, which denotes symmetry between the R and L system,  one may see  that $I({\cal B}_{R},\bar{\cal B}_{R})=I({\cal B}_{L},\bar{\cal B}_{L})$ is also time-independent, since the entanglement entropy of one side to the other, $S({\cal B}_{R}\cup \bar{\cal B}_{R})=S({\cal B}_{L}\cup \bar{\cal B}_{L})$, is time-independent in the TFD construction.  Recalling that $I({\cal B}_{R},\bar{\cal B}_{R})=I({\cal B}_{L},\bar{\cal B}_{L})$ was  initially close to zero (See \eqref{eq: mutual info br brb}),   one may note that ${\cal B}_{R/L}$ would be nearly disentangled from $\bar{\cal B}_{R/L}$ at any time.  By dividing the quadripartite state to bipartite ones, we can see that
\begin{align}    \label{}
S(\bar{\cal B}_{R}) &= S({\cal B}_{L}\cup \bar{\cal B}_{L} \cup {\cal B}_{R}) =  S({\cal B}_{R})+ S({\cal B}_{L}\cup \bar{\cal B}_{L}) - I({\cal B}_{R}, {\cal B}_{L}\cup \bar{\cal B}_{L}) \nonumber \\
&=  S({\cal B}_{R}) + S({\cal B}_{L}) + S(\bar{\cal B}_{L}) - I({\cal B}_{L},\bar{\cal B}_{L}) -  I({\cal B}_{R}, {\cal B}_{L}\cup \bar{\cal B}_{L})\,,
\end{align}
which,  together with the RL symmetry, leads to
\begin{equation} 
2S({\cal B}_{R}) =  2 S({\cal B}_{L})  =  I({\cal B}_{L},\bar{\cal B}_{L}) +  I({\cal B}_{R}, {\cal B}_{L}\cup \bar{\cal B}_{L})  \simeq  I({\cal B}_{R}, {\cal B}_{L}\cup \bar{\cal B}_{L})  \,.
\end{equation}
%
Due to the purity of the whole state $ {\cal B}_{R} \cup \bar{\cal B}_{R} \cup {\cal B}_{L}\cup \bar{\cal B}_{L}$, nearly vanishing entanglement of ${\cal B}_{R}$  with $\bar{\cal B}_{R}$ would lead to  nearly  maximal entanglement of ${\cal B}_{R}$ with ${\cal B}_{L}\cup \bar{\cal B}_{L}$.   Together with this, we assume that the entanglement structure  of our system  has a property that the strong subadditivity of the following form is nearly saturated\footnote{Interestingly, this can explicitly be written in terms of the so-called conditional mutual information as $I({\cal B}_{L}, \bar{\cal B}_{L}\, | \, {\cal B}_{R}) = I({\cal B}_{R}, \bar{\cal B}_{R}\, | \, {\cal B}_{L}) \simeq 0$  in our case. Or it may be rephrased that the density matrix for ${\cal B}_{R}$ is decomposed nearly into the direct sum of tensor products in an appropriate way   (See~\cite{Hayden} for a rigorous mathematical explanation).}
\begin{equation} \label{}
 S({\cal B}_{R}\cup {\cal B}_{L}) + S({\cal B}_{R}\cup \bar{\cal B}_{L})  \ge S({\cal B}_{R}) + S( {\cal B}_{R}\cup {\cal B}_{L} \cup \bar{\cal B}_{L})\,.
\end{equation}
This means that  $S({\cal B}_{R}\cup {\cal B}_{L}) + S({\cal B}_{R}\cup \bar{\cal B}_{L}) \simeq S({\cal B}_{R}) + S(\bar{\cal B}_{R})=S({\cal B}_{L}) + S(\bar{\cal B}_{L})$ in our setup. By using these relations, we deduce that
\begin{equation} \label{}
I({\cal B}_{R}, {\cal B}_{L}\cup \bar{\cal B}_{L})  \simeq  I({\cal B}_{R}, {\cal B}_{L})+ I({\cal B}_{R},\bar{\cal B}_{L}) \simeq 2S({\cal B}_{R}) =  2 S({\cal B}_{L})\,.
\end{equation}
Now, one can see that the decrease of $I({\cal B}_{R}, {\cal B}_{L})$ leads to the increase of $I({\cal B}_{R},\bar{\cal B}_{L})$ while their sum remains constant. 
As a result, the large initial entanglement between ${\cal B}_{R}$ and ${\cal B}_{L}$ is transferred to  that between ${\cal B}_{R}$ and $\bar{\cal B}_{L}$ (or R $\leftrightarrow$ L vice versa).  This tells us in our setup the entanglement or information transfer between ${\cal B}$ and $\bar{\cal B}$.

\section{Outside-horizon description of 2d gravities}\label{sec6}
In this section, we would like to provide a  2d effective description for our 3d bulk dynamics. This description should be  equivalent  to  the 3d bulk counterpart by definition. Except for the bulk CFT$_{2}$ part of our ICFT, all the remaining degrees of freedom (that are mostly associated with the interface) are described by effective 2d gravities. For this purpose, we propose the following procedure to obtain the effective 2d gravities. First, we assume the separation of $+$ and  $-$ sides in the  large $\ln A$ limit (See Figures~\ref{fig3} and~\ref{LRgeo}), which will be justified further later on.  In this limit, let us note that the shadow region, specified by $-\mu_I < \mu < \mu_I$, becomes  large. We shall remove this shadow region and, instead,
introduce  two branes at $\mu=\pm \mu_{I}$.   We then obtain  the 2d  
actions by following the standard Randall-Sundrum scenario~\cite{Randall:1999vf} where the corresponding brane dynamics play the role of replacing  that of the shadow region.  Some details of this procedure are  presented in Appendix~\ref{AppC}.



 As a summary of Appendix~\ref{AppC},   we obtain the 2d gravities ${\rm Grav}_\pm$, whose solutions  are given by the AdS$_2$ metrics
\be 
ds^2_\pm =\left[
 -(w_\pm^2-1) \frac{(2\pi)^2}{\beta^2}dt^2 + \frac{dw_\pm^2}{w_\pm^2-1}\right]\ell^2_2\,,
 \label{rindler}
\ee
where $\ell_2 =\ell \sqrt{f(\pm \mu_I)}$.
One also  notes that the Grav$_{\pm}$ actions imply that  some CFT's with  central charges $c$ should be present on the $\mu=\pm \mu_{I}$ surfaces~\cite{Chen:2020uac}. We view 
these CFT's 
as originating from the coupling to the CFT$_{\pm}$ living on the boundary of our 3d spacetime. The coupling is made through the AdS$_{2}$ boundary  cut-off surface with a transparent boundary condition as further specified below. As is given in Appendix~\ref{AppC}, one has a solution of  pure AdS$_{2}$ with a vanishing dilaton; this can also be checked from our 3d description. Namely in the 3d description of entropy, for instance, given in (\ref{3entropy}), there is no interface contribution that is linear in the temperature. 
Of course, the first term on the right hand side of  (\ref{3entropy}) is linear in the temperature and extensive  in the system size $L_s$, but this has nothing to do with our interface degrees of freedom.

The key point in our 2d interpretation stems from the fact that our 2d background is pure AdS$_{2}$ with a constant dilaton, not  nearly AdS$_{2}$ with so-called Schwarzian dynamics in~\cite{Maldacena:2016upp}.  In our case, the reparametrization modes of the AdS$_{2}$ cutoff surface  are fixed by the cutoff condition of our 3d bulk. This  gives us  the relation of the AdS$_{2}$ cutoff surface time $\tau$ to the CFT$_{2}$ time $t$. From 
our 2d perspectives, the transparent coupling between the two CFT's 
does not allow any non-trivial reparametrization modes. Therefore, in our setup there would be no stability issue~\cite{Almheiri:2014cka} of pure AdS$_{2}$  which arises mainly due to the back-reaction of  a dynamical dilaton. 
This is then a new type of 2d gravity which is certainly 
different from the conventional 
JT theory. 

We note that the AdS$_{2}$ radius $\ell_{2}$ is  of the same order as  its 3d counterpart $\ell$ as is given in Appendix~\ref{AppC}. Hence the short distance cutoff scale of our 2d description will be set by the scale $\ell_2 \sim \ell$, which  in turn is translated into the length scale $\beta$ in our ICFT on $\mathbb{R}^{1,1}$.
As depicted in Figure~\ref{LRgeo}, there is a non-trivial time evolution of the shadow region. This 3d bulk phenomenon may be interpreted as an RG flow of the AdS$_{2}$ dynamics and an emergence of new degrees of freedom in AdS$_{2}$ in the low energy regime, whose  transitional behaviors shall be explored in Section~\ref{sec8}.

In the above, we have taken the large $\ln A$ limit in such a way that two branes at $\mu=\pm \mu_{I}$ are treated as separate objects. On the other hand, at the length scale $\tilde{L} \gg \ell_{2} \ln A$, two branes lose their separate identities and act as a single brane, whose behavior shall be further described in Section~\ref{sec8}. Especially in this regime of extremely low energies or $t > t_P$, one finds that only the sum of the $+/-$ topological contributions, in the $\rm RR$ or $\rm LL$ entanglement entropies, 
will be fixed to be $S_I$  (See Sections \ref{sec7} and \ref{sec8}).

\subsection{Comparisons }   
\begin{figure}[htbp]  
\begin{center}
\includegraphics[width=0.7\textwidth]{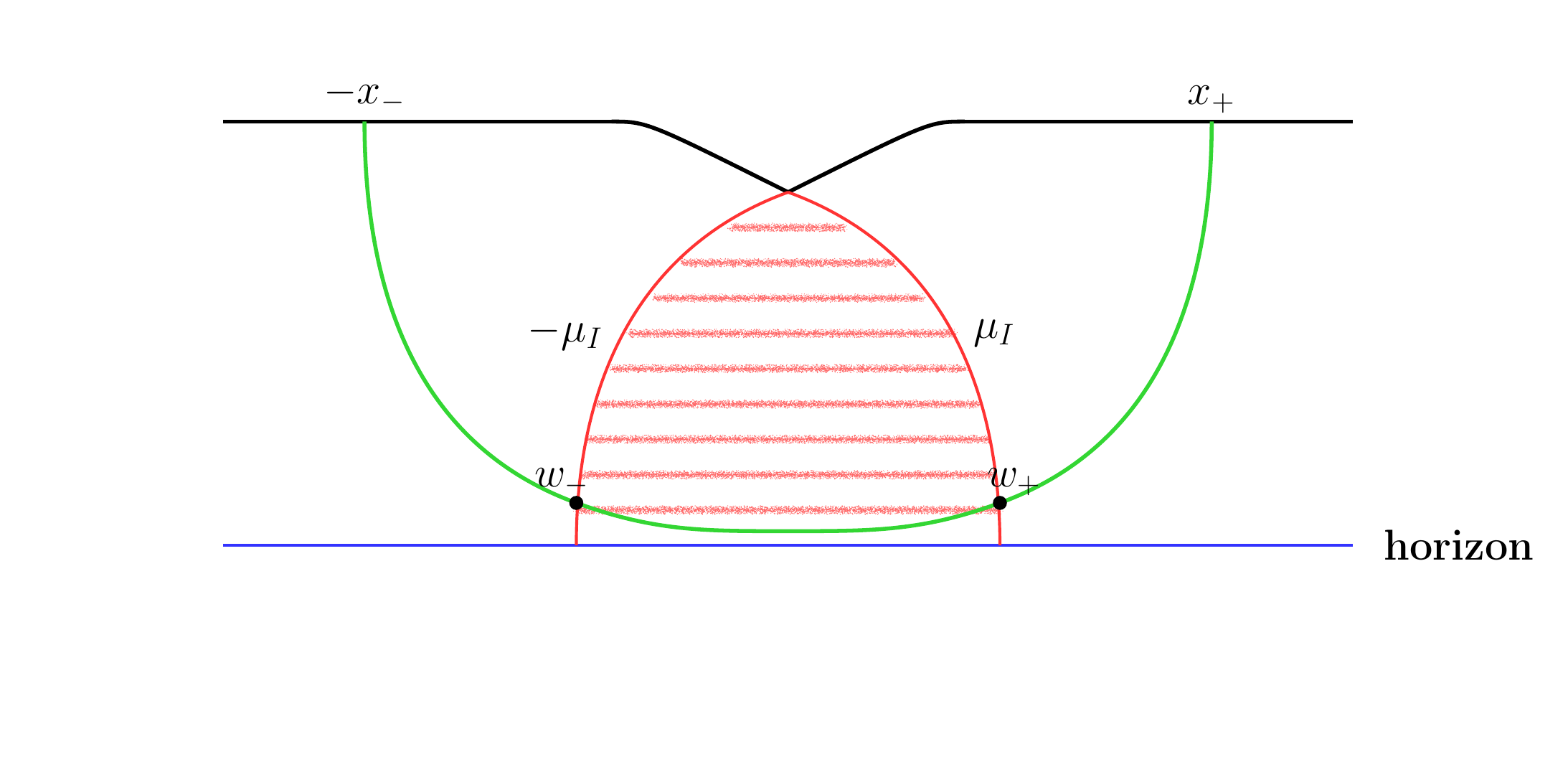}
\vskip-1cm
\caption{\small We draw here  a single-sided boundary-to-boundary extremal geodesic curve starting from $(-x_-, t)$ and ending on
$(x_+, t)$.  This entangling geodesic  intersects the $\pm \mu_I$ surfaces at $(w_\pm, t)$ respectively.}
\label{fonshell}
\end{center}
\end{figure}

In this subsection, we reproduce some 3d bulk results from the 2d perspective. Let us consider a single-sided boundary-to-boundary extremal 
curve starting from $(-x_-, t)$ ending on
$(x_+, t)$ where we take $x_\pm > 0$ such that  its trajectory passes through the shadow region. As drawn in Figure \ref{fonshell}, this entangling geodesic cuts the $\pm \mu_I$ surfaces at $(w_\pm, t)$ respectively where
$w_\pm$ will be a function of $x_\pm$ in general. When $\gamma$ is small, it is clear that $w_+(x_\pm)=w_-(x_\pm) +O(\gamma^2)$, which shows that ${\rm Grav}_\pm$ are strongly coupled  to each other. 

When $A$ becomes large, the value $w_\pm$ induced on the $\pm\mu_I$ surface approaches $w_{\pm \infty}$ as
\be
w_\pm=w_{\pm\infty}(x_\pm)+ O(1/\ln A)\,,
\label{onshell}
\ee
with the boundary  value $w_{\pm\infty}(x_\pm) = {\rm coth} \frac{2\pi}{\beta} x_\pm$, respectively.
To show this, we first note that the relevant geodesic equations in (\ref{eq: geodesic eq}) can be
integrated to give
\begin{align}    \label{}
\rho (y) &= 2 {\cal E}  \int^y_0 dy \frac{A}{\sqrt{\cosh 2y +A}\sqrt{\cosh 2y +A- 2A{\cal E}^2}}\,, \nonumber \\
 s(y) &= \int^y_0 dy \frac{\sqrt{\cosh 2y +A}}{\sqrt{\cosh 2y +A- 2A{\cal E}^2}}\,,
\label{geosol}
\end{align}
together with $t= constant$. 
In order to make the geodesic stay outside horizon, we will require 
${\cal E}^2 < \frac{1}{2} \left(1+1/A\right)$. 
Further assuming ${\cal E} \ll 1$, we may expand the above expressions with respect to $\cal E$ leading to 
\begin{align}    \label{}
\rho_{\infty}- \rho_{-\infty}&=  4 {\cal E} {\cal Q} (A) + O({\cal E}^3)\,, \nonumber \\
s_{\infty}-s_{-\infty}&= y_{\infty}-y_{-\infty} +2 {\cal E}^2 {\cal Q} (A) +O({\cal E}^4) \,,
\end{align}
where 
\be
{\cal Q}(A) \equiv  \int^\infty_0 dy \frac{A}{\cosh 2y +A} 
=\frac{1}{2}\ln 2A +O(A^{-2})\,.
\ee
Therefore, the integration constant $\cal E$ can be fixed as
\be
{\cal E}= \frac{\rho_\infty - \rho_{-\infty}}{4 {\cal Q}(A)}+O({\cal E}^3)\,,
\label{cale}
\ee 
and the corresponding renormalized geodesic distance becomes
\be
s_R= \ln A + \ln 2 \sinh \frac{2\pi}{\beta} x_+ +\ln 2 \sinh \frac{2\pi}{\beta} x_- +\frac{(\rho_\infty - \rho_{-\infty})^2}{8 {\cal Q}(A)} +O({\cal E}^4) \,,
\ee
where $\cosh \rho_{\pm\infty}= {\rm coth}\frac{2\pi}{\beta} x_\pm$. Thus one finds 
our assumption ${\cal E} \ll 1$ is fulfilled for any  choice of finite $x_\pm$ 
since the factor ${\cal Q}(A)$ in the denominator of (\ref{cale}) becomes large when $A$ becomes 
large enough. 
We conclude that the resulting entanglement  entropy for the interval $\bf [\tth\tth[-x_-,x_+]\tth\tth]$ becomes
\be
S =  S_{(+)}(x_+) +S_{(-)}(x_-) + O(1/\ln A)\,,
\label{3dentropy}
\ee
where
\be 
S_{(\pm)} (x_\pm)=\frac{c}{6}\ln 2 \sinh \frac{2\pi}{\beta} x_\pm +S^{(\pm)}_{\phantom{i}I}\,,
\ee
with $S^{(+)}_{\phantom{i}I}+S^{(-)}_{\phantom{i}I}=S_I$.
This shows an effective decoupling of the $(+)$ and $(-)$ theories when $\ln A \gg 1$. However the decoupling has a subtlety since the interface degrees of freedom 
will be shared by the $(+)$ and $(-)$ theories
at the same time. At the moment one may regard the interface contributions $S^{(\pm)}_{\phantom{i}I} (\ge 0)$ to be arbitrary once their sum is fixed to be $S_I$.
To complete our discussion here, we now compute the differences
\be
\rho_{\pm \infty}- \rho({\pm y_I})= \pm \frac{\rho_\infty - \rho_{-\infty}}{2 {\cal Q}(A)} {\cal Q}_I(A)+ O({\cal E}^3)\,,
\ee
where the coordinate values $\pm y_I$ again referring to the $\pm$ surfaces are defined by 
$\pm y_I = y(\pm \mu_I)$ and 
\be
{\cal Q}_I(A) \equiv  \int^\infty_{y_I} dy \frac{A}{\cosh 2y +A} 
=1 +O(A^{-2})\,.
\ee
Therefore the differences are of order $1/\ln A$, which demonstrates our claim in (\ref{onshell}).
The resulting  value $w_\pm (x_\pm)$ may be considered as on-shell  solution of the ${\rm Grav}_{\pm}$
 theory, respectively. Since ${\rm Grav}_\pm$  is coupled to ${\rm CFT}_\pm$ and $x_\pm$ represents coordinate  value in
${\rm CFT}_\pm$ respectively, the above result strongly suggests that ${\rm Grav}_+$/${\rm Grav}_-$ is coupled only to  ${\rm CFT}_+$/${\rm CFT}_-$ respectively. Hence we  conclude that ${\rm Grav}_+\times {\rm CFT}_+$ and ${\rm Grav}_-\times {\rm CFT}_-$ are effectively decoupled from each other as $\ln A \gg 1$. 
Below we shall focus on the nature 
of the $(\pm)$ theories in the limit $\ln A \gg 1$ safely ignoring any possible interactions between them.

We posit here one possible description of ${\rm Grav}_\pm \times {\rm CFT}_\pm$ for the region outside horizon, which is based on straightforward re-interpretation of our 3d bulk computation. We shall check our proposal in various limiting cases later on. For the $(+)$ theory of ${\rm Grav}_+ \times {\rm CFT}_+$, we choose the following coordinate system. We first introduce spatial coordinate
$a_\pm > 0$ by $w_\pm= {\rm coth} \frac{2\pi}{\beta} a_\pm$ in the gravity side. For the AdS Rindler wedge of the black hole spacetime, let us introduce coordinates $\sigma^\pm_{(+)}=
t \mp a_+$ with a restriction $\sigma^+_{(+)} < \sigma^-_{(+)}$. The  metric in (\ref{rindler}) becomes
\be
ds^2_{(+)} =
 -\frac{ d\sigma^+_{(+)}d\sigma^-_{(+)}}{\sinh^2 \frac{\pi }{\beta}(\sigma^+_{(+)}-\sigma^-_{(+)})}\left(\frac{2\pi \ell_2}{\beta}\right)^2 \,.
 \label{rindler+}
\ee
For the flat spacetime region of ${\rm CFT}_+$, we introduce 
the coordinates  $\sigma^\pm_{(+)}=t \pm x_+$ with the flat metric
\be
ds^2_{(+)} =
 -{ d\sigma^+_{(+)}d\sigma^-_{(+)}}\,,
 \label{flat+}
\ee
with the range $\sigma^+_{(+)} > \sigma^-_{(+)}$. These two  charts are joined through the surface
$\sigma^+_{(+)} =\sigma^-_{(+)}$ and then the whole coordinate range of $(\sigma^+_{(+)},\sigma^-_{(+)})$
covers the entire planar region of $\mathbb{R}^2$. 

For the $(-)$ theory, one has 
$\sigma^\pm_{(-)}=t \pm a_-$ for the black hole part with the restriction
$\sigma^+_{(-)} > \sigma^-_{(-)}$ and $\sigma^\pm_{(-)}=t \mp x_-$ for the flat space of ${\rm CFT}_-$ with 
$\sigma^+_{(-)} < \sigma^-_{(-)}$. The metric in the black hole/the flat region is respectively   given by (\ref{rindler+})/(\ref{flat+}) 
with all the subscripts  $(+)$ replaced by $(-)$.

As we described earlier, 
our original  CFT (with the central charge $c$) on the flat region of the 2d 
spacetime is extended into an outside-horizon region of the black hole spacetime\footnote{As will be clarified in Section \ref{sec8}, they may be further extended into the behind-horizon region excluding any  such region where extra AdS$_2$ matter is excited.}.
This determines basically the coupling between Grav$_\pm$ and ${\rm CFT}_\pm$. 
Recall that our pure gravity part is solely given by the topological contribution. 
The total topological contribution is non-dynamical and shared by Grav$_+$ and Grav$_-$. 
In this sense, the $(\pm)$  theories are not completely decoupled from each other. 

With these preliminaries, the generalized entropy for the interval $[-a_+,x_+ ]\cup [-x_-,a_- ]$, which  includes quantum matter contribution, can be identified as \cite{Almheiri:2019yqk}
\be
S_{\rm gen}(a_\pm,x_\pm)= S_+(a_+,x_+)+S_-(a_-,x_-) \,,
\label{gen}
\ee
where
\be
 S_{(\pm)}(a_\pm, x_\pm)=\frac{c}{6} \ln \frac{2\sinh^2 \frac{\pi}{\beta} (a_\pm+x_\pm)  }{\sinh \frac{2\pi}{\beta} a_\pm }+S^{(\pm)}_{\phantom{i}I} \,.
\ee
Note that  each $S^{(\pm)}_{\phantom{i}I}$  is  from the topological contribution of the pure gravity part. 

\subsection{ Some checks}
Let us now justify the above expression of the generalized entropy. First of all, its extremization 
with respect to $a_\pm$ leads to the conditions
\be
{\rm coth}  \frac{\pi}{\beta} (a_\pm+x_\pm) ={\rm coth} \frac{2\pi}{\beta} a_\pm  \,.
\ee
Their solutions are simply given by
\be
a_\pm = x_\pm \,,
\ee
which lead to the quantum extremal entropies
\be
S_{\rm ext}^{(\pm)}=  \frac{c}{6}\ln 2 \,{\sinh \frac{2\pi}{\beta} x_\pm  }
+S^{(\pm)}_{\phantom{i}I} \,.
\ee
The solutions and  the resulting extremal entropies perfectly agree with those from the 3d gravity 
in (\ref{onshell}) and (\ref{3dentropy}).  Thus we have checked  the  validity of the 2d description at least on-shell. 

\begin{figure}[htbp]  
\begin{center}
\includegraphics[width=0.7\textwidth]{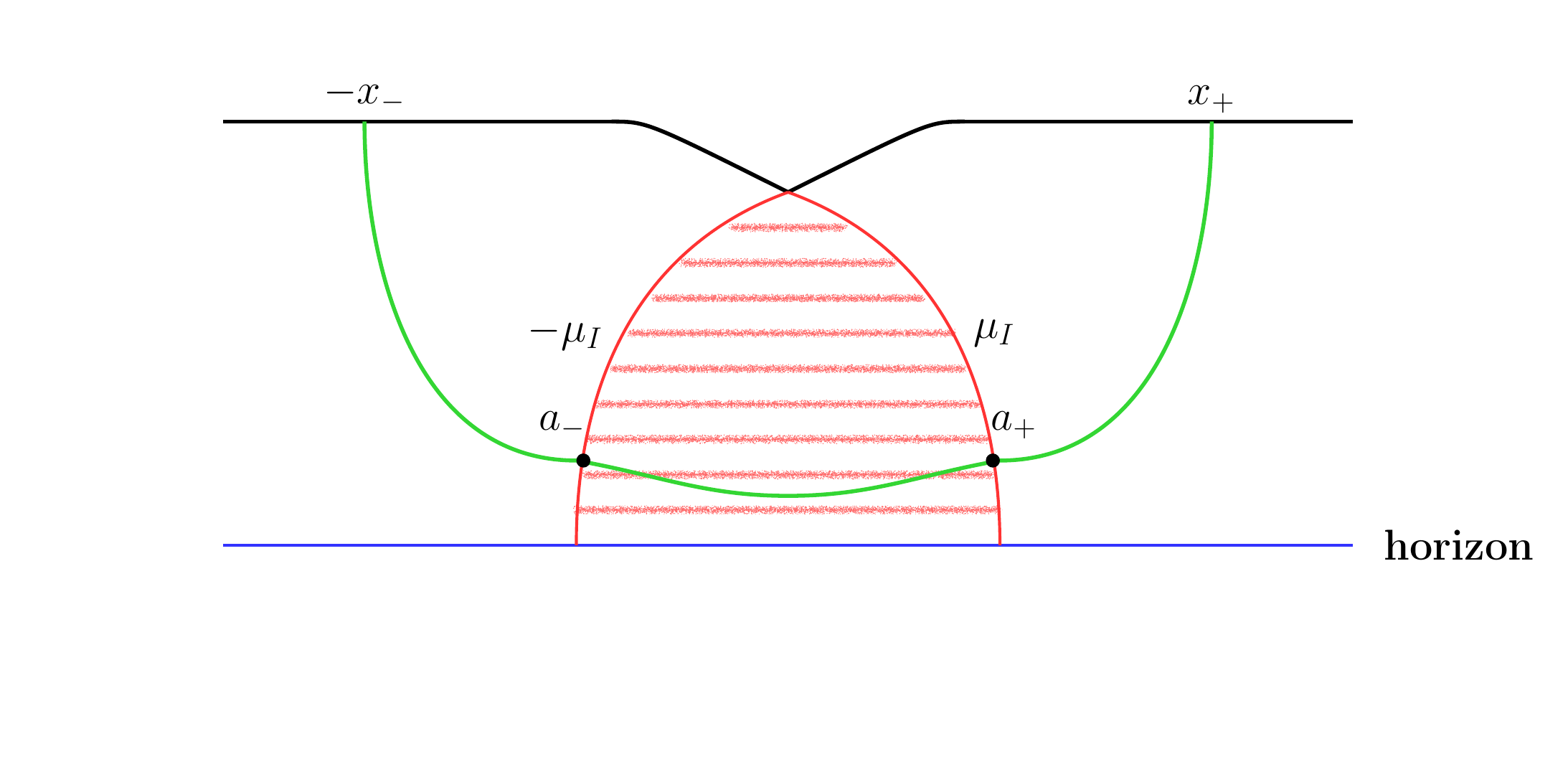}
\vskip-1cm
\caption{\small In this figure, we depict an off-shell configuration where one connects $(x_+,t)$ to $(a_+,t)$, 
 $(a_+,t)$ to $(a_-,t)$, 
 and $(a_-,t)$ to $(-x_-,t)$ with each segment  connected extremally.}
\label{foffshell}
\end{center}
\end{figure} 

We now check the  generalized entropy in (\ref{gen}) at its off-shell level. For this let us consider
an off-shell configuration where one connects $(x_+,t)$ to $(a_+,t)$ on the $+\mu_I$ surface,
 $(a_+,t)$ to $(a_-,t)$ on the $-\mu_I$ surface,  and $(a_-,t)$ to $(-x_-,t)$ with each segment  connected extremally. See its illustration in Figure \ref{foffshell}.
The configuration in total will be geodesic when $a_\pm = x_\pm +O(1/\ln A)$ as mentioned previously. For each segment, we apply the solution in (\ref{geosol}) by matching 
the starting  and the ending values of $\rho$ coordinate, which will fix the integration constant 
$\cal E$ uniquely.   We first consider the case where $\frac{2\pi}{\beta} x_\pm \gg 1$ and $\frac{2\pi}{\beta} a_\pm \gg 1$. In this case, one finds that ${\cal E} \ll 1$ for each segment
and the solution in (\ref{geosol}) can be expanded with respect to $\cal E$ as was done previously. For the extremal curve connecting $(a_\pm,t)$ to $(\pm x_\pm,t)$, we note that 
\begin{align}    \label{}
\rho_{\pm\infty} &= 2 e^{- \frac{\pi}{\beta} x_\pm} +O( e^{- \frac{3\pi}{\beta} x_\pm} ) \,, \nonumber \\
\rho (\pm y_I)  &= 2 e^{- \frac{\pi}{\beta} a_\pm} +O( e^{- \frac{3\pi}{\beta} a_\pm} )\,,
\end{align}
and then 
\be
{\cal E}= \frac{e^{- \frac{\pi}{\beta} x_\pm} - e^{- \frac{\pi}{\beta} a_\pm}}{ {\cal Q}_I(A)}+
O\left[\bigl({\scriptsize e^{- \frac{\pi}{\beta} x_\pm} \tth +\tth e^{- \frac{\pi}{\beta} a_\pm} }\bigr)^3\right] \,.
\label{caleoff}
\ee 
The resulting renormalized  extremal distance becomes
\be
s_{a_\pm \rightarrow x_\pm}= \frac{1}{2}\ln A -y_I + \ln 2 \sinh \frac{2\pi}{\beta} x_\pm  +\frac{(e^{- \frac{\pi}{\beta} x_\pm} -e^{- \frac{\pi}{\beta} a_\pm} )^2}{ {\cal Q}_I(A)} +O\left[\bigl({\scriptsize e^{- \frac{\pi}{\beta} x_\pm} \tth +\tth e^{- \frac{\pi}{\beta} a_\pm} } \bigr)^4\right]\,.
\ee
The extremal distance from  $(a_-,t)$ to $(a_+,t)$ can also be computed in a similar way 
 leading to
\be
s_{a_- \rightarrow a_+}= 2 y_I + \frac{(e^{- \frac{\pi}{\beta} a_+} -e^{- \frac{\pi}{\beta} a_-} )^2}{ {\cal Q}_{-+}(A)} +O\left[\bigl({\scriptsize e^{- \frac{\pi}{\beta} a_+} \tth +\tth e^{- \frac{\pi}{\beta} a_-} } \bigr)^4\right]\,,
\label{length-+}
\ee
where
\be 
{\cal Q}_{-+}(A)  \equiv  \int^{+y_I}_{-y_I} dy \frac{A}{\cosh 2y +A} \,.
\ee
The first term in this expression is independent of $a_\pm$ and gives the topological contribution of $S_I$ if one includes the constant terms of the remaining segments. Noting
\be 
{\cal Q}_{-+}(A)= \ln A +\cdots  \,,
\ee 
one may ignore the second term of (\ref{length-+}) in the limit $\ln A \gg 1$.
Therefore one finds the total contribution to the generalized entropy
becomes 
\be
S_{\rm tot}=S_{\rm gen}(a_\pm, x_\pm) 
 + O\left[\bigl({\scriptsize e^{-\frac{\pi}{\beta} a_+}\tth +\tth e^{- \frac{\pi}{\beta} a_-} } \bigr)^2/\ln A\right] +
O\left[\bigl({\scriptsize e^{- \frac{\pi}{\beta} a_+} \tth +\tth e^{- \frac{\pi}{\beta} a_-}}\tth +\tth e^{- \frac{\pi}{\beta} x_\pm}\bigr)^4\right] \,.
\ee 
Hence we have an agreement with (\ref{gen}) ignoring the higher order correction terms.
Finally we consider the off-shell configuration where
$a_\pm = x_\pm + \delta a_\pm$ but with no further assumption on $x_\pm > 0$. It is straightforward to show that
\be
S_{\rm tot}=S_{\rm gen}(x_\pm +\delta a_\pm, x_\pm) 
+O(\delta a_\pm^2 /\ln A)+
O(\delta a_\pm^4)\,.
\ee
Hence, one  has again a perfect agreement with (\ref{gen}) up to the 
 order of  $\delta a_\pm^2$. 

\section{ICFT description of entanglement entropy 
}\label{sec7}
Before going on, we would like to explain the ICFT computation of the entanglement entropy and its relation to our HEE in the previous sections rather 
schematically\footnote{We are working with our Janus ICFT which has the corresponding dual gravity description. Therefore, note that  some results in this section depend on the microscopic details of underlying AdS/CFT correspondence.}. The main 
object we are interested in is 
the reduced density matrix $\rho_{{\cal I}{\cal I}}$ of the  Janus TFD state (\ref{tfda}) over the RL intervals 
 ${\cal I}{\cal I}\equiv {\cal I}_{L}\cup {\cal I}_{R}={\bf [\tth\tth[-x_-,x_+]\tth\tth]_L \cup [\tth\tth[-x_-,x_+]\tth\tth]_R}$ at time $t_L=t_R=t\ge 0$.  As before, the trace of its $n$-th power can be computed using the R/L twist operators $\Phi^\pm_{nR/L}(t,x)$
  by
\bea
{\rm tr}\rho^n_{{\cal II}}=\langle  \Phi^+_{nR}(t,x_+) \Phi^-_{nR}(t,-x_-) \Phi^+_{nL}(t,-x_-)\Phi^-_{nL}(t,x_+)\rangle_{\rm JTFD} \,.
\label{tracerhon}
\eea 
Then the corresponding entanglement entropy is given by
\bea
S^{EE}_{\,\, {\cal I}{\cal I}}(t)= -\lim_{n\rightarrow 1} \frac{\partial}{\partial n}{\rm tr}\rho^n_{{\cal I}{\cal I}} \,.
\eea
The above four-point function on the Janus TFD can be mapped to a four-point correlation function on a single $\mathbb{R}^2$ by the exponential map \cite{Hartman:2013qma}
\be
\pm X^\pm = e^{\pm \frac{2\pi}{\beta} x^\pm_{R/L}} \,.
\ee
where
\bea
x_R^\pm &=& \ \ \  t \pm x \,, \cr
x_L^\pm &=& -\Bigl(t+\frac{\beta}{2}i\Bigr)\pm x  \,.
\eea
Namely the trace in (\ref{tracerhon}) can be mapped to
\bea
{\rm tr}\rho^n_{{\cal II}}=\langle  \Phi^+_{n}(X^\pm_1) \Phi^-_{n}(X^\pm_2) \Phi^+_{n}(X^\pm_3)\Phi^-_{n}(X^\pm_4)\rangle_{\rm ICFT} \,.
\label{tracerhona}
\eea 
with 
\bea
x_1^\pm &=& \ \ \  t \, \pm  \, x_+ \,, \cr
x_2^\pm &=&  \ \ \  t \pm (-x_-) \,, \cr
x_3^\pm &=& -\Bigl(t+\frac{\beta}{2}i\Bigr) \pm (-x_-)  \,, \cr
x_4^\pm &=& -\Bigl(t+\frac{\beta}{2}i\Bigr) \pm \, x_+  \,,
\eea
where the expectation value of operators is taken over
the ICFT vacuum state on $\mathbb{R}^{1,1}$. Hence the HEE computation of Sections 
\ref{sec4}
and \ref{sec5} should be understood as the above four-point function with 
the choice $x_+= x_-= x > 0$. In this case, the remaining $SO(2,1)$  symmetry of the ICFT dictates the general form of the four-point function to be
\bea
{\rm tr}\rho^n_{{\cal II}}|_{x_\pm=x}
=\frac{1}{\Bigl(2\sinh \frac{2\pi}{\beta}x\Bigr)^{4\Delta_n}} \Bigl(G_n (\xi) \Bigr)^2 \,,
\eea
where $\xi$ is the cross ratio given by
\bea
\xi =\frac{\cosh^2 \frac{2\pi}{\beta}t}{\sinh^2 \frac{2\pi}{\beta}x} \,.
\label{xixt}
\eea
Note that the holographic counterpart of $\ln\sqrt{\xi}$ is the function $P(A,B)$.

The function $Q(A,B)$ of HEE side defined over $0\le \xi \le
\xi(t_P)$ is then related to $G_n$ by
\bea
\frac{c}{3} \bigg[Q\Big(A, B(\xi) \Big)+\ln \sqrt{A}\bigg] = -\lim_{n\rightarrow 1} \frac{\partial}{\partial n} G_n(\xi)  \,,
\eea
where $B(\xi)$ is defined by the relation $\xi= 
\sinh^2 q_\infty (A,B)
$ together with $q_\infty (A,B)$ in (\ref{qinfty}).
The $\xi \rightarrow 0$ limit is the so-called bulk OPE limit
where the presence of our interface can be ignored. Namely, when $\xi \ll 1$, one has
\bea
G_n \simeq 
G_0 \, \xi^{-\Delta_n}  \,,
\label{gblimit}
\eea
where $G_0$ is  a constant independent of $n$. This basically follows from the bulk OPE limit since the inserted points are relatively far away from the interface  and thus the presence of the interface can be safely ignored. 
From this, one may recover the small $\xi$ behavior
\bea
S^{EE}_{{\cal II}} 
\simeq 
\frac{2c}{3}\ln 2\cosh \frac{2\pi}{\beta}t \,,
\label{xitozero}
\eea   
which agrees with our HEE result given in (\ref{LREEa}) and (\ref{LREE}). On the other hand,  the transition occurs at  $t=t_P$; when $\xi \ge \xi(t_P)$, the corresponding expression of $[G_n]^2$ in the strongly coupled regime becomes
\be
\big[G_n (\xi)\big]^2 ={\widetilde{G}}^2_0\, A^{\frac{c}{3}(1-n)}  \,,
\label{adep}
\ee 
whose  $n$-dependence is  determined from the HEE expression in (\ref{LLEE}). 
The $A$ dependence  comes from the interface identity operator. Namely, $e^{\frac{c}{3}(1-n)\ln A}$  stems from the degeneracy factor of the interface ground states in  $n$ copies of replicas of the ICFT. 
In the intermediate region of $0 \le \xi \le \xi(t_P)$, the detailed dynamics of the
RL extremal curve plays a role, which was discussed briefly in Sections \ref{sec4} and \ref{sec5}.
In the limit $A \rightarrow 1$, the interface degrees of freedom disappear completely and one regains the full conformal symmetry out of $SO(2,1)$.
In this case, the behavior in (\ref{xitozero}) will be valid over the full region of $0 
\le \xi \le \xi(t_P)$ if one assumes   the large $c$ limit of holographic theories \cite{Hartman:2013qma}.  

Below we shall be mainly concerned with the large deformation limit, $\ln A \gg 1$, with general $x_\pm \gg \beta$. In this case, we again have an
effective ($\pm$) 
separation of the Janus TFD theory. Namely, one has an effective factorization\footnote{This factorization fails in a subtle manner when $t \ge t_P$ because the interface degrees of freedom are shared by the ($\pm$) theories. We shall clarify this subtlety later on.}
\bea
{\rm tr}\rho^n_{{\cal II}}\simeq \langle  \Phi^+_{nR}(t,x_+) \Phi^-_{nL}(t,x_+)\rangle_{\rm TFD} \,\langle \Phi^-_{nR}(t,-x_-) \Phi^+_{nL}(t,-x_-) \rangle_{\rm TFD}   \,.
\label{separation}
\eea 
The resulting ($\pm$) dynamics has the interpretation of 
${\rm Grav}_{\pm} \times {\rm CFT}^R_{\pm} \times {\rm CFT}^L_{\pm}$ respectively.  Furthermore each $(\pm)$ theory has a corresponding BCFT interpretation where some part of the interface degrees of freedom play  the roles of boundary degrees of freedom. Again the $(\pm)$ two-point functions have the general forms
\bea
\langle  \Phi^+_{nR}(t, \pm x_\pm) \Phi^-_{nL}(t,\pm  x_\pm)\rangle_{\rm TFD} =
\frac{1}{
\Bigl(2\sinh \frac{2\pi}{\beta}x_\pm
\Bigr)^{2\Delta_n}
} 
 G^{(\pm)}_n (\xi_\pm)   \,,
\eea
where the $(\pm)$ cross ratios are respectively given by
 \bea
\xi_\pm =\frac{\cosh^2 \frac{2\pi}{\beta}t}{\phantom{i}\sinh^2 \frac{2\pi}{\beta}x_\pm} \,.
\eea
Then, the function $Q(A,B)$ of the HEE side defined over $0\le \xi_\pm \le
\xi_\pm(t_P)$ is again  related to $G^{(\pm)}_n$ by
\bea
\frac{c}{3}\Big[Q(A, B(\xi_\pm))+\ln\sqrt{A}\Big] = - \lim_{n\rightarrow 1} \frac{\partial}{\partial n} G^{(\pm)}_n(\xi_\pm) \,.
\label{pmg}
\eea
However, as we shall clarify below, there remain  some subtle dynamical correlations between the $(\pm)$ theories since the interface degrees of freedom are shared by the $(\pm)$ theories. 

\section{Islands and behind-horizon dynamics}\label{sec8}
In this section, we shall  be mainly concerned with the behind-horizon dynamics of the region
$0\le \xi_\pm \le \xi_\pm(t_P)$, which is described by the RL extremal curves holographically. We again
assume $\ln A \gg 1$ such that one may 
trust our 2d gravity description of the  ${(\pm)}$ 
theories. In this section, we shall omit any possible corrections of order $1/\ln A$ for the simplicity of our presentation.

\begin{figure}[thb!]
\vskip-0.5cm
\centering  
\includegraphics[height=6.3cm]{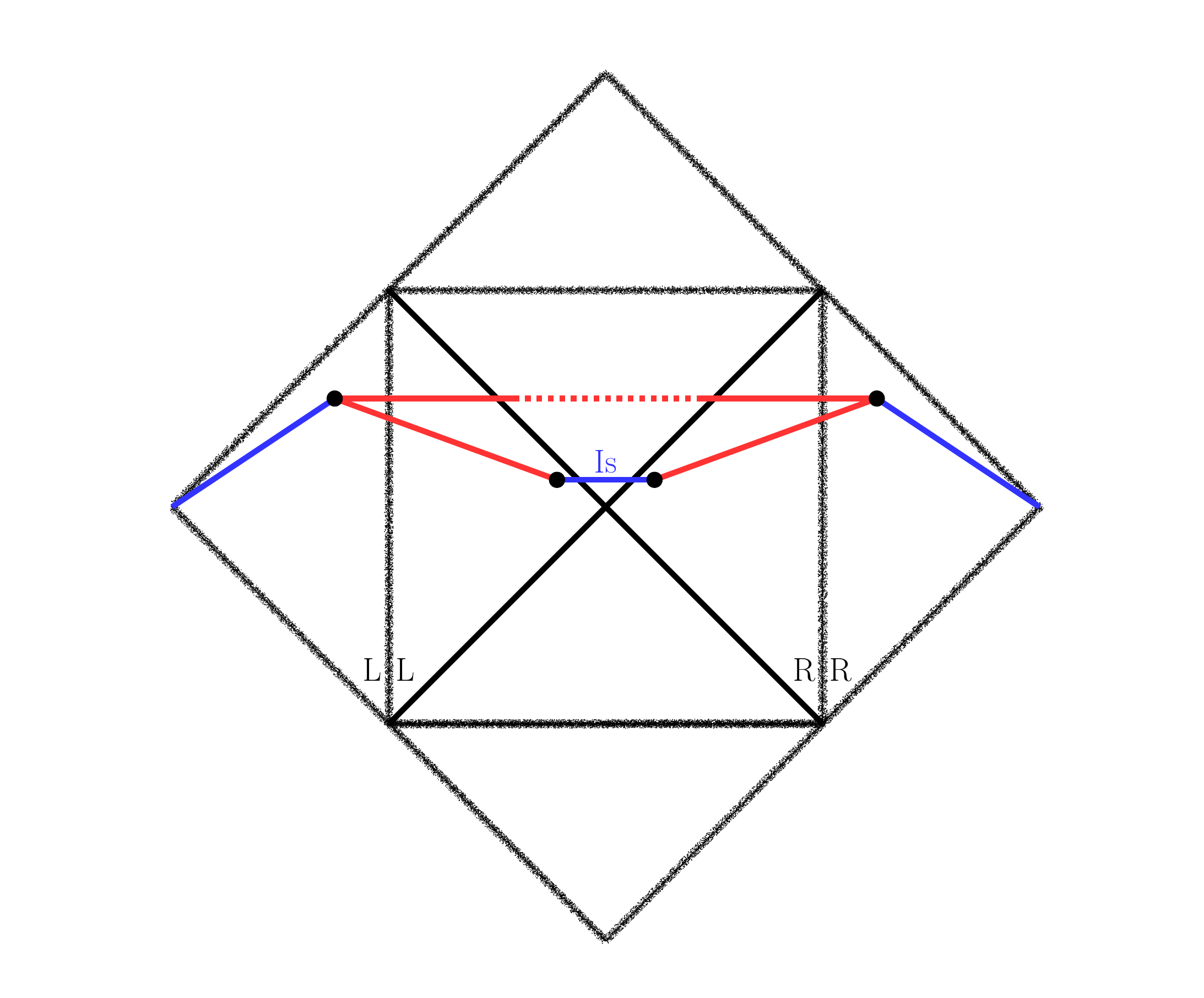} 
\vskip-0.5cm
\caption{\small 
We draw the Penrose diagram of the full 2d spacetime. One two-sided AdS$_2$ black hole spacetime
is joined to the R/L flat regions of $\sigma^+_{R/L} > \sigma^-_{R/L}$  along $\sigma^+_{R/L} = \sigma^-_{R/L}$. The upper R/L point has coordinates
$(t_{R/L},x_{R/L})=(t, x_+)$ with our choice. As was shown in Section \ref{sec6}, 
before extremization, the lower R/L point has coordinates $(t,-a_{R/L})$, which will be fixed to be $(t,-x_\pm)$ after extremization. The  blue line denoted by ``Is" is for the island configuration.
}
\label{fig10}
\end{figure}

First, let us describe the spacetime picture of ${\rm Grav}_{\pm}\times {\rm CFT}^R_{\pm}\times {\rm CFT}^L_{\pm} $. We present here only the case of the $(+)$ theory
as the $(-)$ theory can be treated in a parallel manner.  Below we basically follow the reference \cite{Almheiri:2019yqk}. Let us begin by introducing two copies of $\mathbb{R}^2$ coordinates $\sigma^\pm_R=t_R\pm x_R$ and $\sigma^\pm_L=-t_L\pm x_L$ covering R/L AdS$_2$ Rindler wedge  for $x_{R/L} <0$ joined to the R/L flat spacetime   $x_{R/L}>0$. The metric for the R/L Rindler wedge is given by
\be
ds^2_{R/L} =
 -\frac{ d\sigma^+_{R/L}d\sigma^-_{R/L}}{\sinh^2 \frac{\pi }{\beta}(\sigma^+_{R/L}-\sigma^-_{R/L})}\left(\frac{2\pi \ell_2}{\beta}\right)^2\,,
 \label{rindlerRL}
\ee
for the region $\sigma^+_{R/L} <\sigma^-_{R/L}$ and the one for
the R/L flat region  $\sigma^+_{R/L} > \sigma^-_{R/L}$  by
\be
ds^2_{R/L} =
 - d\sigma^+_{R/L}d\sigma^-_{R/L} \,.
 \label{flatRL}
\ee
Those two regions in each set are joined along $\sigma^+_{R/L} = \sigma^-_{R/L}$ as described by vertical lines in Figure \ref{fig10}.

The above two  copies of $\mathbb{R}^2$ can be mapped to a single 
$\mathbb{R}^2$ with coordinates $U^\pm$ by the exponential map,
$U^\pm =\pm e^{\pm \frac{2\pi }{\beta}\sigma^\pm_{R}}$ and 
$U^\pm =\mp e^{\pm \frac{2\pi }{\beta}\sigma^\pm_{L}}$ \cite{Almheiri:2019yqk}. The R/L flat regions specified by $U^+U^- <-1$ have the metric
\be
ds^2= \,\,\frac{ dU^+dU^-}{U^+ U^-}
\frac{\beta^2}{4\pi^2}
\label{flatmetric}\,,
\ee 
whereas the two-sided black hole spacetime specified by $  -1 <U^+ U^- $ has the metric
 \be
ds^2= -\frac{ 4dU^+dU^-}{(1+U^+ U^-)^2}\ell_2^2 \,.
\ee 
In this 
 coordinate system,  the surface $U^+U^-=-1$ is the junction of
the black hole and the R/L flat regions. 

\begin{figure}[htb!]
\vskip-0.5cm
\centering  
\includegraphics[height=5cm]{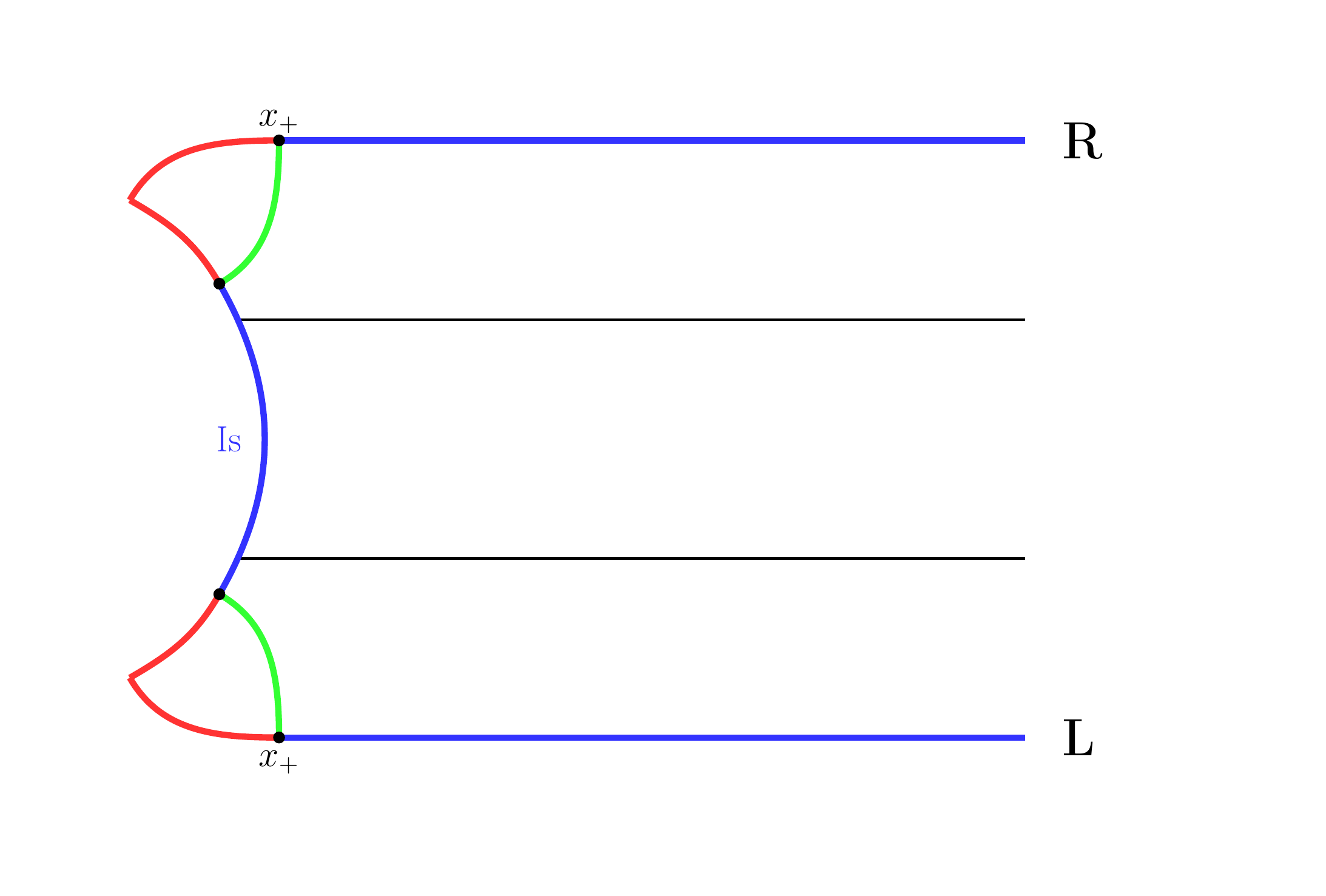}
 \vskip-0.5cm
\caption{\small 
The  bulk picture is given for  $t \ge t_P$. The time slice of the configuration is chosen as follows; 
except the island plus its bulk extension, which is in the constant $\tau$ slice, all the remaining regions are in constant $t$ slice. The green-colored curves represent the bulk extremal curves and the blue curve denoted by  ``Is''   stands for the island configuration.}
\label{fig11}
\end{figure}

As was mentioned already, at $t=t_P$, there will be a transition from the RL connecting extremal curves to the RR/LL extremal curves in the bulk side. After the transition, the bulk picture is given 
in Figure \ref{fig11}.  
The time slice of the configuration is chosen as follows; 
except for the island  plus its bulk extension which is in the constant $\tau$ slice, all the remaining regions are in constant $t$ slice.
The green curves represent the relevant part of the RR/LL bulk 
extremal curves. The 2d
boundary of the relevant 
bulk spacetime is given in Figure  \ref{fig10}.
In this 2d picture, the whole configuration after the  Page time consists of two blue curves connecting  $(t,\infty)_{R/L}$ and  $(t,x_+)_{R/L}$, two red curves connecting $(t,x_+)_{R/L}$ and 
$(t, -a_{R/L})_{R/L}$ and the so-called island  curve connecting $(t, -a_R)_{R}$ and 
$(t, -a_L)_{L}$. 
In Section \ref{sec6}, we have shown that the corresponding generalized entropy in (\ref{gen}) is
minimized with $a_{R/L}=x_+$. The bulk extremal curves are then
represented by the two
red curves connecting $(t,\,x_+)_{R/L}$ and 
$(t, \tth\tth-\tth\tth x_+)_{R/L}$ as depicted in Figure \ref{fig10}. 
Note that,  except for the island  which is along the corresponding 
constant $\tau$ slice, all the remaining  curves are along the constant $t$ slice upon extremization. 
Adding the contribution of the ($-$) theory,   we have
\bea\label{llrr}
S_{\cal II}\ \ \ \ 
&=&\  S_{(+)} +S_{(-)}  \,, 
\\
S_{(\pm)}(t \tth \ge\tth  t_P)
&=& \frac{c}{3} 
\ln 2\sinh \frac{2\pi}{\beta} x_\pm +S^{(\pm)}_{\phantom{\,}I}  \,,
\label{afterpage}
\eea
where  the topological contributions $S^{(\pm)}_{\phantom{\,}I}$ are constrained by
 $S^{(+)}_{\phantom{\,}I}+S^{(-)}_{\phantom{\,}I}=2S_I$ as was explained before.
We shall specify the values of $S^{(\pm)}_{\phantom{\,}I}$ later on.

Since the full two dimensional theories are unitary, one may alternatively obtain $S_{(+)}
$ by the QES including island contribution \cite{Almheiri:2019yqk}
\bea
S_{(+)}
={\rm min}\,{\rm ext} \Big[S^{(+)}_{\phantom{\,}I}+ S^{\rm matter}_{[x_+,\infty)_L \cup[x_+,\infty)_R\cup Is}\, \Big] \,,
\eea
where the topological term $S^{(+)}_{\phantom{\,}I}$ is  the geometric contribution from the end points of the island and the second term  from the 2d matter contribution of the relevant intervals. (Of course, one has a parallel story for the ($-$) theory.)
Therefore we conclude that the island is formed after the Page time and the degrees of freedom in the island region are entangled with radiation of
the region $[x_+,\infty)_R \cup [x_+,\infty)_L$. Since the island is connected to the radiation through the 3d bulk, the development of entanglement between them seems rather clear.  Also note that the island contribution should be included in the original ICFT computation of the entanglement entropy of the intervals
$\bf [\tth\tth[x_+, \infty)\tth\tth)_L \cup [\tth\tth[x_+, \infty)\tth\tth)_R \cup (\tth\tth(-\infty,- x_-]\tth\tth]_L \cup (\tth\tth(-\infty,- x_-]\tth\tth]_R$.  Hence, its appearance is solely due to our  effective 2d gravity description.   
\subsection{3d description and a new phase in entanglement evolution}
\begin{figure}[htb!]
\vskip-0.5cm
\centering  
\includegraphics[height=5.2cm]{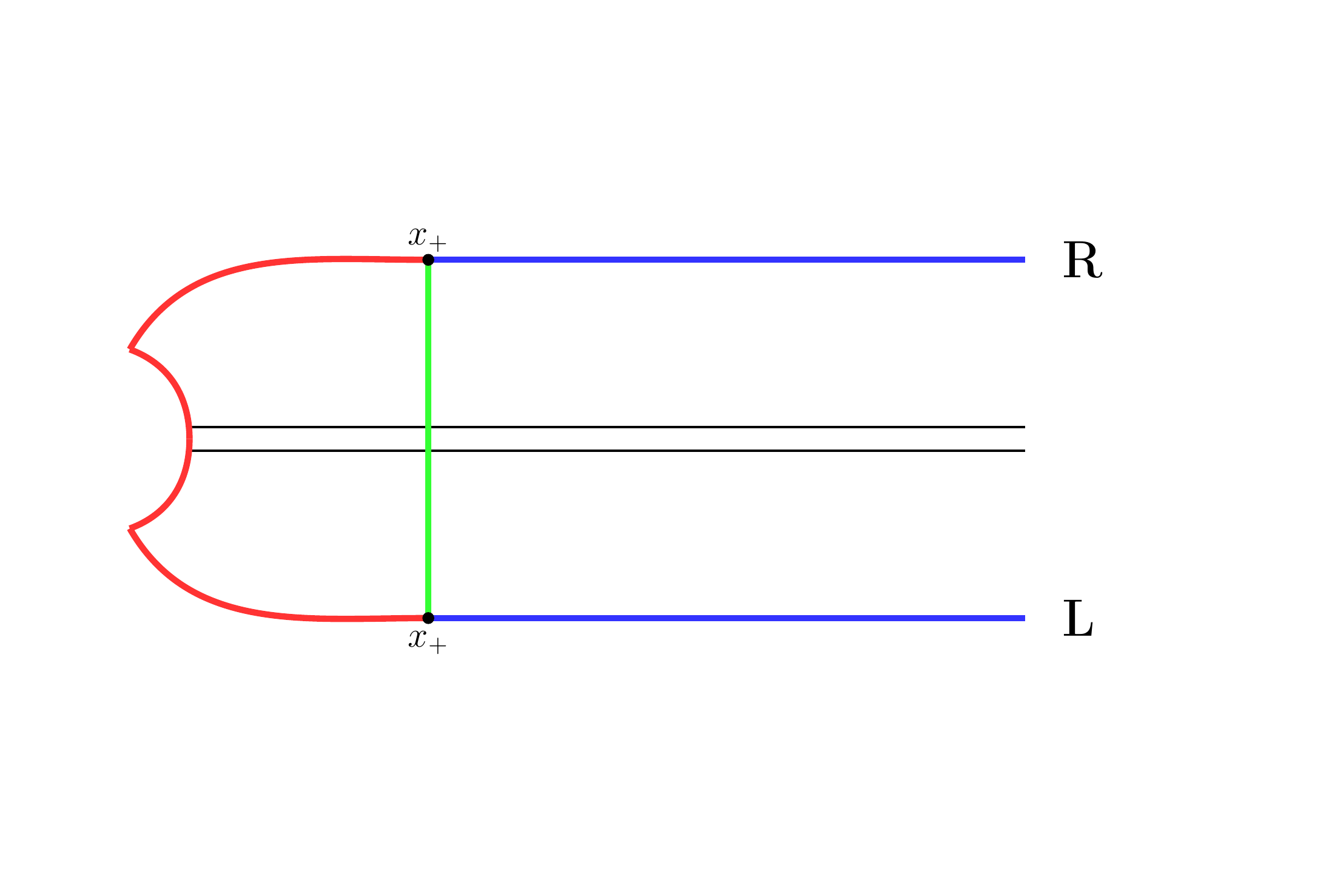}\includegraphics[height=5.2cm]{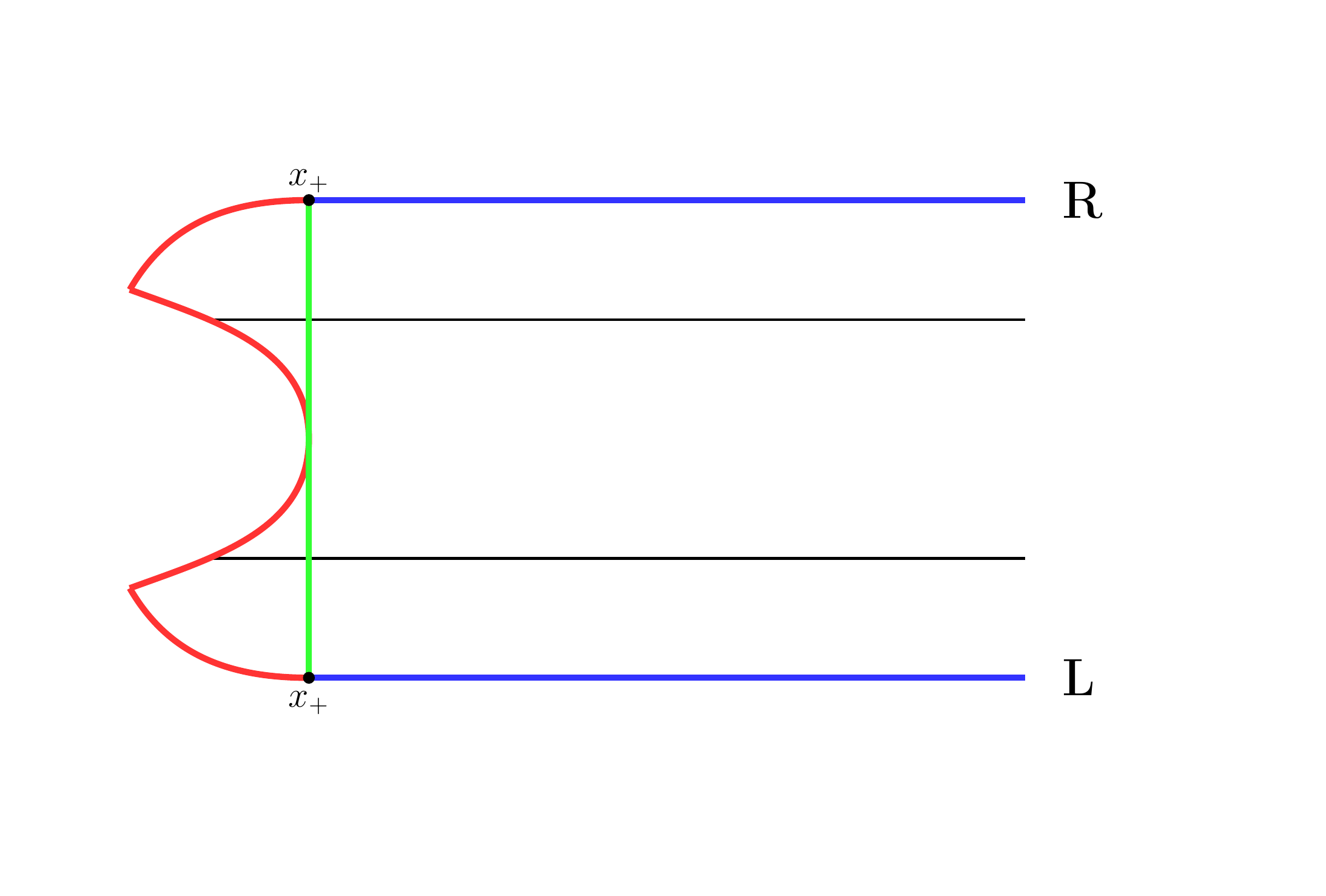}
 \vskip-0.5cm
\caption{\small 
The 3d bulk picture of constant $\tau$ slice is given for $0\le \xi_+ \le \xi_I$. On the left side, we draw the RL extremal curve with $\xi_+ \ll 1$. On the right side we depict the RL extremal curve just 
touching the $\mu=\mu_I$ surface.
}
\label{fig12}
\end{figure}

In this subsection, we  explain how the entanglement is developed in time  from the view points of 3d bulk and ICFT. It will be mainly accounted for by the behind-horizon dynamics of the RL extremal curves in the region 
$0 \le \xi_\pm \le \xi_\pm(t_P)$.
As was mentioned, the dynamics of QES before the transition is rather
complicated, whose details are mainly based on our holographic 
computation of the RL extremal curves.   It basically shows how degrees of freedom 
in $\bf [\tth\tth[-x_-,x_+]\tth\tth]_L \cup \bf [\tth\tth[-x_-,x_+]\tth\tth]_R$, which in particular include the R/L interface degrees of freedom, are entangled with the rest (called as radiation) as time goes by. The first is the so-called bulk OPE 
limit where $\xi_+ \ll 1$, i.e. 
the RL extremal curve is  relatively far away from the surface 
$\mu=\mu_I$.
We depict the corresponding configuration on the left side of Figure 
 \ref{fig12}. In this regime, 
one has
\be
S_{{\cal I}{\cal I}}=S_{(+)}+S_{(-)} \,,
\ee
where
\bea
S_{(\pm)}=\frac{c}{3}\ln 2\cosh \frac{2\pi}{\beta}t +O(\xi_\pm) \,.
\label{bulklimit}
\eea
For the entire region before the transition $0\le \xi_+< \xi_+(t_P)$,
the contribution from the  (green-colored) bulk extremal curve 
can be recovered from the  red-colored region which is connecting $(t,x_+)_R$ to $(t,x_+)_L$ through the black hole 
spacetime as depicted in Figure \ref{fig12}. The corresponding curve is also depicted in Figure \ref{fig10} by the single red line connecting 
$(t,x_+)_R$ to $(t,x_+)_L$. In the bulk OPE limit of 
$\xi_+ \ll 1$, the contribution from the interface degrees of freedom can be ignored and the original CFT with  central charge $c$ will be responsible for the dynamics even including  behind-horizon region. This contribution has been identified in \cite{Almheiri:2019yqk}, which  precisely agrees with the expression in (\ref{bulklimit}). See Appendix \ref{AppB} for its detailed computation using the two-point function of the twist operators.

In this bulk OPE limit, the form of $G^{(\pm)}_n$ is known to have a general form  
\bea
G^{(\pm)}_n \simeq 
G_0 \, \xi_\pm^{-\Delta_n}  \,,
\label{gpmbulk}
\eea
which is a straightforward generalization of (\ref{gblimit}). Of course the entanglement entropy in (\ref{bulklimit}) then follows from the formula (\ref{pmg}) and (\ref{gpmbulk}).

We now turn to a general holographic expression valid for the region 
$0 \le \xi_\pm \le \xi_\pm(t_P)$ with $\ln A \gg 1$. The corresponding behaviors are basically described by (\ref{q1}) and  (\ref{q2}). In terms 
of $\xi_\pm$, $S_{(\pm)}$ is identified as
\bea
S_{(\pm)}=\frac{c}{3}\left(\ln 2\sinh \frac{2\pi}{\beta}x_\pm +
\frac{\ln \big(\sqrt{\xi_\pm}\tth +\tth\sqrt{\xi_\pm \tth+\tth 1}\big)}{\sqrt{2}\sqrt{1\tth +\tth\alpha_\pm}}
  -\frac{1}{2}\ln {\frac{1\tth +\tth \alpha_\pm}{2}}
  \right)  \,,
  \label{lrholo}
\eea
where $\alpha_\pm$ is related to $\xi_\pm$ by
 \bea
\ln \big(\sqrt{\xi_\pm}\tth +\tth\sqrt{\xi_\pm \tth+\tth 1}\big) =
 \frac{\sqrt{2}\sqrt{1\tth+\tth\alpha_\pm}}{\sqrt{\alpha_\pm}}{\rm arcsin}
 \frac{1}{\sqrt{1+\alpha_\pm}} \,.
\eea
The  small $\xi_\pm$ behavior of (\ref{bulklimit}) is following from
the limit $\alpha_\pm \gg 1$. Another well known regime of interest is
the so-called boundary (interface in our case) OPE limit, {\it i.e.} $\xi_\pm \gg 1$. The  transition from the bulk to the interface limit occurs around $\xi_\pm=1$, which corresponds to $\alpha_\pm(\xi_\pm=1)\simeq 2.83586$. In the regime of
$0 \le \xi_\pm \le 1$, the radiation of bulk RL entanglement 
$\ln 2 \cosh  \frac{2\pi}{\beta} t$ (via the bulk channel of operator $\Phi^\pm_n$) 
plays a dominant role. Of course the outgoing  and ingoing components 
are balanced with each other such that the spacetime outside horizon 
remains stationary\footnote{The outgoing and the ingoing components of radiation are between ${\cal B}_{L}$ and $\bar{\cal B}_{R}$ or between ${\cal B}_{R}$ and $\bar{\cal B}_{L}$ in the notation of Section \ref{sec5}.}.

At $\alpha_\pm = \alpha_I =\frac{1}{e^2-1}$, the extremal curves begin to touch the surface $\mu= \pm \mu_I$ where our Grav$_{\pm}$ is defined respectively. 
At this point, one has $\xi_\pm =\xi_I \simeq 18841.1$. The corresponding configuration is drawn on the right hand side of Figure \ref{fig12}. 

\begin{figure}[htb!]
\vskip-0.1cm
\centering  
\includegraphics[height=5.5cm]{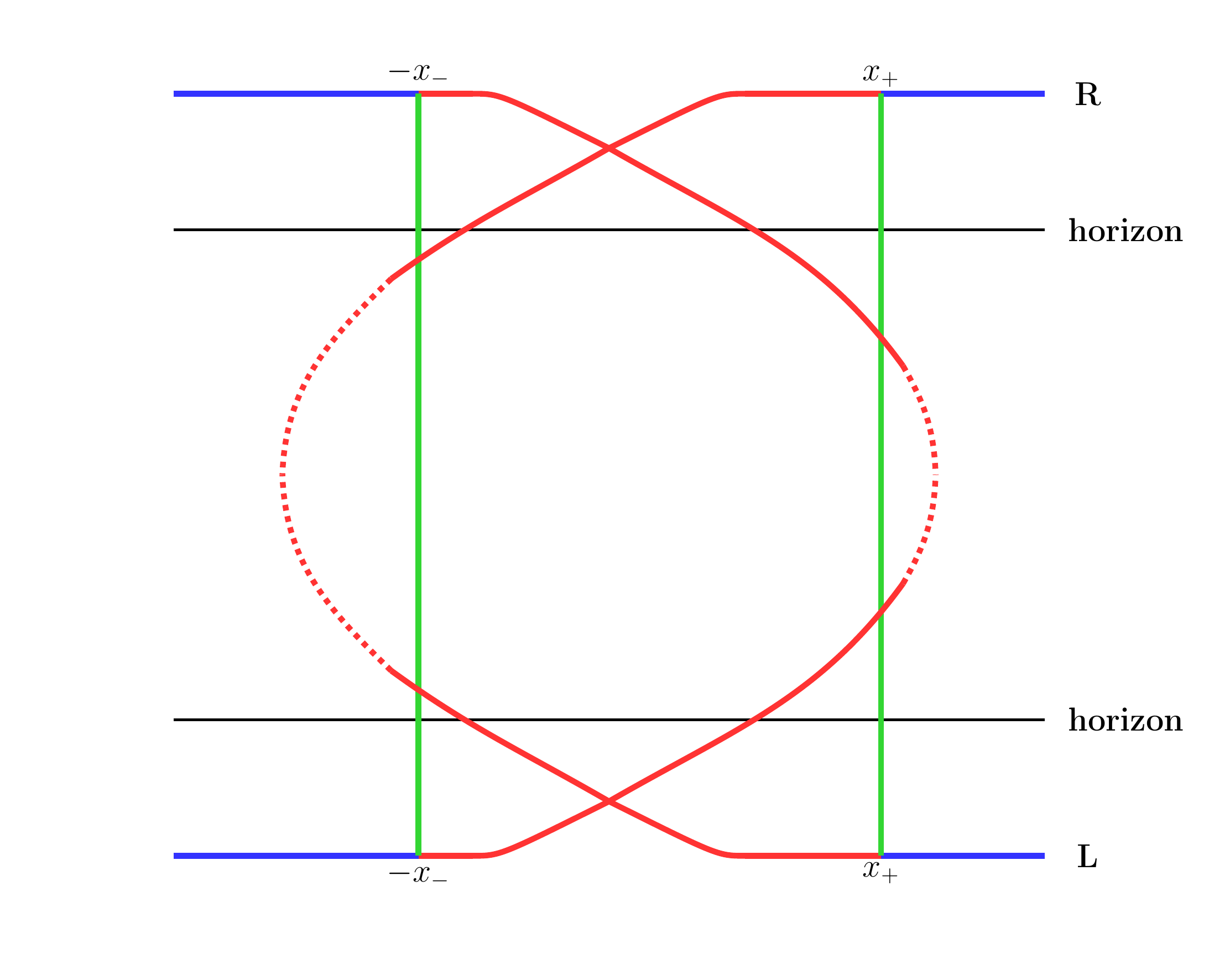}
 \vskip-0.5cm
\caption{\small 
The 3d bulk picture of constant $\tau$ slice is given for $\xi_I \le \xi_\pm \le \xi_\pm(t_P)$. In this figure, we choose the case $x_+ > x_-$ for the sake of illustration. The dotted red lines represent
the behind-horizon regions where the induced AdS$_2$ matter is excited.  
On the remaining part of the 2d spacetimes,  
the original CFT matter with central charge $c$ propagates.
}
\label{fig13}
\end{figure}

After then one begins to see details of the shadow region. As time goes by, the shadow region behind horizon is getting bigger and bigger as drawn
in Figure \ref{fig13}. When $\xi_\pm \gg 1$, the $S_{(\pm)}$ becomes
\bea
S_{({\pm})}=\frac{c}{3}\Big(\ln 2\sinh \frac{2\pi}{\beta}x_\pm +
\frac{1}{\sqrt{2}}\ln \frac{\cosh \frac{2\pi}{\beta} t\,\,}{\sinh \frac{2\pi}{\beta}x_\pm}
  +\frac{1\tth+\tth\sqrt{2}}{2}\ln {2}
  \tth+\tth O\big((\ln \xi_\pm)^{-2}\big)\Big) \,.
\label{largeA}  
\eea
As $\xi_\pm$ is getting bigger, one is probing deeper region of the shadow. This implies that the radiation and the degrees of freedom in the deeper region of the shadow are entangled more and more as time goes by. It clearly shows an appearance of new phase in our behind-horizon dynamics. Especially, in this new phase (with $ A\gg 1$), the slope in our entanglement time evolution approaches $\frac{\beta}{2\sqrt{2}\pi}$ before the Page transition (See Figure \ref{PageC}).  Below we shall identify the corresponding boundary operator from the view point of ICFT.

Now comparing the above with the entanglement entropy at the Page time in (\ref{llrr}), one finds the Page time satisfies
\bea
\frac{1}{2}\ln \xi_+(t_P)\xi_-(t_P)=  \sqrt{2}\ln \frac{A}{2} -2\ln 2  \,,
\label{page}
\eea
where we assume $\xi_\pm(t_P)\gg 1$ for the simplicity of our presentation.
When $x_\pm \gg \beta$ and $  \ln A \gg |x_+\tth-\tth x_-|/\beta$, one has a solution
\bea
t_P= \frac{x_++ x_-}{2} +\frac{\beta}{2\pi}\Big[\frac{1}{\sqrt{2}}\ln \frac{A}{2} -\ln 2\Big]  \,, 
\eea
where we ignore any exponentially small  correction.

The entropy $S_{(\pm)}$ developed up to the Page time reads
\bea
S_{(\pm)}(t_P) =\frac{c}{3}
\ln 2\sinh \frac{2\pi}{\beta}x_\pm +S^{(\pm)}_{\phantom{i}I} \,,
\label{pageent}
\eea  
where
\bea
S^{(\pm)}_{\phantom{i}I}=
\frac{c}{6}\Big[
\mp
{\sqrt{2}}\ln \frac{\sinh \frac{2\pi}{\beta} x_+\,\,}{\sinh \frac{2\pi}{\beta}x_-}
  +\ln A
  \Big] \,.
\eea
The first term  on the right hand side of (\ref{pageent})
shows the  entanglement between the bulk CFT degrees of freedom in $[-x_-, x_+]_{R/L}$ and
the $(\pm)$ radiation. The remaining term $S^{(\pm)}_{\phantom{i}I}$ represents the entanglement between the interface degrees of freedom and the $(\pm)$ radiation respectively.   
Hence  for $x_+ > x_-$, one can see that the 
($-$) radiation ($[-\infty, -x_-]_L\cup [-\infty, -x_-]_R$) is more entangled with the interface degrees of freedom than the ($+$) radiation ($[x_+,\infty]_L\cup [x_+,\infty]_R$), and vice versa. In Figure \ref{fig13},
we draw the shape of the configuration in the regime $\xi_I \le \xi_\pm \le 
\xi_\pm(t_P)$ with  $x_+ > x_-$. Thus we conclude that the $(+)$ and $(-)$
theories are dynamically correlated with each other even in the limit $\ln A \gg 1$.
\subsection{ICFT description and extra boundary operators}
The behavior for the regime of $\xi \gg 1$ in the previous section  can be summarized 
in terms of the function ${\rm tr}\rho^n_{{\cal II}}$ by
\bea
{\rm tr}\rho^n_{{\cal II}}\simeq \frac{
1}{\Big(4\sinh \frac{2\pi}{\beta} x_+ \sinh \frac{2\pi}{\beta}x_-\Big)^{2\Delta_n}}\left[G_n^{(\pm)}(\xi_\pm)G_n^{(\pm)}(\xi_\pm) +G^2_I A^{-\frac{c}{3}(n-1)} \right] \,, 
\label{bope}
\eea
 where 
\bea
G_n^{(\pm)}(\xi_\pm)=G_I \Big[  \sum_k g_{n,k}\xi_{\pm}^{- \hat\Delta_{n,k} }+2^{-(1+\sqrt{2})\delta_n}\xi^{-
\hat\Delta_n }_\pm 
\Big]\,,
\label{bope1}
\eea
with $\hat\Delta_{n,k} > 
\hat\Delta_n $ for $n > 1$, $\hat\Delta_{1,k}=\hat\Delta_{1}=0$ and $\delta_1=0$. Here, $G_I$
is  a constant independent of $n$.
From  the behavior of the entanglement entropy in the previous section,
one finds 
$
{\partial_n}\delta_n|_{n\rightarrow 1}=\sqrt{2}{\partial_n}\hat\Delta_n|_{n\rightarrow 1}=\frac{c}{6}$.
This form is consistent with the boundary OPE limit in \cite{Chiodaroli:2016jod,Mazac:2018biw}.
From this one may find  the transition of the entanglement entropy at the Page time $t_P$ in (\ref{page}).
It is also consistent with the requirement 
 ${\rm tr}\,\rho_{{\cal II}}=1$. To recover (\ref{largeA}) in the regime $1\ll \xi_\pm \le \xi_\pm(t_P)$,
we assume the last term in the bracket of (\ref{bope1}) dominates over the remaining terms once $\xi_\pm \gg 1$.
It also gives us the 
desired form of the entanglement entropy in (\ref{afterpage}).  The second term 
in the bracket of (\ref{bope})  comes from the boundary OPE between bulk and boundary identity operator. The $n$-dependence of its coefficient is explained  below (\ref{adep}). Since the interface degrees of freedom are shared by the ($\pm$) theories, the corresponding interface ground states, on which  ($\pm$) boundary operators including the interface identity are acting, are shared by the ($\pm$) theories as well. This is the reason why the factorization fails with the interface identity operator 
in (\ref{bope}). It is clear that the assumption of vacuum block dominance in \cite{Sully:2020pza} is not respected in our holographic interface theory. 
\subsection{Emergence of new AdS$_2$ matter}
All the above  boundary (interface) operators of dimensions $\hat{\Delta}_{n,k}$ and 
$\hat{\Delta}_{n}$, which are induced by $\Phi_n^\pm$, are responsible for the behind-horizon dynamics of generalized entropy. 
The corresponding AdS$_2$ matter contribution should be included when we are dealing with the generalized entropy using the 2d gravity theory. For the illustration, see Figures \ref{fig10} and \ref{fig13}. The dotted red lines represent regions where the extra AdS$_2$ matter propagates. On the remaining part of the 2d 
spacetime, the original CFT matter with central charge $c$ propagates. The transition between  them 
occurs roughly where the bulk extremal curves are touching the $\mu=\pm \mu_I$ surfaces. 

Based on this observation, we carry out the 2d  CFT computation of the generalized entropy $\hat{S}_{\rm gen}$ in Appendix \ref{AppB}. In this 2d setup, we consider the causal diamond $D_{RL}$ defined
by the two points $P_{R/L}$ with coordinates $U^{\pm}_{R}= \pm e^{\frac{2\pi}{\beta}(\pm t+ x )}=\tan \frac{\tau\pm \lambda }{2}$   and $U^{\pm}_{L}= \mp e^{\frac{2\pi}{\beta}(\mp t +x )}=\tan \frac{\tau\mp \lambda }{2}$, respectively. 
The interval with end points
$P_{R/L}$ will be denoted by ${\cal I}_{RL}$. 
We take $x>0$ and $\pi > \lambda >\frac{\pi}{2}$   such  that these points lie in the flat region of Figure \ref{fig10}. We further introduce an interval ${\cal I}_{rl}$ specified by  two points $P_{r/l}$ with coordinates $U^{\pm}_{r}=\tan \frac{\tau_0 \pm \lambda_0 }{2}$   and $U^{\pm}_{l}=\tan \frac{\tau_0 \mp \lambda_0 }{2}$, respectively, where we used the  RL symmetry of our model. 
We  require $\frac{\pi}{2} > \lambda_0 \ge 0$ such 
that the points $P_{r/l}$ lie within the AdS$_2$ 
region (of the diamond $D_{RL}$). 
The induced boundary (interface) operator $\hat{O}_n$ of dimension $\hat{\Delta}_n$  is assumed to be  excited within the interval ${\cal I}_{rl}$ of the AdS$_2$ region\footnote{Upon extremization, the two points $P_{r/l}$ roughly become the end points of each dotted red line in Figures \ref{fig10} and \ref{fig13}. }. 
 We further denote the interval defined by   $\{P_{R}, P_{r}\}$/$\{P_{L}, P_{l}\}$ 
as  ${\cal I}_{Rr}/{\cal I}_{Ll}$, respectively.

The entanglement entropy $\hat{S}_{Rr}/\hat{S}_{Ll}$ of 
the interval  ${\cal I}_{Rr}/{\cal I}_{Ll}$   can be evaluated from the two-point function of  the twist operators $\Phi^\pm_n$ as usual. Similarly the induced contribution  $\hat{S}_{rl}$ can be computed from the two-point function 
$\langle \hat{O}_n(P_r)\hat{O}_n(P_l)\rangle$. 
We then consider the generalized entropy given by
\bea
\hat{S}_{\rm gen}= {\hat{S}}_{Rr} +{\hat{S}}_{Ll} +{\hat{S}}_{rl}=2{\hat{S}}_{Rr}
+{\hat{S}}_{rl}  \,, 
\eea
where we set ${\hat{S}}_{Rr}={\hat{S}}_{Ll}$ using the  RL symmetry of our problem.  In this computation we assume that  any possible mutual information between the three  intervals can be ignored. One finds $\tau_0=\tau$ upon extremization. Hence the extremum 
is achieved along the constant $\tau$ slice which is in accordance with our holographic computation. 
Upon further extremizing with respect to $\lambda_0$, one finds two solutions 
in the regime $\xi_\pm \gg 1$. Note that the points $P_{r/l}$ in these solutions all lie behind the horizon of AdS$_2$. Another relevant configuration for the entanglement entropy of ${\cal I}_{RL}$ is the one  without any extra  AdS$_2$ matter contribution.  This becomes
\bea
\hat{S}_{RL}=\frac{c}{3} \ln 2\cosh \frac{2\pi}{\beta}t  \,.
\label{srl}
\eea
Choosing the minimum  among  those three and setting $\hat{q}=\frac{1}{\sqrt{2}}$ (See (\ref{qhat}) for its definition), one finds 
the entanglement entropy  from the 2d  perspective as
\bea
\hat{S}_{(\pm)}=\frac{c}{3}\bigg[\ln 2\sinh \frac{2\pi}{\beta}x_\pm +
\frac{1}{\sqrt{2}}\ln \frac{2\cosh \frac{2\pi}{\beta} t\,\,}{\sinh \frac{2\pi}{\beta}x_\pm}
  +2
\ln {2} \tth-\tth \sqrt{2}\ln(1\tth+\tth\sqrt{2})
  \tth+\tth O\big( \xi_\pm^{-1}\big)\bigg]\,,
\label{hatrl}
\eea
for the regime $\xi_\pm \gg 1$. 
The details of the computation will be relegated to Appendix \ref{AppB}.  Note  that the above entropy $\hat{S}_\pm$ agrees with   
 our HEE in (\ref{largeA}) up to constant terms: The difference reads
\bea
S_{(\pm)}-\hat{S}_{(\pm)} \simeq \frac{c}{3}\Big[  \sqrt{2}\ln(1\tth+\tth\sqrt{2}) -\frac{3}{2}\ln 2\Big]
\simeq 0.069 \, c   \,.
\eea
Of course, we do not expect any precise agreement since our discussion is based on a number of 
approximations. In particular
 the transitional region between 
the original CFT and the induced AdS$_2$ matter excitation is not so sharply defined in our original HEE configuration in Figure \ref{fig13}.

This behind-horizon matter will be responsible 
for the outgoing and ingoing components of radiation in the regime $\xi_\pm \gg 1$.
Hence the extra AdS$_2$ matter is contributing to the entanglement 
evolution in the behind-horizon region in 
addition to the original  CFT with central charge $c$. 
Further study is required in this direction.

\section{Conclusions}\label{sec9}

In this work, we have investigated the   entanglement entropy/information evolution of black holes  surrounded with radiations from three different perspectives and showed that it follows the anticipated Page curve. Firstly, we have evaluated holographically the entanglement entropy of the boundary intervals by using the geodesic distance in the 3d Janus black holes. Secondly, we have made the boundary ICFT interpretation of this HEE in Janus black holes. And then we  have provided the effective 2d gravity realization dual to the interface degrees of freedom, which is coupled to CFT$_2$. In this reduced gravity,  we have also confirmed that the QES with island picture can reproduce the HEE computed in our 3d gravity. All of these perspectives lead to a consistent picture and  confirm the unitary evolution of entanglement entropy.

In the 3d Janus black hole, the conventional Ryu-Takayanagi surface (geodesic in our case) can give the entanglement entropy of two intervals located in the left and the right  boundary, respectively. As usual, the change of the topology of the Ryu-Takayanagi surface leads to the phase transition of the entanglement at the Page time $t_P$ which is increasing as the number of the interface degrees of freedom, represented by $\ln A$,  gets larger. When the Page time is large enough, we found an additional phase transition before the Page time. This new phase can naturally be understood from the point of view of the effective 2d gravity dual to the interface degrees of freedom coupled to CFT$_2$. As the interface degrees of freedom are mixed with CFT$_2$, the surface where the 2d gravity lives expands. If $\ln A$ is large enough, this surface for the 2d gravity intersects with the Ryu-Takayanagi surface connecting the left and right intervals before the Page time. And, this  is responsible for the new phase transition.

When   $\ln A$ is sufficiently large, one can view our system as two {\it nearly decoupled} BCFT's. From the BCFT point of view, the entanglement entropy can be evaluated by the two point function of the twist operators. In early time, the bulk OPE channel dominates to the first phase. As time passes, we consider the boundary OPE channel which is mediated by  boundary operators in 2d gravity induced by the twist operator. Note that the broken conformal symmetry to SO$(1,2)$ in the 2d gravity leads to the effective conformal dimension of the induced operator. This reproduces the second phase obtained by the 3d Ryu-Takayanagi surface.

Following the island conjecture, we computed the generalized entropy in the 2d gravity coupled to a CFT$_2$ system, and its extremization agrees with the 3d gravity calculation. Also, for  given points $x_+$ and $-x_-$ on the CFT$_2$, we considered a geodesic connecting  them with $a_\pm$ on the 2d gravity surface (See Figure~\ref{foffshell}), which was found to agree with the generalized entropy in our 2d system. This explains why we need to extremize the generalized entropy to get the correct answer because the geodesic connecting  $\pm x_\pm$ and $a_\pm$ naturally leads to the Ryu-Takayanagi surface in 3d gravity by its extremization with respect to $a_\pm$ by definition of the geodesic.

Our top-down approach in this work
 can give us concrete answers on  the 2d gravities. Starting from the 3d Janus black hole for instance, we obtain the effective 2d gravity directly by integrating out the bulk degrees of freedom. 
Also, it is intriguing to investigate the entanglement wedge reconstruction. From the point of view of 2d effective gravity coupled to a CFT$_2$, one can study the Petz map which reconstructs operators behind the horizon~\cite{Penington:2019kki} to see the effect of the interface degrees of freedom on the reconstruction. And, one might be able to reinterpret the entanglement wedge reconstruction of the 2d system from the 3d point of view.  

In this work, our study is focused on the entanglement evolution of the 3d Janus black hole.
Its higher dimensional generalization will be of interest.  It might also be interesting to consider  higher derivative corrections in the gravity action~\cite{Alishahiha:2020qza} and  the flat space adaptation~\cite{Krishnan:2020oun}.


\subsection*{Acknowledgement}
We would like to thank Andreas Gustavsson for careful reading of the manuscript.
DB was
supported in part by
NRF Grant 2020R1A2B5B01001473, by  Basic Science Research Program
through National Research Foundation funded by the Ministry of Education
(2018R1A6A1A06024977), and  by the 2020 sabbatical year research grant of the University of Seoul.
 C.K.\ was supported by NRF Grant 2019R1F1A1059220.
S.-H.Y. was supported by 
NRF Grant  
2018R1D1A1A09082212 and supported  by Basic Science Research Program through the NRF funded by the Ministry of Education(NRF-2020R1A6A1A03047877).
JY was in part supported by a KIAS Individual Grant (PG070101) at Korea Institute for Advanced Study, and by the National Research Foundation of Korea(NRF) grant funded by the Korea government(MSIT) (2019R1F1A1045971).

\appendix 

\section{Elliptic integrals } \label{AppA}
\renewcommand{\theequation}{A.\arabic{equation}}
  \setcounter{equation}{0}

In this section, we summarize various formulae about elliptic integrals, which are used in the main text.  (See~\cite{NIST} and references therein for a more detailed information about elliptic integrals.)
The  incomplete elliptic integral of the first kind $F(\varphi\, |\, m)$ is defined as
\begin{equation} \label{}
F(\varphi\,|\, m) =  \int^{\varphi}_{0}  \frac{d\theta}{\sqrt{1-m\sin^{2}\theta}}\,, 
\end{equation}
and the incomplete elliptic integral of the third kind is defined as
\begin{equation} \label{}
\Pi(\nu\,;\,\varphi\,|\, m) =  \int^{\varphi}_{0} \frac{1}{1-\nu\sin^{2}\theta}\frac{d\theta}{\sqrt{1-m\sin^{2}\theta}}\,.
\end{equation}
These become  the complete elliptic integrals in the case of $\varphi=\frac{\pi}{2}$ as 
\begin{equation} \label{}
{\bf K}(k) = F\Big(\frac{\pi}{2}\, \Big|\, k^{2}\Big)\,, \qquad {\bf \Pi}(\nu\,;\, k) = \Pi\Big(\nu\,;\,\frac{\pi}{2}\, \Big|\, k^{2} \Big)\,.
\end{equation}

The symmetric elliptic integrals are defined as
\begin{align}    \label{}
R_{F}(x,y,z) &= \frac{1}{2}\int^{\infty}_{0} \frac{dt}{\sqrt{t+x}\sqrt{t+y}\sqrt{t+z}}\,,    \\
R_{J} (x,y,z,p) &= \frac{3}{2}\int^{\infty}_{0} \frac{dt}{ (t+p)\sqrt{t+x}\sqrt{t+y}\sqrt{t+z}}\,,  
\end{align}
and 
\begin{equation} \label{Rred}
R_{C}(x,y)  \equiv R_{F}(x,y,y)\,, \qquad R_{D}(x,y,z) \equiv R_{J}(x,y,z,z) \,.
\end{equation}
These symmetric forms are more useful for obtaining the asymptotic expansion and for providing more efficient numerical computation. In particular, the symmetric integral $R_{C}$ can be written in terms of elementary functions as 
\begin{equation} \label{Rcexp}
R_{C}(x,y) =%
\left\{ \begin{array}{ll}  %
\frac{1}{\sqrt{y-x}} \arccos \textstyle{ \sqrt{\frac{x}{y}} } &,  \qquad x < y    \\
\frac{1}{\sqrt{x-y}} \textrm{arccosh} \sqrt{\frac{x}{y}} = \frac{1}{\sqrt{x-y}}  \ln \frac{\sqrt{x} + \sqrt{x-y}}{\sqrt{y}} &, \qquad x> y 
\end{array}  \right.   \,.
\end{equation}
The symmetric forms have the following scaling properties:
\begin{equation} \label{}
R_{F}(\lambda x, \lambda y, \lambda z) = \lambda^{-\frac{1}{2}}R_{F}(x,y,z)\,, \qquad R_{J}(\lambda x, \lambda y , \lambda z, \lambda p) = \lambda^{-\frac{3}{2}}R_{J}(x,y,z,p)\,.
\end{equation}
Another useful relation is
\begin{equation} \label{RJtoRC}
R_{J}(x,y,y,p) = \frac{3}{p-y}\Big[R_{C}(x,y) -R_{C}(x,p) \Big]\,, \qquad p \neq y\,.
\end{equation}
Asymptotic expansion formulae useful in the main text are
\begin{align}    \label{}
 R_{F}(x,y,z) &= \frac{1}{2\sqrt{z}}\ln \frac{16z}{(\sqrt{x} + \sqrt{y})^{2}} + {\cal O}\Big(\textstyle{\frac{x+y}{2}, \sqrt{xy}}\Big) \qquad \textrm{for}  \quad  x,y \ll z\,.
 \\
R_{J}(x,y,z,p) &= \textstyle{\frac{3}{2}\frac{1}{\sqrt{xyz}}\ln \frac{4xyz}{p \sigma^{2}} }+2 R_{J}(x+\sigma,y+\sigma,z+\sigma, \sigma)  \nonumber  \\
& \qquad \qquad \qquad  \quad + {\cal O}\Big(p\ln p\Big)   ~~ \qquad \qquad  \qquad  \textrm{for} \quad   p \ll x,y,z \,, \label{ASEJ}
\end{align}
where  $\sigma \equiv \sqrt{xy} + \sqrt{yz} + \sqrt{zx}$. 

Note that  the incomplete elliptic integrals may be represented by the symmetric forms as 
\begin{align}    \label{EllipticF}
F(\varphi\,|\, m) &= R_{F} \Big(\textstyle{\frac{\cos^{2}\varphi}{\sin^{2}\varphi},~ \frac{1}{\sin^{2}\varphi}-1,~ \frac{1}{\sin^{2}\varphi} } \Big) \,, \\
\Pi(\nu\,;\,\varphi\,|\, m) &= \sin\varphi~ R_{F}(\cos^{2}\varphi,~1-m\sin^{2}\varphi,~ 1) \nonumber \\
& \qquad + \frac{\nu}{3}\sin^{3}\varphi~ R_{J}(\cos^{2}\varphi,~ 1-m\sin^{2}\varphi,~1,~1-\nu \sin^{2}\varphi)\,. 
\end{align}
%

\section{2d computation with AdS$_2$  matter contribution} \label{AppB}
\renewcommand{\theequation}{B.\arabic{equation}}
  \setcounter{equation}{0}
 We begin with the entanglement entropy of the interval ${\cal I}_{RL}$ without any extra AdS$_2$ matter propagation. The two-point function with the twist operator insertion may be evaluated as
 \be
 G_n^{RL}=\langle  \Phi_n^+ (P_R)  \Phi_n^- (P_L) \rangle_{CFT}
 = \left[  \frac{\sqrt{U_R^+U_R^-U_L^+U_L^-}}{(U_R^+-U_L^+)(U_L^--U_R^-)}\right]^{\Delta_n}\,.
 \ee 
In this expression, the numerator inside the bracket comes from the Weyl factor at each point in the 
Weyl transformation from the trivial flat metric $ds_f^2= -dU^+ dU^-$ to our metric in  
\eqref{flatmetric}. In this appendix we shall omit the discussion involved with the issue of
 regularization and renormalization. Once we have two point function $G_n$, the corresponding  entanglement entropy will be evaluated by
\be
\hat{S}=-\lim_{n\rightarrow 1} \partial_n G_n\,.
\ee 
Thus one finds that the entanglement entropy is given by $\hat{S}_{RL}$ in \eqref{srl}.
Similarly, ${\hat{S}}_{Rr}$ can be evaluated using the two-point function 
 \be
 G_n^{Rr}=\langle  \Phi_n^+ (P_R)  \Phi_n^- (P_r) \rangle_{CFT}
 = \left[  \frac{\big(1+U^+_r U^-_r\big)\sqrt{-U_R^+U_R^-}}{2 (U_R^+-U_r^+)(U_r^--U_R^-)}\right]^{\Delta_n} \,.
 \ee 
This leads to
\bea
\hat{S}_{Rr}= \frac{c}{6} \left[ \ln \frac{\cos(\tau\tth-\tth\tau_0)\tth-\tth\cos(\lambda\tth-\tth\lambda_0) }{\cos \lambda_0 \,\, \cos \lambda}+\ln  2 \sinh \frac{2\pi}{\beta}x\right]\,.
\eea
By the same way, one may check that $\hat{S}_{Rr}=\hat{S}_{Ll}$, which may be understood from the 
left right symmetry of our configuration. For the AdS$_2$ matter contribution of the interval ${\cal I}_{rl}$, we use the 
two-point function of the boundary (interface) operator $\hat{O}_n$
\bea
 G_n^{rl}=\langle  O_n (P_r)  O_n (P_l) \rangle_{CFT}
 = \left[  \frac{\big(1+U^+_r U^-_r\big)\big(1+U^+_l U^-_l\big)}{4 (U_r^+-U_l^+)(U_l^--U_r^-)}\right]^{\hat\Delta_n} \,.
\eea 
This leads to
\be
\hat{S}_{rl}= \frac{c \,\hat{q}}{3} \ln 2 \tan \lambda_0\,,
\ee
where we introduce $\hat{q}$ by
\bea
\hat{q} = \frac{6}{c} \partial_n \hat\Delta_n|_{n=1}\,.
\label{qhat}
\eea
We assume $0 <\hat{q} < 1$. 
Then the generalized entropy including the AdS$_2$ matter contribution is given by
\bea
\hat{S}_{gen}=\frac{c}{3} \left[ \hat{q}\ln  2 \tan \lambda_0 + \ln \frac{\cos(\tau\tth-\tth\tau_0)\tth-\tth\cos(\lambda\tth-\tth\lambda_0) }{\cos \lambda_0 \,\, \cos \lambda}+\ln  2 \sinh \frac{2\pi}{\beta}x\right]\,.
\eea
Its extremization with respect to $\tau_0$ is solved by $\tau_0=\tau$. Then the extremization 
condition with respect to $\lambda_0$ becomes 
\be
 \frac{1+\cos (\lambda\tth-\tth\lambda_0)}{\sin (\lambda\tth-\tth\lambda_0)}= \hat{q}\big( \tan \lambda_0 + \cot \lambda_0\big) +\tan \lambda_0\,.
 \label{extcon}
\ee
Let us first consider the case where $|\tan \lambda|^2=\xi \gg 1$ with $\xi$ defined in (\ref{xixt}).
Then there are two solutions for the range $0\le \lambda_0 <\frac{\pi}{2}$. One is
\bea
\tan \lambda^{(1)}_0 =\frac{1-\hat{q}}{1+\hat{q}} \,|\tan \lambda|\left(
1+O(\xi^{-1}) \right)\,,
\eea
which leads to the extremal value
\be
\hat{S}_{(1)} =\frac{c}{3}\left[ \hat{q}\ln 
\frac{2(1-\hat{q})\cosh \frac{2\pi}{\beta}t}{(1+\hat{q})\,\,\sinh \frac{2\pi}{\beta}x} 
 +\ln \frac{4 \sinh \frac{2\pi}{\beta}x}{1-{\hat{q}}^2} +O(\xi^{-1})\right] 
 \label{s1}\,.
\ee
The other solution is
\bea
\sin \lambda^{(2)}_0 =\hat{q}
+O(\xi^{-\frac{1}{2}}) \,,
\eea
and the corresponding extremal value becomes
\bea
\hat{S}_{(2)} =\frac{c}{3}\left[ \frac{\hat{q}}{2}\ln 
\frac{4{\hat{q}}^2 }{1-{\hat{q}}^2} 
 -\frac{1}{2}\ln \frac{1+\hat{q}}{1-\hat{q}} +\ln  2 \cosh \frac{2\pi}{\beta}t+O(\xi^{-1})\right] \,.
 \label{s2}
\eea
The minimum of \eqref{s1}, \eqref{s2} and \eqref{srl} gives us the true entanglement entropy. Note that
in these solutions, the points $P_{r/l}$ lie in the behind-horizon region. 
Thus
we find, for $\xi \gg 1$, 
$\hat{S}= \hat{S}_{(1)} $
which involves the behind-horizon AdS$_2$ matter contribution.

One may also consider $\xi \ll 1$. In this case, one finds no solution of  the extremal condition \eqref{extcon} within the range $0\le \lambda_0 <\frac{\pi}{2}$. Hence,  for $\xi \ll 1$,
$
\hat{S}= \hat{S}_{RL}
$ 
which does not involve any extra  AdS$_2$ matter contribution.

\section{Effective 2d description} \label{AppC}
\renewcommand{\theequation}{C.\arabic{equation}}
  \setcounter{equation}{0}

In this appendix, we provide some details of an effective two-dimensional `theory of gravity' description presented in Sections~\ref{sec6} and~\ref{sec8}.  Our approach follows the spirit of the Randall-Sundrum construction~\cite{Randall:1999vf} and is closely related to the construction given in Ref.~\cite{Chen:2020uac}. However, our  construction is simply intended to reproduce our 3d bulk results   
while keeping our original 3d bulk  intact.

Let us recall that the bulk gravity action of the asymptotically AdS space with a boundary surface may be written as follows:
\begin{equation} \label{HJact}
I = \int_{z > z_{0} } d^{d+1}x \sqrt{-g}  {\cal L} (g,\psi) + I_{B}[h,\psi,z=z_{0}]\,,
\end{equation}
where $\psi$ is a generic matter field $\psi$, $z$ is the inverse of AdS-radial coordinate  and $h$ is a boundary metric at   a boundary surface, $z=z_{0}$. Here, $I_{B}$ is a boundary action which gives us an appropriate boundary condition at $z=z_{0}$ upon variation.  This boundary action $I_{B}$ satisfies the so-called Hamilton-Jacobi~(HJ) equation and may be obtained by integrating the functional derivative equation appropriately.  

In our case,
we would like to place the boundary surfaces at $\mu=\pm \mu_{I}$ and obtain the effective description of the shadow region specified by $-\mu_I <\mu  < \mu_I$   (See Figure~\ref{LRgeo}). To achieve this description in Secion~\ref{sec6}, we borrow the Randall-Sundrum construction and replace the bulk part in 
(\ref{HJact}) (or the shadow region in the present context) by  brane actions in the form of
%
\begin{equation} \label{}
I^\pm_{br} = -T_{2}\int_{\mu=\pm\mu_{I}} d^{2}x \sqrt{-h}  - S_{0}\int_{\mu=\pm\mu_{I}} d^{2}x\sqrt{-h} (\phi - C_\pm)\,,
\end{equation}
where $C_\pm$ are constants
and  would be taken by the value, $\bar{\phi}_{I}=\phi(\pm\mu_{I})$, 
of the scalar field $\phi$ bulk solution at $\mu=\pm\mu_{I}$ in the following. The brane tension $T_{2}$ and the scalar source coefficient $S_{0}$ should be chosen appropriately to match the boundary condition at $\mu=\pm \mu_{I}$. 
 Now, the total 2d effective action at each $\mu=\pm \mu_{I}$ may be taken as 
\begin{equation} \label{}
I^{\pm}_{2} = I^{\pm}_{br}  + I^{\pm}_{B}\,,
\end{equation}
where $I_{B}$ is a solution to HJ  equation. Since $\pm$ branches take the same form, we focus on the $+$ branch with $\mu=\mu_{I}$, for simplicity.

Ignoring any 2d derivatives  
on 
$R_{h}$ and $\phi$, 
a general solution to HJ equation 
may be obtained in the form of~\cite{Ioannis note} (See also Ref.~\cite{Papadimitriou:2016yit})
\begin{align}    \label{}
I_{B} &= \frac{1}{8\pi G_{3}\, \ell}\int d^{2}x\sqrt{-h} \bigg[ \sqrt{1+\frac{\ell^{2}}{2}R_{h} +\frac{\gamma^{2}\ell^{4}}{8}R^{2}_{h}} + \frac{\ell^{2}}{4}R_{h}~\text{arctanh}\bigg(\frac{1+\frac{\ell^{2}}{4}R_{h}}{\sqrt{1+\frac{\ell^{2}}{2}R_{h} + \frac{\gamma^{2}\ell^{4}}{8}R^{2}_{h}}}\bigg)  \nonumber \\
& \qquad \qquad -\frac{\gamma \ell^{2}}{2\sqrt{2}}R_{h}~ \text{arctanh} \bigg(\frac{\sqrt{2}\gamma \sqrt{1+\frac{\ell^{2}}{2}R_{h} + \frac{\gamma^{2}\ell^{4}}{8}R^{2}_{h}} }{1-\frac{\gamma^{2}\ell^{2}}{2} R_{h}}\bigg) + \frac{\ell^{2}}{2}(\gamma\,\phi +\alpha) ~R_{h}  \bigg]\,,
\end{align}
where $R_{h}$ denotes the 2d Ricci scalar  on $\mu=\mu_{I}$ and $\alpha$ is an integration constant in the HJ equation. 
Further ignoring higher order corrections in 
$1/A$, 
one can show that the  above action reduces to the following form
\begin{align}    \label{}
I_{B} = &\frac{\ell}{16\pi G_{3}}\int d^{2}x\sqrt{-h}\bigg[\frac{2}{\ell^{2}} + \frac{1}{2}R_{h}\, (1+\alpha+\sqrt{2}\bar{\phi}_{I}) -\frac{1}{2}R_{h}\ln \Big[ -\frac{\ell^{2}}{4}R_{h}\Big]  \nonumber  \\
&\qquad \qquad  \qquad  \qquad \quad    + \frac{1}{\sqrt{2}} R_{h}\, (\phi-\bar{\phi}_{I})\bigg] + {\cal O}\Big(\frac{1}{A^{2}}\Big)\,.
\end{align}

One may fix brane parameters $T_{2}$ and $S_{0}$ by using the on-shell solution in (\ref{3djanus}) with the condition $\frac{\delta I_{2}}{\delta h_{ij}} = \frac{\delta I_{2}}{\delta \phi } =0$ at $\mu=\mu_{I}$, and then one obtains the effective 2d action as
\begin{equation} \label{}
I_{2} = \tilde{I}_{top} + \frac{1}{16\pi G_{2}} \int d^{2}x\sqrt{-h} \bigg[\frac{2}{\ell^{2}_{2}} - R_{h} \ln \Big[ -\frac{\ell^{2}_{2}}{2}R_{h}\Big]  + \varphi\Big(R_{h} + \frac{2}{\ell^{2}_{2}}\Big) \bigg]\,,
\end{equation}
where  $G_{2} \equiv 2G_{3}/\ell$  and $\ell^{2}_{2}$ is defined by $-2/\ell^{2}_{2} \equiv R_{\bar{h}_{I}}$ as the 2d Ricci scalar value of the on-shell solution on $\mu=\mu_{I}$. It may be useful to recall that the same expression is given in the form of $\ell_{2} = \ell \sqrt{f(\pm \mu_I)}$ below (\ref{rindler}).  Here, $\varphi$ is defined by $\varphi \equiv \frac{1}{\sqrt{2}}(\phi -\bar{\phi}_{I})$, while $\tilde{I}_{top}$ denotes the topological term in two dimensions defined as
\begin{equation} \label{}
\tilde{I}_{top} \equiv  \frac{\alpha_{I}}{16\pi G_{2}}\int d^{2} x \sqrt{-h}\, R_{h}\,,
\end{equation}
where  the constant $\alpha_{I}$ is given by $\alpha_{I} \equiv 1+\alpha-\frac{\ell^{2}}{4}R_{\bar{h}_{I}}$. 

Equations of motion for the 2d metric $h$ and the dilaton $\varphi$ may be written as
\begin{align}    \label{}
0 &= R_{h} + \frac{2}{\ell^{2}_{2}}\,,  \\
0&=  \Big(\nabla_{i}\nabla_{j} -h_{ij}\nabla^{2} + \frac{1}{\ell^{2}_{2}}h_{ij} \Big)\varphi \,.
\end{align}
The solution to the metric equations of motion gives us the anticipated AdS$_{2}$ space at $\mu=\mu_{I}$, which is consistent with the on-shell AdS$_{3}$ bulk solution, while the dilaton equations of motion seem to allow some non-trivial solution. However, the boundary condition at the boundary of AdS$_{3}$ or our  cutoff condition at $\mu=\frac{\pi}{2} -\epsilon$ in the AdS$_{3}$ bulk  suggests that  this kind of solution is not allowed in our setup.  Concretely speaking, our 3d bulk cutoff implies the reparametrization modes of 2d cutoff trajectory are fixed by 
\begin{equation} \label{}
ds_{AdS_{2}}^{2}\Big|_{\text{cut-off}}= -\frac{dt^{2}}{\epsilon^{2}}\,,
\end{equation}
where $t$ is the boundary time matched with the bulk boundary CFT$_{2}$ time. 
Though there are no fluctuating degrees of freedom on the surface $\mu=\mu_{I}$ and the graviton localization to that surface is obscured, we  call  this 2d description  as  a `theory of gravity' in our main text.

Note also that the 3d bulk topological term $I_{top}$ related to  the interface degrees of freedom is determined by 
\begin{equation} \label{eqc11}
I_{top} = \tilde{I}^{+}_{top} + \tilde{I}^{-}_{top} \,,
\end{equation}
which tells us that $+$ and $-$ sides are far from the complete disentanglement.  It is quite notable to observe that the separate description in terms of $I^{\pm}$ would become very good only when the length scale $\tilde{L}$ of our interest is much smaller than $\ell \ln A $ ( $\gg \ell$). 



\begin{thebibliography}{100}


\bibitem{Hawking:1974sw}
  S.~W.~Hawking,
  ``Particle Creation by Black Holes,''
  Commun.  Math. Phys.  {\bf 43} (1975) 199
   Erratum: [Commun. Math.\ Phys.  {\bf 46} (1976) 206].


\bibitem{Hawking:1974rv}
  S.~W.~Hawking,
  ``Black hole explosions,''
  Nature {\bf 248} (1974) 30.

\bibitem{Wald:1975kc}
R.~M.~Wald,
``On Particle Creation by Black Holes,''
Commun. Math. Phys. \textbf{45} (1975), 9-34

\bibitem{tHooft:1996rdg}
G.~'t Hooft,
``The Scattering matrix approach for the quantum black hole: An Overview,''
Int. J. Mod. Phys. A \textbf{11} (1996), 4623-4688
[arXiv:gr-qc/9607022 [gr-qc]].


\bibitem{Mathur:2005zp}
  S.~D.~Mathur,
  ``The Fuzzball proposal for black holes: An Elementary review,''
  Fortsch.\ Phys.\  {\bf 53} (2005) 793


\bibitem{Polchinski:2016hrw}
  J.~Polchinski,   ``The Black Hole Information Problem,'' 
  arXiv:1609.04036 [hep-th].


\bibitem{Unruh:2017uaw}
  W.~G.~Unruh and R.~M.~Wald,
  ``Information Loss,''
  Rept.\ Prog.\ Phys.\  {\bf 80} (2017) no.9,  092002
  [arXiv:1703.02140 [hep-th]].


\bibitem{Ashtekar:2020ifw}
  A.~Ashtekar,
  ``Black Hole evaporation: A Perspective from Loop Quantum Gravity,''
  Universe {\bf 6} (2020) no.2,  21
  [arXiv:2001.08833 [gr-qc]].
  
  
\bibitem{Maldacena:2020ady}
J.~Maldacena,
``Black holes and quantum information,''
Nature Rev. Phys. \textbf{2} (2020) no.3, 123-125


\bibitem{Almheiri:2020cfm}
A.~Almheiri, T.~Hartman, J.~Maldacena, E.~Shaghoulian and A.~Tajdini,
``The entropy of Hawking radiation,''
[arXiv:2006.06872 [hep-th]].

\bibitem{Hawking:1982dj}
S.~Hawking,
``The Unpredictability of Quantum Gravity,''
Commun. Math. Phys. \textbf{87} (1982), 395-415

\bibitem{Jacobson:2003vx}
T.~Jacobson,
``Introduction to quantum fields in curved space-time and the Hawking effect,''
[arXiv:gr-qc/0308048 [gr-qc]].


\bibitem{Banks:1983by}
T.~Banks, L.~Susskind and M.~E.~Peskin,
``Difficulties for the Evolution of Pure States Into Mixed States,''
Nucl. Phys. B \textbf{244} (1984), 125-134

\bibitem{Almheiri:2012rt}
  A.~Almheiri, D.~Marolf, J.~Polchinski and J.~Sully,
  ``Black Holes: Complementarity or Firewalls?,''
  JHEP {\bf 1302} (2013) 062
  [arXiv:1207.3123 [hep-th]].


\bibitem{Almheiri:2013hfa}
  A.~Almheiri, D.~Marolf, J.~Polchinski, D.~Stanford and J.~Sully,
  ``An Apologia for Firewalls,''
  JHEP {\bf 1309} (2013) 018
  [arXiv:1304.6483 [hep-th]].


\bibitem{Maldacena:2013xja}
  J.~Maldacena and L.~Susskind,
  ``Cool horizons for entangled black holes,''
  Fortsch.\ Phys.\  {\bf 61} (2013) 781
  [arXiv:1306.0533 [hep-th]].


\bibitem{Page:1993wv}
  D.~N.~Page,
  ``Information in black hole radiation,''
  Phys.\ Rev.\ Lett.\  {\bf 71} (1993) 3743
  [hep-th/9306083].


\bibitem{Jackiw:1984je}
  R.~Jackiw,
  ``Lower Dimensional Gravity,''
  Nucl.\ Phys.\ B {\bf 252} (1985) 343.


\bibitem{Teitelboim:1983ux}
  C.~Teitelboim,
  ``Gravitation and Hamiltonian Structure in Two Space-Time Dimensions,''
  Phys.\ Lett.\  {\bf 126B} (1983) 41.


\bibitem{Almheiri:2019psf}
  A.~Almheiri, N.~Engelhardt, D.~Marolf and H.~Maxfield,
  ``The entropy of bulk quantum fields and the entanglement wedge of an evaporating black hole,''
  JHEP {\bf 1912} (2019) 063
  [arXiv:1905.08762 [hep-th]].


\bibitem{Penington:2019npb}
  G.~Penington,
  ``Entanglement Wedge Reconstruction and the Information Paradox,''
  arXiv:1905.08255 [hep-th].


\bibitem{Lewkowycz:2013nqa}
A.~Lewkowycz and J.~Maldacena,
``Generalized gravitational entropy,''
JHEP \textbf{08} (2013), 090
[arXiv:1304.4926 [hep-th]].

\bibitem{Faulkner:2013ana}
T.~Faulkner, A.~Lewkowycz and J.~Maldacena,
``Quantum corrections to holographic entanglement entropy,''
JHEP \textbf{11} (2013), 074
[arXiv:1307.2892 [hep-th]].



\bibitem{Engelhardt:2014gca}
N.~Engelhardt and A.~C.~Wall,
``Quantum Extremal Surfaces: Holographic Entanglement Entropy beyond the Classical Regime,''
JHEP \textbf{01} (2015), 073
[arXiv:1408.3203 [hep-th]].





\bibitem{Almheiri:2019hni}
  A.~Almheiri, R.~Mahajan, J.~Maldacena and Y.~Zhao,
  ``The Page curve of Hawking radiation from semiclassical geometry,''
  JHEP {\bf 2003} (2020) 149
  [arXiv:1908.10996 [hep-th]].

\bibitem{Ryu:2006bv}
  S.~Ryu and T.~Takayanagi,
  ``Holographic derivation of entanglement entropy from AdS/CFT,''
  Phys.\ Rev.\ Lett.\  {\bf 96} (2006) 181602
  [hep-th/0603001].
  
\bibitem{Hubeny:2007xt}
V.~E.~Hubeny, M.~Rangamani and T.~Takayanagi,
``A Covariant holographic entanglement entropy proposal,''
JHEP \textbf{07} (2007), 062
[arXiv:0705.0016 [hep-th]].
  
\bibitem{Bak:2003jk}
D.~Bak, M.~Gutperle and S.~Hirano,
``A Dilatonic deformation of AdS(5) and its field theory dual,''
JHEP \textbf{05} (2003), 072
[arXiv:hep-th/0304129 [hep-th]].
  

\bibitem{Bak:2011ga}
  D.~Bak, M.~Gutperle and R.~A.~Janik,
  ``Janus Black Holes,''
  JHEP {\bf 1110} (2011) 056
  [arXiv:1109.2736 [hep-th]].
  
\bibitem{Calabrese:2004eu}
P.~Calabrese and J.~L.~Cardy,
``Entanglement entropy and quantum field theory,''
J. Stat. Mech. \textbf{0406} (2004), P06002
[arXiv:hep-th/0405152 [hep-th]].
  

\bibitem{Rozali:2019day}
M.~Rozali, J.~Sully, M.~Van Raamsdonk, C.~Waddell and D.~Wakeham,
``Information radiation in BCFT models of black holes,''
JHEP \textbf{05} (2020), 004
[arXiv:1910.12836 [hep-th]].
  
\bibitem{Balasubramanian:2020hfs}
V.~Balasubramanian, A.~Kar, O.~Parrikar, G.~Sárosi and T.~Ugajin,
``Geometric secret sharing in a model of Hawking radiation,''
[arXiv:2003.05448 [hep-th]].

  
\bibitem{Sully:2020pza}
J.~Sully, M.~Van Raamsdonk and D.~Wakeham,
``BCFT entanglement entropy at large central charge and the black hole interior,''
[arXiv:2004.13088 [hep-th]].

\bibitem{Geng:2020qvw}
H.~Geng and A.~Karch,
``Massive islands,''
JHEP \textbf{09}, 121 (2020)
[arXiv:2006.02438 [hep-th]].

\bibitem{Chen:2020uac}
  H.~Z.~Chen, R.~C.~Myers, D.~Neuenfeld, I.~A.~Reyes and J.~Sandor,
  ``Quantum Extremal Islands Made Easy, Part I: Entanglement on the Brane,''
  arXiv:2006.04851 [hep-th].




\bibitem{Maldacena:2016upp}
J.~Maldacena, D.~Stanford and Z.~Yang,
``Conformal symmetry and its breaking in two dimensional Nearly Anti-de-Sitter space,''
PTEP \textbf{2016} (2016) no.12, 12C104
[arXiv:1606.01857 [hep-th]].

\bibitem{Almheiri:2014cka}
A.~Almheiri and J.~Polchinski,
``Models of AdS$_{2}$ backreaction and holography,''
JHEP \textbf{11} (2015), 014
[arXiv:1402.6334 [hep-th]].



\bibitem{Bak:2007jm}
  D.~Bak, M.~Gutperle and S.~Hirano,
  ``Three dimensional Janus and time-dependent black holes,''
  JHEP {\bf 0702} (2007) 068
  [hep-th/0701108].

\bibitem{Banados:1992wn}
M.~Banados, C.~Teitelboim and J.~Zanelli,
``The Black hole in three-dimensional space-time,''
Phys. Rev. Lett. \textbf{69} (1992), 1849-1851
[arXiv:hep-th/9204099 [hep-th]].


\bibitem{Hartman:2013qma}
  T.~Hartman and J.~Maldacena,
  ``Time Evolution of Entanglement Entropy from Black Hole Interiors,''
  JHEP {\bf 1305} (2013) 014
  [arXiv:1303.1080 [hep-th]].


  
\bibitem{Affleck:1991tk}
I.~Affleck and A.~W.~Ludwig,
``Universal noninteger 'ground state degeneracy' in critical quantum systems,''
Phys. Rev. Lett. \textbf{67}, 161-164 (1991)
  
  
  
  

\bibitem{Maldacena:2001kr}
  J.~M.~Maldacena,
  ``Eternal black holes in anti-de Sitter,''
  JHEP {\bf 0304} (2003) 021
  [hep-th/0106112].



\bibitem{Almheiri:2019yqk}
  A.~Almheiri, R.~Mahajan and J.~Maldacena,
  ``Islands outside the horizon,''
  arXiv:1910.11077 [hep-th].


\bibitem{Almheiri:2019qdq}
  A.~Almheiri, T.~Hartman, J.~Maldacena, E.~Shaghoulian and A.~Tajdini,
  ``Replica Wormholes and the Entropy of Hawking Radiation,''
  JHEP {\bf 2005} (2020) 013
  [arXiv:1911.12333 [hep-th]].

\bibitem{Penington:2019kki}
G.~Penington, S.~H.~Shenker, D.~Stanford and Z.~Yang,
``Replica wormholes and the black hole interior,''
[arXiv:1911.11977 [hep-th]].




\bibitem{Liu:2020gnp}
  H.~Liu and S.~Vardhan,
  ``A dynamical mechanism for the Page curve from quantum chaos,''
  arXiv:2002.05734 [hep-th].

\bibitem{Randall:1999vf}
L.~Randall and R.~Sundrum,
``An Alternative to compactification,''
Phys. Rev. Lett. \textbf{83} (1999), 4690-4693
[arXiv:hep-th/9906064 [hep-th]].



\bibitem{Hayden}
Hayden, P., Jozsa, R., Petz, D. et al. 
``Structure of States Which Satisfy Strong Subadditivity of Quantum Entropy with Equality'', Commun. Math. Phys. 246, 359–374 (2004). 



\bibitem{Caputa:2015waa}
  P.~Caputa, J.~Simón, A.~Štikonas, T.~Takayanagi and K.~Watanabe,
  ``Scrambling time from local perturbations of the eternal BTZ black hole,''
  JHEP {\bf 1508} (2015) 011
  [arXiv:1503.08161 [hep-th]].


\bibitem{Faulkner:2010jy}
  T.~Faulkner, H.~Liu and M.~Rangamani,
  ``Integrating out geometry: Holographic Wilsonian RG and the membrane paradigm,''
  JHEP {\bf 1108} (2011) 051
  [arXiv:1010.4036 [hep-th]].


\bibitem{Chiodaroli:2016jod}
  M.~Chiodaroli, J.~Estes and Y.~Korovin,
  ``Holographic two-point functions for Janus interfaces in the $D1/D5$ CFT,''
  JHEP {\bf 1704} (2017) 145
  [arXiv:1612.08916 [hep-th]].


\bibitem{Mazac:2018biw}
  D.~Mazáč, L.~Rastelli and X.~Zhou,
  ``An analytic approach to BCFT$_{d}$,''
  JHEP {\bf 1912} (2019) 004
  [arXiv:1812.09314 [hep-th]].







\bibitem{Alishahiha:2020qza}
M.~Alishahiha, A.~Faraji Astaneh and A.~Naseh,
``Island in the Presence of Higher Derivative Terms,''
[arXiv:2005.08715 [hep-th]].


\bibitem{Krishnan:2020oun}
C.~Krishnan, V.~Patil and J.~Pereira,
``Page Curve and the Information Paradox in Flat Space,''
[arXiv:2005.02993 [hep-th]].


\bibitem{NIST} 
  Frank W. Olver, Daniel W. Lozier, Ronald F. Boisvert, Charles W. Clark,
  ``NIST Handbook of Mathematical Functions,'' Cambridge University Press, New York, NY,
  2010.



\bibitem{Ioannis note} I. Papadimitriou, `Janus effective action', unpublished note.



\bibitem{Papadimitriou:2016yit}
I.~Papadimitriou,
``Lectures on Holographic Renormalization,''
Springer Proc. Phys. \textbf{176} (2016), 131-181.



\end{thebibliography}
\end{document}